\documentclass[a4paper,12pt, epsfig]{article}
\def\letter{0}\def\pr{0}
\usepackage{epsfig}
\usepackage{epstopdf}
\usepackage{textcomp}
\usepackage{graphicx}
\usepackage{ifthen}
\usepackage{ulem}

\usepackage{amsmath}
\usepackage[psamsfonts]{amssymb}
\usepackage{euscript}

\usepackage{latexsym}
\usepackage[arrow,matrix,curve]{xy}

\usepackage[hypertexnames=false]{hyperref}
\hypersetup{
    colorlinks=true,
    linkcolor=blue,
    filecolor=magenta,   
    citecolor=black,   
    urlcolor=cyan,
    }

\newskip\humongous \humongous=0pt plus 1000pt minus 1000pt

\newif\ifdtup

\def\,{\hspace{-.1cm}}
\def\hsp{,\hspace{.7cm}}

\def\fc#1#2 {\frac{n}{q}#1\frac{n}{q}#2}

\def\kt{\kappa}
\def\kt{\mathfrak{K}}

\def\hp{H^{\prime(t)}}

\newcommand{\vac}{\ensuremath{|0\rangle}}

\renewcommand{\tanh}{\textrm{tanh}}
\newcommand{\sech}{\textrm{sech}}
\newcommand{\csch}{\textrm{csch}}

\def\exp#1{\hbox{\rm exp}\left[#1\right]}

\renewcommand{\theequation}{\arabic{section}.\arabic{equation}}
\renewcommand{\(}{\begin{equation}}
\renewcommand{\)}{end{equation} \vspace{-.05in}\linebreak}

\newcounter{saveeqn}
\newcounter{savealpheqn}

\newcommand{\alpheqn}{\setcounter{saveeqn}{\value{equation}}%
  \stepcounter{saveeqn}\setcounter{equation}{0}%
  \renewcommand{\theequation}{\mbox{\arabic{section}.\arabic{saveeqn}
\alph{equation}}}
  \renewcommand{\)}{\end{equation}}}
\def\part#1{\frac{\partial}{\partial{#1}}}%
\def\group#1{\refstepcounter{equation}\setcounter{saveeqn}
 {\value{equation}}%
  \label{#1}\setcounter{equation}{0}%
\renewcommand{\theequation}{\mbox{\arabic{section}.\arabic{saveeqn}
\alph{equation}}}
  \renewcommand{\)}{\end{equation}}}
\newcommand{\reseteqn}{\setcounter{equation}{\value{saveeqn}}%
  \renewcommand{\theequation}{\arabic{section}.\arabic{equation}}%
  \renewcommand{\)}{\end{equation}}}

\newcommand{\aalpheqn}{\setcounter{saveeqn}{\value{equation}}%
  \stepcounter{saveeqn}\setcounter{equation}{0}%
  \renewcommand{\theequation}{\mbox{
        \Alph{subsection}.\arabic{saveeqn}\alph{equation}}}
   \renewcommand{\)}{\end{equation}}}
\newcommand{\areseteqn}{\setcounter{equation}{\value{saveeqn}}%
  \renewcommand{\theequation}{\Alph{subsection}.\arabic{equation}}%
  \renewcommand{\)}{\end{equation}}}

\renewcommand{\thefootnote}{\alph{footnote}}
\renewcommand{\(}{\begin{equation}}
\renewcommand{\)}{\end{equation}}
\newcommand{\ba}{\begin{eqnarray}}
\newcommand{\ea}{\end{eqnarray}}

\renewcommand{\b}{\beta}

\renewcommand{\sl}{{\sqrt{\lambda}}}

\newcommand{\cbp}{\mathop{\vtop{\ialign{##\crcr
   $\hfil\displaystyle{}\hfil$\crcr\noalign{\kern-13pt\nointerlineskip}
   \BIG{)}\hskip0pt\crcr\noalign{\kern3pt}}}}}
\newcommand{\pa}{\mathop{\vtop{\ialign{##\crcr

$\hfil\displaystyle{\oplus}\hfil$\crcr\noalign{\kern+1pt\nointerlineskip
}
   \hspace{.08in}$^{\alpha=0}$\hskip6pt\crcr\noalign{\kern3pt}}}}}
\renewcommand{\hsp}{,\hspace{.3in}}
\newcommand{\p}{^\prime}
\newcommand{\pp}{^{\prime\prime}}

\catcode`\@=11
\def\vereq#1#2{\lower3pt\vbox{\baselineskip1.5pt \lineskip1.5pt
\ialign{$\m@th#1\hfill##\hfil$\crcr#2\crcr\sim\crcr}}}
\catcode`\@=12

\renewcommand{\(}{\begin{equation}}
\renewcommand{\)}{\end{equation}}

\def\vx{{\vec{x}}}
\def\vp{{\vec{p}}}

\def\vk{{\vec{k}}}

\def\pin#1{\int \frac{d#1}{2\pi}}
\def\ppin#1{\int\hspace{-17pt}\sum \frac{d#1}{2\pi}}
\def\ppink#1{\int\hspace{-17pt}\sum\frac{d^{#1} k}{(2\pi)^{#1}}}

\def\ppinkp#1{\int\hspace{-17pt}\sum\frac{d^{#1}k\p}{(2\pi)^{#1}}}
\def\dint{\int\hspace{-12pt}\sum }
\def\pink#1{\int \frac{d^{#1}k}{(2\pi)^{#1}}}

\def\kinv#1#2{\int\hspace{-17pt}\sum \frac{d^{#1}\vec{#2}}{(2\pi)^{#1}}}
\def\pinv#1#2{\int \frac{d^{#1}\vec{#2}}{(2\pi)^{#1}}}

\def\Bd#1{B^\ddag_{k_{#1}}}

\def\cc{\mathcal{C}}
\def\df{\mathcal{D}_{f}}

\def\dx{\mathcal{D}_{x_0}}
\def\mx{\left(\frac{mx}{2}\right)}

\def\up{|\,\uparrow\rangle}
\def\dn{|\,\downarrow\rangle}
\def\upx#1{|\,\uparrow,#1\rangle}
\def\dnx#1{|\,\downarrow,#1\rangle}

\def\I{\mathcal{I}}

\def\red#1{\textcolor{red}{Jarah: #1}}

\usepackage[dvipsnames]{xcolor}

\newcommand{\beas}{\begin{eqnarray*}}
\newcommand{\eeas}{\end{eqnarray*}}

\newcommand{\bquo}{\begin{quote}}
\newcommand{\enqu}{\end{quote}}

\def\lim#1{\stackrel{\rm{lim}}{{}_{#1}}}

\newcommand{\g}{\mathfrak g}
\newcommand{\mm}{\mathfrak m}
\newcommand{\nn}{\mathfrak n}

\def\ch{{\mathcal{H}}}
\def\co{{\mathcal{O}}}

\def\ok#1{\omega_{k_{#1}}}
\def\okp#1{\omega_{k\p_{#1}}}

\def\nohat#1{#1}
\def\V#1{V^{(#1)}(\sqrt{\lambda}f(x))}

\def\ck{\csch\left(\frac{\pi k}{2\beta}\right)}
\def\ckm{\csch\left(\frac{\pi k}{m}\right)}

\def\gt{\tilde{\g}}

\def\uu{{\uparrow\uparrow}}
\def\ud{{\uparrow\downarrow}}
\def\du{{\downarrow\uparrow}}
\def\dd{{\downarrow\downarrow}}
\def\ua{\uparrow}
\def\da{\downarrow}
\def\ga{A}
\def\sk{\sech\left(\frac{\pi k}{2\beta}\right)}
\def\skm{\sech\left(\frac{\pi k}{m}\right)}
\def\kkk{\sum_{i}^3k_i}
\def\csk{\csch\left(\frac{\pi\kkk}{m}\right)}

\def\hp{{\hat{\Phi}}}

\def\XXint#1#2#3{{\setbox0=\hbox{$#1{#2#3}{\int}$}
     \vcenter{\hbox{$#2#3$}}\kern-.5\wd0}}

\newcommand{\beq}{\begin{equation}}
\newcommand{\eeq}{\end{equation}}
\newcommand{\bea}{\begin{eqnarray}}
\newcommand{\eea}{\end{eqnarray}}

\newskip\humongous \humongous=0pt plus 1000pt minus 1000pt

\newif\ifdtup

\catcode`\@=11

\ifthenelse{\equal{\letter}{0}}{ 

\@addtoreset{equation}{section}
\def\theequation{\arabic{section}.\arabic{equation}}

\def\@normalsize{\@setsize\normalsize{15pt}\xiipt\@xiipt
\abovedisplayskip 14pt plus3pt minus3pt%
\belowdisplayskip \abovedisplayskip
\abovedisplayshortskip \z@ plus3pt%
\belowdisplayshortskip 7pt plus3.5pt minus0pt}

\def\small{\@setsize\small{13.6pt}\xipt\@xipt
\abovedisplayskip 13pt plus3pt minus3pt%
\belowdisplayskip \abovedisplayskip
\abovedisplayshortskip \z@ plus3pt%
\belowdisplayshortskip 7pt plus3.5pt minus0pt
\def\@listi{\parsep 4.5pt plus 2pt minus 1pt
      \itemsep \parsep
      \topsep 9pt plus 3pt minus 3pt}}

\relax

\def\section{\@startsection{section}{1}{\z@}{3.5ex plus 1ex minus  .2ex}{2.3ex plus .2ex}{\large\bf}}

\def\thesection{\arabic{section}}
\def\thesubsection{\arabic{section}.\arabic{subsection}}

\def\appendix{\setcounter{section}{0}
 \def\thesection{
 \Alph{section}}
 \def\thesubsection{\Alph{section}.\arabic{subsection}}
 \def\theequation{\Alph{section}.\arabic{equation}}}
\renewcommand{\theequation}{\arabic{section}.\arabic{equation}}

}{
\renewcommand{\theequation}{\arabic{equation}}

} 

\begin{document}
\def\thefootnote{\fnsymbol{footnote}}
\def\thetitle{Krakow Lectures on Scalar Quantum Solitons}
\def\autone{Jarah Evslin}
\def\autthree{Hui Liu}
\def\autfour{Baiyang Zhang}
\def\auttwo{Hengyuan Guo}
\def\affd{Yerevan Physics Institute, 2 Alikhanyan Brothers St., Yerevan, 0036, Armenia}
\def\affb{University of the Chinese Academy of Sciences, YuQuanLu 19A, Beijing 100049, China}
\def\affa{Institute of Modern Physics, NanChangLu 509, Lanzhou 730000, China}
\def\affc{School of Physics and Astronomy, Sun Yat-sen University, Zhuhai 519082, China}
\def\affe{Institute of Contemporary Mathematics, School of Mathematics and Statistics, Henan University, Kaifeng, Henan 475004, P. R. China}


\ifthenelse{\equal{\pr}{1}}{
\title{\thetitle}
\author{\autone}
\author{\auttwo}
\author{\autthree}
\affiliation {\affa}
\affiliation {\affb}
\affiliation {\affc}

}{

\begin{center}
{\large {\bf \thetitle}}

\bigskip

\bigskip


{\large 
\noindent  \autone{${}^{12}$}
\footnote{jarah@impcas.ac.cn}
}


\vskip.7cm

1) \affa\\
2) \affb\\

\end{center}

}

\begin{abstract}
\noindent
We give a pedagogical introduction to Linearized Soliton Perturbation Theory (LSPT), a new and efficient tool for calculations involving quantum solitons.  It is a Hamiltonian approach with a focus on explicitly constructing the soliton states.  These states are squeezed, coherent states plus perturbative corrections.  We will describe multi-loop corrections to states and their masses.  An inner product suitable for non-normalizable momentum eigenstates will be introduced and applied to kink-meson scattering.  We will also discuss domain wall solitons.

\end{abstract}


%
\setcounter{footnote}{0}
\renewcommand{\thefootnote}{\arabic{footnote}}

\ifthenelse{\equal{\pr}{1}}
{
\maketitle
}{}

 \tableofcontents
 
\section{Solitons}

\subsection{Classical Solitons}

Consider a classical field theory with scalar fields $\phi_i(\vx,t)$ whose dual momenta are $\pi_i(\vx,t)$.  Let the theory be described by the classical Hamiltonian density
\beq
\ch=\sum_i\left[\frac{\pi_i^2+\left|\nabla \phi_i\right|^2}{2}\right]
+\frac{V(\sl\vec\phi)}{\lambda}.
\eeq
Here the coupling, $\lambda$, is a parameter. The notation $\vec{\phi}$ denotes the collection of all fields $\phi_i$, and $V$ is a potential which depends on some or all of these fields.  

We will refer to the minima of the potential as vacua.  Expanding the potential about these minima, there will be small excitations that sit close to each minimum.  These perturbative excitations, when quantized, are called mesons.  They remain in the spectrum even in the limit $\lambda\rightarrow 0$, in other words, in the free theory.

These lectures will be concerned with another kind of excitation, which explores the field space $\vec{\phi}$ far from the vacua and disappears from the spectrum as $\lambda\rightarrow 0$.  There are actually several such kinds of solutions.  Some, such as oscillons are necessarily time-dependent, and  surely but slowly they decay into radiation.  Others, such as Q-balls, are time-dependent but only leak their energy during interactions.  Yet others, such as breathers, are time-dependent but leak no energy at all. We will, instead, be interested in solitons whose lowest energy state is time-independent and absolutely stable.

The word {\it{soliton}} refers to a classical solution of a classical field theory.  There are many conventions in the literature for just which solutions may be called solitons.  Following the common convention in high energy physics, we will refer to a solution as a soliton if it has the following properties:
\begin{enumerate}
  \item \hypertarget{first}{Localized: At least in some directions the fields should tend to vacua.}
  \item \hypertarget{second}{Stable: It should remain localized forever, and its mass or tension should be bounded from below by some positive value.} 
  \item \hypertarget{third}{Intrinsically nonlinear: It should disappear from the spectrum as $\lambda\rightarrow 0$.}
\end{enumerate}
Solitons first appeared in hydrodynamics, as stable waves.  However they now also play important roles in optics, condensed matter physics, astrophysics and particle physics, besides being a mobile excitation in the conducting plastic polyacetylene~\cite{poly} that won the Nobel Prize in Chemistry in 2000.

\subsection{Quantum Solitons}

Classical field theory is an approximation to quantum field theory.  And so what becomes of the solitons when lifted to quantum field theory?  In quantum field theory, in addition to the dimensionful constants $\lambda$ and $m$, there is an additional parameter $\hbar$.  If there are $d$ spatial dimensions, then the combination $\lambda\hbar/m^{3-d}$ is dimensionless.  Treating solitons in quantum field theory is difficult, as the main tool in quantum field theory is perturbation theory, which is an expansion in $\lambda\hbar/m^{3-d}$ is about zero.  However, at $\lambda=0$ there are no solitons.

Nonetheless (setting now $\hbar$=1 and ignoring powers of $m$), it turns out the observables related to solitons can be calculated as a series in $\lambda$, albeit often a series that starts at order $O(1/\lambda)$ and so is only defined for $\lambda>0$.  We will always consider limits $\lambda\rightarrow 0^+$ such that $\lambda$ is smaller than other scales in the problem except for those related to the initial conditions and the dimensions of the system.

The key to such calculations is a conjecture made by Skyrme \cite{skyrme} that solitons in classical field theory correspond to states in quantum field theory.  The correspondence is imperfect.  There are classical solutions with no quantum lift or multiple quantum lifts \cite{wclas}, and there are quantum states that are similar to such lifts but actually do not correspond to any classical solution \cite{del18,dav19}.  However, there is a systematic semiclassical expansion in which a quantum state corresponding to a soliton may be constructed from the classical solution together with a series of quantum corrections arising from loops\footnote{This is not to say that the corrections are always small.  For example, while classical solitons break translation-symmetry, the symmetry is restored by quantum solitons in their ground state.}. 

In fact, there are many such expansions.  At one-loop, there are simple and powerful formalisms such as spectral methods \cite{sal11,gw22} and the classical-quantum correspondence \cite{cq1,cq2}.  If one is interested in arbitrary numbers of loops, then over the past half century the main tool has been the collective coordinate approach of Refs.~\cite{gs74,cl75,tom75}.  Here, one elevates the position $\vec{x}_0$ of the soliton, as well as any other variables which do not affect its classical energy, to operators called collective coordinates.  The commutation relations obeyed by the collective coordinates are quite complicated, and so one obtains a quite complicated and sometimes subtle \cite{gj76,bh24,ga24} formalism which has therefore had limited applications over the years, although recently it has led to some interesting results \cite{ito23,andy24}.

The present review will instead focus on a new formalism \cite{mekink,me2loop} called Linearized Soliton Perturbation Theory (LSPT).  This begins with the observation that the collective coordinate $\vec{x}_0$ can be expanded in terms of the displacement of the soliton from a base point, and that the linear term $\phi_0$ in this expansion in fact does satisfy canonical commutation relations.  This allows for an expansion of all quantities in terms of operators, including $\phi_0$, with simple commutation relations.  It turns out that only a finite number of terms appear at each power of the coupling $\lambda$.  

This expansion in the displacement necessarily sacrifices manifest translation-invariance, as one has chosen a base point in moduli space.  However it provides an accurate description of states localized near the base point $\vec{x}_0$, and imposing translation-invariance order by order allows us to efficiently recover translation-invariance for the entire state.  In fact, imposing translation-invariance order by order, instead of complicating matters, will greatly simplify a number of calculations, as it removes most of the Hilbert space.

We will see in this review that the quantum states corresponding to solitons are squeezed, coherent states plus perturbative corrections.  The {\it{coherent}} part is a nonperturbative correction, corresponding to the action of a displacement operator which is an exponential of the inverse coupling $e^{-1/\sl}$.  Its role is to make the expectation values of the fields nearly equal to the classical solution.  The {\it{squeeze}} is also nonperturbative, but is independent of the coupling and so order $O(\lambda^0)$.  We will see that its role is to de-excite the normal modes.  The perturbative corrections are all suppressed by positive powers of $\sl$.

\section{Quantum Mechanics - The Displacement Operator} \label{qmsez}

The critical ingredient in our construction will be the displacement operator $\mathcal{D}$.  It contains the most nonperturbative information about the state, and yet can be determined entirely from the classical solution.  This decomposition of the state into a nonperturbative but known piece and a largely perturbative but unknown piece will allow us to treat quantum solitons using pertubation theory.  Roughly speaking, the displacement operator allows us to factor out the nonperturbative part of the state.

\subsection{The Quantum Harmonic Oscillator}

All physics lectures begin by solving the quantum harmonic oscillator \cite{cho,h25}
\beq
H\p_2=\frac{p^2}{2}+\frac{\omega^2x^2}{2}\hsp [x,p]=i. \label{qhoh}
\eeq
Here $x$ and $p$ are two operators which satisfy the canonical commutation relations
.  Together with the unit, which we will ignore, they are a basis of all operators in the model.

However, there is another basis, the operators
\beq
a=\frac{\omega x+ip}{\sqrt{2\omega}}\hsp a^\dag=\frac{\omega x-ip}{\sqrt{2\omega}}\hsp [a,a^\dag]=1. \label{qhob}
\eeq
Any operator in the theory can be expressed in terms of $x$ and $p$, and it can alternatively be expressed in terms of $a$ and $a^\dag$.  

For example, in Eq.~(\ref{qhoh}) the Hamiltonian is expressed in the $\{x,p\}$ basis.  However, in the $\{a,a^\dag\}$ basis it is
\beq
H\p_2=\omega a^\dag a+\frac{\omega}{2}.
\eeq
This form of the Hamiltonian is convenient because one can easily solve the eigenvalue problem
\beq
H\p_2\Psi_n=E^{(0)}_n\Psi_n. \label{qhose}
\eeq
As in any quantum theory, the eigenvectors $\Psi_n$ of the Hamiltonian are the time-independent states, and their eigenvalues $E_n^{(0)}$ are the corresponding energies.

As the eigenvalues of $a^\dag a$ are real and nonnegative, the ground state is the state with eigenvalue equal to zero.  This is the state $\vac_0$ which is annihilated by $a$
\beq
a\vac_0=0.
\eeq

Alternately, we may write this state in the Schrodinger representation as a wavefunction.  If we define the basis of states $|y\rangle$ by 
\beq
x|y\rangle=y|y\rangle
\eeq
where $y$ runs over the real numbers, then any state $|\tilde\psi\rangle$ in the theory may be written
\beq
|\tilde\psi\rangle=\int dy \tilde\psi(y)|y\rangle
\eeq
for some function $\tilde\psi(y)$.  Then $x$ and $p$ act on $|\tilde\psi\rangle$ as
\beq
x|\tilde\psi\rangle=\int dy y\tilde\psi(y)|y\rangle\hsp
p|\tilde\psi\rangle=-i \int dy \partial_y \tilde\psi(y)|y\rangle. \label{qhoaz}
\eeq
Therefore $a$ acts as
\beq
a|\tilde\psi\rangle=\frac{1}{\sqrt{2\omega}}\int dy \left(\omega y\tilde\psi(y)+\partial_y\tilde\psi(y)\right)|y\rangle.
\eeq
The wavefunction $\tilde\psi_0(y)$ corresponding to the state $\vac_0$ then satisfies
\beq
\omega y \tilde\psi_0(y)+\partial_y\tilde\psi_0(y)=0
\eeq
identifying it, up to a multiplicative constant, as
\beq
\tilde\psi_0(y)=e^{-\frac{\omega}{2}y^2}\hsp \vac_0=\int dy \tilde\psi_0(y)|y\rangle.
\eeq

The commutation relation
\beq
[H\p_2,a^\dag]=\omega a^\dag
\eeq
allows us to find the rest of the spectrum.  The $n$th excited state is
\beq
|n\rangle_0=a^{\dagger n}\vac_0 \hsp H\p_2|n\rangle_0=\left(n+\frac{1}{2}\right)\omega |n\rangle_0.
\eeq
These are a basis of all finite energy states.  To write them in Schrodinger form is conceptually easy, as we know how $a^\dag$ is written in terms of $x$ and $p$ (\ref{qhob}) and we know how $x$ and $p$ act on states (\ref{qhoaz}).

\subsection{Perturbation Theory}

Let us deform the potential by including an interaction $H\p_I$ into our Hamiltonian
\beq
H\p=H\p_2+H\p_I.
\eeq
For example, we might consider a quartic interaction
\beq
H\p_I=\lambda x^4
\eeq
or a more general polynomial $V$
\beq
H\p_I=\frac{V(\sl x)}{\lambda}\hsp V(0)=V\p(0)=V\pp(0)=0. \label{genpoly}
\eeq
Here the coupling constant $\lambda$ parametrizes the perturbation so that $H\p_I$ vanishes when $\lambda=0$.

We will decompose $H\p$ by the powers of the operators $x$ and $p$ in each monomial
\beq
H\p=\sum_{n=2}^\infty H\p_n \label{hdec}
\eeq
where $H\p_n$ is $n$th order in $x$ and $p$ and so of order $O(\lambda^{n/2-1})$.  Explicitly the interaction terms are
\beq
H\p_{n>2}=\frac{\lambda^{\frac{n}{2}-1}}{n!}V^{(n)}(0)x^n
\eeq
where the notation $V^{(n)}$ denotes the $n$th derivative of $V$ with respect to its argument.

For each eigenvector $|n\rangle_0$ of $H\p_2$, we can write an eigenvector $|n\rangle$ of $H\p$
\beq
H\p|n\rangle=E_n|n\rangle \label{qms}
\eeq
such that, at leading order, $|n\rangle$ is just $|n\rangle_0$.  More precisely, we will decompose $|n\rangle$ and $E_n$ in powers of the coupling
\beq
|n\rangle=\sum_{i=0}^\infty |n\rangle_i\hsp E_n=\sum_{i=0}^\infty E_n^{(i)} \label{ndec}
\eeq
where $|n\rangle_i$ is of order $O(\lambda^{i/2})$ and $E_n^{(i)}$ is of order $O(\lambda^i)$, where we have used the usual fact that perturbation theory only leads to energy corrections at even orders\footnote{This is a result of fact at $x$ has an odd number of raising and lowering operators, and so an even number of powers of $x$ are needed to take a state to itself, and so to contribute to the energy.}.  The case $i=0$ of course is just that of the quantum harmonic oscillator.

Substituting the decompositions (\ref{hdec}) and (\ref{ndec}) into the eigenvalue equation (\ref{qms}) one arrives at a series of equations, one at each order in the coupling.  The leading order equation is just (\ref{qhose}) which we have already solved.  

For example, at order $O(\sl)$ there is no contribution to $E_n$ and so one finds
\beq
H\p_3|n\rangle_0+\left(H\p_2-E_n^{(0)}\right)|n\rangle_1=0.
\eeq
The inverse of $H\p_2-E_n^{(0)}$ is unique up to terms in its kernel, which consists of $|n\rangle_0$.  Therefore $|n\rangle_1$ is fixed up to a term proportional to $|n\rangle_0$.  This ambiguity is just the choice of normalization of the state, which we can fixed as desired.  We can proceed similarly, to find $|n\rangle_i$ at each order.

Note that our perturbation theory in powers of $\lambda$ is also a perturbation theory in powers of $x$, indeed we have noted that $H\p_i$ is proportional to $x^i$.  Therefore, at each order we arrive at corrections which are higher powers of $x$.  Overall the wavefunctions nonetheless remain bounded, as these polynomials are multiplied by $\tilde\psi_0(x)$ which falls as $e^{-x^2}$ at large $x$.

\subsection{The Shifted Harmonic Oscillator}

Now let us consider another theory, the shifted harmonic oscillator, described by the Hamiltonian
\beq
\nohat{H}_2=\frac{p^2}{2}+\frac{\omega^2(x-x_0)^2}{2}\hsp [x,p]=i. \label{qhos}
\eeq
We will not assume $x_0$ to be in any sense small, in fact we will be ultimately interested in the case in which it is of order $O(1/\sl)$, which would be the generic result of a potential $V(\sl x)$ of the kind presented in Eq.~(\ref{genpoly}) expanded about a minimum.  How can we find the spectrum of $\nohat{H}_2$?  These are the states $|\nohat{n}\rangle^S_0$ that satisfy
\beq
\nohat{H}_2|\nohat{n}\rangle^S_0=E_{\nohat{n}}^{S(0)}|\nohat{n}\rangle^S_0. \label{npse}
\eeq

Our perturbative expansion is a moment expansion in $x$.  Here, the $i$th moment will be of order $x_0^i$, and so the moments, if anything, get larger at higher orders.  This means that a perturbative approach, expanding the corresponding wavefunctions $\psi_{\nohat{n}}$
\beq
|n\rangle_0^S=\int dx \psi_n(x) |x\rangle
\eeq
in powers of $x$, will be a poor approximation.

The expansion in the moments of $x$ fails because, at each finite $n$, the corresponding wavefunction $\psi_{\nohat{n}}$ is genuinely nonperturbative.  The key to our approach is that the nonperturbative part can be read off of the Hamiltonian and is fixed, and so can be factored out.  

This can be done with the unitary displacement operator
\beq
\dx=e^{-ipx_0}.
\eeq
We recall that $p$ is an operator but $x_0$ is a $c$-number, appearing as a parameter in the Hamiltonian.  It is a simple exercise to show that the displacement operator $\dx$ commutes with $p$ and shifts $x$
\beq
\dx^\dag x\dx=x+x_0.
\eeq
More generally, let $F(x,p)$ be an operator.  Like all operators in the theory, it is constructed from our basis of operators $x$ and $p$.  Then the displacement operator acts on it as
\beq
\dx^\dag F(x,p)\dx= F(x+x_0,p). \label{dxs}
\eeq
Our Hamiltonian is such an operator, although we have left the $x$ and $p$ dependence implicit to avoid clutter.  The property (\ref{dxs}) then implies
\beq
\dx^\dag\nohat{H}_2\dx=H\p_2. \label{qhohp}
\eeq

We conclude that our Hamiltonian $\nohat{H}_2$ is unitarily equivalent to the quantum harmonic oscillator Hamiltonian $H\p_2$.  How does this help us?

Let us consider an eigenstate $|n\rangle_0$ of $H\p_2$.  Then we have the following sequence of equalities
\beq
\nohat{H}_2\dx|n\rangle_0=\dx H\p_2|n\rangle_0=\dx E_n^{(0)} |n\rangle_0=E_n^{(0)}\dx|n\rangle_0.
\eeq
In other words, we learn that $\dx|n\rangle_0$ is an eigenstate of $\nohat{H}_2$ and the eigenvalue is $E_n^{(0)}$.  This implies that
\beq
|\nohat{n}\rangle^S_0=\dx |n\rangle_0\hsp E_{\nohat{n}}^{S(0)}=E_{n}^{(0)}.
\eeq
We have learned that our nonperturbative state $|\nohat{n}\rangle^S_0$ can be decomposed into a perturbative part $|n\rangle_0$ and a nonperturbative but known operator $\dx$.  The state $|n\rangle_0$ is perturbative because the moments of its wavefunction are all small.  The operator $\dx$ is nonperturbative because it has $x_0$, which is large and in general proportional to the inverse coupling, in the exponential.  

The displacement operator therefore simplifies the problem of finding the spectrum of time-independent states, reducing the nonperturbative problem (\ref{npse}) to the perturbative problem (\ref{qhose}).

\subsection{Amplitudes}

Besides the spectral problem, there is another problem of interest in physics, the initial value problem.  Here one starts with a state $|\nohat\psi\rangle^S$, evolves it in time from time $0$ to time $t$, and then computes the amplitude for it to be the state $|\nohat\phi\rangle^S$.  This amplitude can be written in the Schrodinger picture as
\beq
A={}^S\langle\nohat\phi|e^{-i\nohat{H}_2t}|\nohat\psi\rangle^S.
\eeq
Again this is nonperturbative, because $\nohat{H}_2$ contains $x_0$ and it is exponentiated.  

However, as a result of (\ref{dxs}) applied to the evolution operator $e^{-i\nohat{H}_2t}$ we find
\beq
\dx^\dag e^{-i\nohat{H}_2t}\dx=e^{-iH\p_2t}.
\eeq
Let us define
\beq
|\nohat\phi\rangle^S=\dx|\phi\rangle\hsp |\nohat\psi\rangle^S=\dx|\psi\rangle
\eeq
for some $|\phi\rangle$ and $|\psi\rangle$.  In fact, if $|\nohat{\phi}\rangle^S$ and $|\nohat{\psi}\rangle^S$ are localized close to the minimum of the potential, so that they are linear combinations of $|\nohat{n}\rangle^S_0$ for finite $n$ at large $x_0$, then $|\psi\rangle$ and $|\phi\rangle$ will similarly be linear combinations of $|\nohat{n}\rangle_0$ for finite $n$.  
In short, $|\phi\rangle$ and $|\psi\rangle$ can be constructed perturbatively, with moments that remain finite at large $x_0$, unlike $|\nohat{\phi}\rangle^S$ and $|\nohat{\psi}\rangle^S.$

With these identities in hand, we may rewrite the amplitude as
\beq
A=\langle\phi|\dx^\dag e^{-i\nohat{H}_2t}\dx |\psi\rangle=\langle\phi|e^{-iH\p_2t}|\psi\rangle.
\eeq
In other words, to compute amplitudes, we can forget about the nonperturbative $\nohat{H}_2$ and about its nonperturbative eigenstates $|\nohat{n}\rangle^S_0$.  The amplitude can be calculated using $|\psi\rangle$ and $|\phi\rangle$ which in general will be constructed from  $H\p_2$ together with its eigenstates.

\subsection{Frames}

We have argued that we can calculate spectra and amplitudes using $|\phi\rangle$ and $|\psi\rangle$ to represent the states that one would usually represent by $|\nohat{\phi}\rangle^S$ and $|\nohat{\psi}\rangle^S$.  This trick can be formalized as follows.  

The abstract Hilbert space vector $|\nohat\psi\rangle^S$ is identified with the wavefunction $\nohat\psi(x)$ via the decomposition
\beq
|\nohat\psi\rangle^S=\int dx \nohat\psi(x)|x\rangle.
\eeq
The wavefunction $\nohat\psi(x)$ characterizes the physical state.  We refer to it using the abstract vector $|\nohat\psi\rangle^S$.  We call the identification between the ket $|\nohat\psi\rangle^S$ and the wavefunction $\nohat\psi(x)$ the defining frame.  It gives us a name for each state, just as a coordinate system gives a name to each point.  In the defining frame, time evolution is generated by the operator $\nohat{H}_2$.

However there is another identification that we have seen simplifies calculations.  We may decompose $|\psi\rangle$ as
\beq
|\psi\rangle=\dx^\dag|\nohat{\psi}\rangle^S=\int dx \nohat{\psi}(x)\dx^\dag |x\rangle=\int dx \nohat{\psi}(x) |x-x_0\rangle.
\eeq
This decomposition, identifies the ket $|\psi\rangle$ with the state $\nohat\psi(x)$.  We call this the shifted frame.  

The physical state is still that given by the wavefunction $\nohat\psi(x)$, for example it is still localized near $x=x_0$.  So we have not changed any physical properties.  We simply have given a new name, $|\psi\rangle$ to this state.  This is similar to a coordinate transformation in general relativity, in which we change the names of the points.  

Now while $|\psi\rangle$ in the shifted frame and $|\nohat{\psi}\rangle^S$ in the defining frame both describe the same state, they are nonetheless different vectors in the Hilbert space.  They are different because
\beq
|\nohat\psi\rangle^S=\dx|\psi\rangle.
\eeq
This means that, as is always the case for passive transformations, the operators acting on these states transform.  Whereas time translation in the defining frame was generated by $\nohat{H}_2$, in the shifted frame time translation is generated by $H\p_2$.  In fact, this is the reason that we have introduced the frames.  Now the calculation of amplitudes is reduced to a perturbative problem in which the time evolution operator is $e^{-iH\p_2t}$.

\begin{table}
\begin{center}
\begin{tabular}{|l|l|l|}
\hline
&Defining Frame&Shifted Frame\\
\hline\hline
wavefunction (state)&$\nohat\psi(x)$&$\nohat\psi(x)$\\
\hline
Hilbert space vector&$|\nohat\psi\rangle^S=\dx|\psi\rangle$&$|\psi\rangle$\\
\hline
time translation generator&$\nohat{H}_2=\dx  H\p_2 \dx^\dag$&$H\p_2$\\
\hline
time-independent states&$|\nohat{n}\rangle^S_0=\dx|n\rangle_0$&$|n\rangle_0$\\
\hline
amplitude&$\langle \phi|\dx^\dag e^{-iH_2 t}\dx|\psi\rangle=\langle \phi|e^{-iH\p_2t}|\psi\rangle$&$\langle \phi|e^{-iH\p_2t}|\psi\rangle$\\
\hline
\end{tabular}
\end{center}
\caption{Different objects are described in the two frames.  Physical quantities such as (up to phases) wavefunctions and amplitudes are frame-independent.  However, in the shifted frame, the intermediate objects (kets and operators) are simpler because they are those of the unshifted harmonic oscillator theory.} \label{tab}
\end{table}

We stress that one never needs to introduce frames.  One could continue to work in terms of $|\nohat{\psi}\rangle^S$.  It can be time-evolved in the defining frame using
\beq
e^{-i\nohat{H}_2t}|\nohat\psi\rangle^S=
e^{-i\nohat{H}_2t}\dx |\psi\rangle=\dx e^{-i H\p_2t}|\psi\rangle.
\eeq
Therefore time evolution always proceeds by acting $e^{-iH\p_2 t}$ on $|\psi\rangle$ whatever frame you use.  The advantage of using the shifted frame is simply that you avoid having to write a $\dx$ to the left of every expression for a state.

In summary, the change of frame is a passive unitary transformation on the Hilbert space which renames all vectors and conjugates all operators.  However physical quantities correspond to inner products of such vectors, and these are unchanged by the unitary rotation.  The situation is summarized in Table~\ref{tab}.

\subsection{A More Complicated Potential}

Similarly, we may consider the model
\beq
H=H_2+\frac{V(\sl(x-x_0))}{\lambda}.
\eeq
In this case the displacement operator yields $H\p$
\beq
H\p=\dx^\dag H\dx.
\eeq
The same arguments as in the case of $H_2$ now tell us that the eigenvectors of $H$ are $\dx|n\rangle$
\beq
H\dx|n\rangle=E_n\dx|n\rangle.
\eeq

Again, acting with $e^{-iH\p t}$ before the $\dx$  evolves the time from $0$ to $t$, in the sense that
\beq
e^{-iHt}\dx|\psi\rangle=\dx e^{-iH\p t}|\psi\rangle.
\eeq
In other words, while the state $\dx|\psi\rangle$ evolves, like all states in the theory, via the action of the evolution operator $e^{-iHt}$, this same evolution may be obtained by acting $e^{-iH\p t}$ on the ket $|\psi\rangle$.

Thus one may again define a shifted frame, in which one drops the $\dx$ on the left hand side of each state and one evolves time using $H\p$ instead of $H$.  As $H\p$ does not contain the nonperturbative quantity $x_0$, which we have taken to be of order $O(1/\sl)$, time evolution in the shifted frame may be computed perturbatively.   

\subsection{Summary}

Perturbative states $|\psi\rangle$ are those whose wavefunctions $\psi(x)$ have moments that do not diverge as the coupling $\lambda$ tends to zero.  Therefore the finite energy eigenstates $|n\rangle_0$ of the quantum harmonic oscillator $H\p_2$ and $|n\rangle$ of its perturbation $H\p$ are perturbative states, as are wavepackets made from superpositions of these states.

On the other hand, the finite energy eigenstates $|n\rangle^S_0$ and $|n\rangle^S$ of the shifted theories $H_2$ and $H$ are nonperturbative.  Indeed, their wavefunctions are localized around $x_0$, which diverges as the coupling $\lambda$ is taken to zero, and so their moments diverge as powers of $x_0$.

However these states are nonperturbative in a rather trivial way.  The nonperturbative part can be factored out, writing the states as
\beq
|n\rangle^S=\dx |n\rangle\hsp |n\rangle^S_0=\dx |n\rangle_0.
\eeq
In other words, they are constructed as a nonperturbative but known operator $\dx$ acting on a state that can be found in perturbation theory.  As a result, these naively nonperturbative theories can be handled in perturbation theory.  The price is that each state has a $\dx$ on the left.  However, to make equations more compact, one may omit this $\dx$ by working in the shifted frame of the Hilbert space.

The operator $\dx$ is a displacement operator and $|n\rangle^S$ is a coherent state.  We will see below that the case of quantum field theory contains one additional ingredient, the coherent states must also be squeezed.

\section{Quantum Kinks as Coherent States} \label{cosez}

In this section we will show that the construction presented above in quantum mechanics can also be applied to quantum field theory.  In the case of quantum mechanics, the operators were $x$ and $p$ and the shift was in $x$ by a constant $x_0$.  In the case of quantum field theory, the operators are $\phi(\vx)$ and $\pi(\vx)$ and the shift is in $\phi(\vx)$ by a function $f(\vx)$.  In other words, instead of shifting in physical space, we now shift in field space.

\subsection{The Models}

So far we have only applied LSPT to models of a scalar field $\phi(\vx)$ with conjugate momentum $\pi(\vx)$.  We have applied it in 1+1, 2+1 and 3+1 dimensions.  For simplicity, we will now consider the case of 1+1 dimensions, returning to higher dimensions in Sec.~\ref{dimsez}.  The (1+1)-dimensional case is simplest as normal ordering eliminates all ultraviolet divergences.  

Consider the Hamiltonian
\beq
H=\int d x: \mathcal{H}(x):, \quad \mathcal{H}(x)=\frac{\pi^2(x)}{2}+\frac{\left(\partial_x \phi(x)\right)^2}{2}+\frac{V(\sqrt{\lambda} \phi(x))}{\lambda} \label{hdef}
\eeq
where again $\lambda$ is a coupling constant and $V$ is a potential.  We will see shortly that this parametrization ensures that the mass is independent of $\lambda$ and that interactions are suppressed by powers of $\lambda$.  As the function $V$ itself is taken to be independent of the parameter $\lambda$, this parametrization makes all powers of $\lambda$ explicit even when the function $V$ is not specified.

In quantum physics we have a parameter $\hbar$ not present in classical field theory.  The combination $\lambda\hbar/m^2$ is dimensionless, and we will organize all of our computations in expansions in this quantity.  As $\lambda$ and $\hbar$ appear together, an expansion in $\lambda$, called a perturbative expansion, is equivalent to an expansion in $\hbar$, called a semiclassical expansion.  With this understood, we will set $\hbar=1$.

In Eq.~(\ref{hdef}), :: is the usual normal ordering, which we recall removes all diagrams in which a loop begins and ends at the same vertex.  These are the only divergent diagrams in (1+1)-dimensional scalar models of this type.

Consider a potential with two degenerate minima, at $\phi$ equal to $\phi_1$ and $\phi_2$ respectively.  Then the classical equations of motion admit a solution
\beq
\phi(x,t)=f(x)\hsp f(-\infty)=\phi_1\hsp f(\infty)=\phi_2.
\eeq
Such a solution is called a kink \cite{finkink}.

Two examples of potentials are popular in the literature because the corresponding kinks and their normal modes can be found analytically.  The first is the Sine-Gordon potential
\beq
V(\sqrt{\lambda}\phi(x))=m^4\left(1-{\rm{cos}}\left(\frac{\sqrt{\lambda}\phi(x)}{m}\right)\right). \label{sgpot}
\eeq
The corresponding classical equations of motion admit the time-independent kink solution
\beq
f(x)=\frac{4m}{\sl}{\rm{arctan}}\left(e^{mx}\right) \label{sgs}
\eeq
called the Sine-Gordon soliton.  This solution is stationary and centered at $x=0$, but translations and boosts give solutions with other positions and velocities.
It is a soliton in the sense of the word used in mathematical physics, its shape is unchanged during an interaction.  The second is the $\phi^4$ double-well model
\beq\label{kpot}
V(\sqrt{\lambda}\phi(x))=\frac{\lambda\phi^2(x)}{4}\left(\sqrt{\lambda}\phi(x)-\sqrt{2}m\right)^2
\eeq
leading to classical equations of motion which also admit a kink solution
\beq
f(x)=\frac{m}{\sqrt{2\lambda}}\left(1+{\rm{tanh}}\left(\frac{mx}{2}\right)\right). \label{ksol}
\eeq
While these kinks are simple to treat, they are not generic in the sense that they are both reflectionless.  That means that, classically, small amplitude incoming wavepackets travel through the kinks with no reflection.

We will use the notation
\beq
V^{(n)}(\sl\phi(x))=\frac{\partial^n}{\partial(\sl\phi(x))^n }V(\sl\phi(x))
\eeq
for the $n$th derivative of the potential with respect to its argument.   As a result, when the potential $V/\lambda$ is expanded in powers of $n$, the $n$th term is
\beq
\frac{\lambda^{n/2-1}}{n!}V^{(n)}(\sl\phi(x))\phi^n.
\eeq
This identifies the mass $m$ as the positive solution of
\beq
m^2=V^{(2)}(\sl\phi_1)=V^{(2)}(\sl\phi_2).
\eeq
Here $m$ is the mass of the perturbative excitation, which we call a meson, on the two sides of the kink.  

The leading quantum correction to the vacuum energy density on each side of the kink will depend on the meson mass on that side of the kink.  If these masses are not equal, then the vacua on the two sides of the kink will have different energy densities.  As a result, the kink will be a bubble wall around a false vacuum, and so will accelerate towards the true vacuum \cite{wstab}.  As a result, such a kink will not be a Hamiltonian eigenstate.  While it may still be described via our formalism, nonetheless we will not treat this case in our lectures.

While models with sextic and higher potentials are quite popular in the literature on classical kinks, these models tend to have unequal masses on the two sides.  As a result, while the classical equations of motion admit static kinks, lifting these kinks to quantum field theory they will accelerate.

\subsection{Spectra and Sectors}

Our models have rich spectra of excitations.  First, they have perturbative mesons of mass $m$ in each vacuum.  Second, there are kink solutions, whose masses we will see are of order $O(1/\lambda)$ and interpolate between the vacua.  

In a sense, half of the models will admit oscillons, which are time-dependent lumps whose expectation values oscillate with an amplitude of order $O(1/\sl)$.  These also have masses of order $O(1/\lambda)$.  In general they decay, however these decays proceed very slowly.  One exception is integrable models such as the Sine-Gordon model, in which the oscillons are called breathers and they do not decay \cite{lamb,hirotab}.

There are also kink-antikink bound states, which resonate, sometimes passing through each other several times  before eventually decaying to radiation or escaping.  These are often deformations of oscillons.

While LSPT can be applied to all of these objects, we will restrict our attentions to kinks and mesons.  More precisely, we will consider states in two sectors.  First, we will consider the vacuum sector.  This is the Fock space of finite numbers of mesons on top of one of the vacua.  Next, we will consider the kink sector.  This is the Fock space of finite numbers of mesons in addition to a single kink.

\subsection{The Vacuum Sector}

The vacuum sector states can be constructed as usual.  One decomposes the Schrodinger picture field $\phi(x)$ and its conjugate $\pi(x)$ in terms of plane waves, which are the solutions of the linearized classical equation of motion, which is the massive Klein-Gordon equation
\beq
\phi(x,t)=e^{ipx}e^{-i\omega_p t}\hsp \omega_p=\sqrt{m^2+p^2}. \label{pw}
\eeq

The notation $\phi(x,t)$ is reserved for the field in the corresponding classical field theory.  In quantum field theory, we will work exclusively in the Schrodinger picture, where the field $\phi(x)$ and its conjugate $\pi(x)$ are functions of $x$ and not $t$.  Our decompositions will therefore use the constant frequency plane waves $e^{ipx}$ without the time-dependence.  This will be crucial later, as the time-dependence will be different in the kink sector.

We begin our construction by constructing the creation and annihilation operators
\beq
A^\ddag_p=\int dx \left( \frac{\phi(x)}{2}-\frac{i\pi(x)}{2\omega_p}
\right) e^{ipx}\hsp
A_{-p}=\int dx \left( \omega_p\phi(x)+i\pi(x)
\right) e^{ipx} \label{invdec}
\eeq
where we have introduced a weighted adjoint
\beq
A^\ddag_p=\frac{A^\dag_p}{2\omega_p}
\eeq
which will simplify our expressions for states later.

Now, the plane waves $e^{ipx}$ are a basis of the space of functions\footnote{We will not specify which space of functions.  Roughly speaking, we are interested in smooth functions that are defined on an interval, so that the interval may include the spatial region of interest for a given process.}.  That means that this construction can be inverted, in this case by a Fourier transform, to yield the decomposition
\beq
\phi(x)=\pin{p}\left(A^\ddag_p+\frac{A_{-p}}{2\omega_p}\right)e^{-ipx}\hsp \pi(x)=i\pin{p}\left(\omega_p A^\ddag_p-\frac{A_{-p}}{2}\right)e^{-ipx}. \label{adec}
\eeq

This decomposition plays the same role as our decomposition (\ref{qhob}) of $x$ and $p$ in terms of $a^\dag$ and $a$ in the case of the quantum harmonic oscillator.  As in that case, $\{A^\ddag_p,A_p\}$ and $\{\phi(x),\pi(x)\}$ are two bases of our algebra of operators.  And so any operator in the theory can be written in either basis.

Substituting the canonical commutation relations
\beq
[\phi(x),\pi(y)]=i\delta(x-y)
\eeq
into the construction (\ref{invdec}) one concludes that the $A$ operators satisfy the oscillator algebra
\beq
[A_p,A^\ddag_q]=i\int dx \left[ \frac{-i\omega_p}{2\omega_p}-\frac{i}{2}
\right]e^{i(q-p)x}=2\pi\delta(p-q).
\eeq

Proceeding as in the case of the quantum harmonic oscillator, we define the tree-level ground state $|\Omega\rangle_0$ to be the state which is annihilated by all of the annihilation operators
\beq
A_p|\Omega\rangle_0=0.
\eeq
Then we can build the tree-level meson Fock space by acting with the creation operators
\beq
|p_1\cdots p_n\rangle_0=A^\ddag_{p_1}\cdots A^\ddag_{p_n}|\Omega\rangle_0.
\eeq
These of course are not Hamiltonian eigenstates.  The corresponding Hamiltonian eigenstates $|p_1\cdots p_n\rangle$ satisfy
\beq
H|p_1\cdots p_n\rangle=E|p_1\cdots p_n\rangle \label{vacse}
\eeq
where we decompose the states in powers of $\lambda$
\beq
|p_1\cdots p_n\rangle=\sum_{i=0}^\infty |p_1\cdots p_n\rangle_i\hsp |p_1\cdots p_n\rangle_i\sim O(\lambda^{i/2})\hsp
E=\sum_{i=0}^\infty E_i\hsp
E_i\sim O(\lambda^{i}).
\eeq
In particular, the true vacuum of the theory is $|\Omega\rangle$.  The terms $|p_1\cdots p_n\rangle_i$ at $i$th order are found by substituting this series into the eigenvalue equation (\ref{vacse}), decomposing it into one equation at each order and solving it order by order.  At $i$th order one obtains $(H_2-E_0)|p_1\cdots p_n\rangle_i$ as a function of the terms at lower orders which have already been found, and so one needs to invert $H_2-E_0$.  As we will now explain, this inverse is not unique.

Our definition of the state $|p_1\cdots p_n\rangle$ as the Hamiltonian eigenstate that is equal to $|p_1\cdots p_n\rangle_0$ at leading order is ambiguous as there are many such states.  One ambiguity arises from the normalization.  We will fix that ambiguity by demanding
\beq
{}_0\langle p_1\cdots p_n|p_1\cdots p_n\rangle_{0}=1
\hsp{}_0\langle p_1\cdots p_n|p_1\cdots p_n\rangle_{i>0}=0.
\eeq
In other words, when finding the quantum correction at each order, one drops the part proportional to the leading order.  

Another ambiguity arises from the fact that, if $n\geq 2$, there will be multiple states with the same momenta and energies.  This observation will actually be critical to our treatment of scattering based on the Lippmann-Schwinger equations, as scattering conserves the energy and momentum and so, after properly combining these states into localized wave packets, interpolates between these degenerate states.

\subsection{How Not to Construct Kink Sector States}

Above we have constructed the Fock space of multimeson states above a vacuum, which we have called the vacuum sector.  We now wish to repeat this construction as closely as possible for the kink sector.

In classical field theory, one-kink states are field configurations that are close to $f(x)$.  One treats them by inserting the decomposition $\phi(x,t)=f(x)+\eta(x,t)$ into the Hamiltonian $H[\phi,\pi]$ and rewriting the Hamiltonian in terms of $\eta(x,t)$.  The result $H\p[\eta,\pi]$ is usually called the kink Hamiltonian.  Then one quantizes the kink Hamiltonian as if $\eta$ were the dynamical field.

Such a procedure is common for quantum kinks, but it was realized in Ref.~\cite{rebhan} that it can lead to incorrect answers for the following reason.  In quantum field theory there are generally ultraviolet divergences which are handled by regularizing the theory with a regulator $\Lambda$.  What if there are two Hamiltonians, $H$ and $H\p$?  Then each is regularized with a regulator, $\Lambda$ and $\Lambda\p$ respectively.  Let us write the regularized Hamiltonians as $H_\Lambda$ and $H\p_{\Lambda\p}$.

The kink mass is obtained by subtracting the energy of the vacuum state from the energy of the kink ground state.  The former is an eigenvalue of $H_\Lambda$ while the latter is an eigenvalue of $H\p_{\Lambda\p}$.  The mass therefore depends on both $\Lambda$ and $\Lambda\p$.  To get the answer, one needs to take both $\Lambda$ and $\Lambda\p$ to infinity.  The problem observed in Ref.~\cite{rebhan} is that the answer depends on the direction in the $(\Lambda,\Lambda\p)$ plane in which one takes this limit, or equivalently on the matching condition between $\Lambda$ and $\Lambda\p$.  At most one matching condition is correct, and there is no physical principle for knowing which.

One way to understand the problem is as follows.  Perhaps the simplest regulator is a momentum cutoff.  However, a momentum cutoff of $\phi(x)$ is not equivalent to a momentum cutoff of $\eta(x)$, because the high momentum Fourier components of $f(x)$ cause a shift to those of $\eta(x)$.

\subsection{The Displacement Operator and A Coherent State}\label{DfOmega0}

Above we tried to quantize the kink using an active transformation of the field, which is intuitively $\phi(x)\rightarrow f(x)+\phi(x)$.  After the active transformation one needed a new Hamiltonian, which was regularized separately.  However the two regularizations were inequivalent.

Our strategy will instead be to regularize exactly once.  In other words, we begin with a Hamiltonian $H$ which is already regularized.  In the case of 1+1 dimensions, as we have removed the ultraviolet divergences by normal ordering, this is already done.  However, in more dimensions when logarithmic ultraviolet divergences arise, for example in 3+1 dimensions at one loop or 2+1 dimensions at two loops, we will use a momentum cutoff.

We wish to make a kink Hamiltonian $H\p$ which is already regularized and has the same spectrum as the regularized defining Hamiltonian $H$.  The spectra will automatically be the same if $H$ and $H\p$ are related by conjugation by a unitary operator.  That will be our approach.  

What unitary operator could shift the field by $f(x)$?  This can be done with the displacement operator
\beq
\df={{\rm Exp}}\left[-i\int dx f(x)\pi(x)\right]. \label{dfdef}
\eeq
To see this, note that
\beq
\df^\dag \phi(x)\df=\phi(x)+f(x)
\eeq

How does this help us to construct a state?  It is easy to see that
\beq
{}_0\langle\Omega|\phi(x)|\Omega\rangle_0=0
\eeq
because the $A_{-p}$ terms annihilate the $|\Omega\rangle_0$ while the $A^\ddag_p$ terms annihilate the ${}_0\langle\Omega|$.  This is reasonable as $|\Omega\rangle_0$ is in the vacuum sector.  Now let us calculate the expectation value of $\phi(x)$ in the state $\df|\Omega\rangle_0$.  This is
\beq
{}_0\langle\Omega|\df^\dag \phi(x)\df|\Omega\rangle_0={}_0\langle\Omega|\left(\phi(x)+f(x)\right)|\Omega\rangle_0=f(x).
\eeq
This certainly seems like a good clue that our state is some sort of quantum kink.  

This state explicitly breaks the translation symmetry, whereas we would expect the ground state to be an eigenstate of the momentum operator $P$ with zero momentum, as the momentum $P$ and Hamiltonian $H$ commute.  So it is not the ground state.  In fact, it is not even a Hamiltonian eigenstate.  The expectation value of the Hamiltonian
\beq
{}_0\langle\Omega|\df^\dag H[\phi,\pi]\df|\Omega\rangle_0={}_0\langle\Omega| H[\phi+f,\pi]|\Omega\rangle_0={}_0\langle\Omega| H[f,0]|\Omega\rangle_0=H[f,0]
\eeq
is just the classical energy of the kink.  We will see that in 1+1 dimensions this is of order one meson mass higher than the ground state energy of the quantum kink, and so such a kink will radiate a meson or two and then decay to a ground state kink\footnote{If the kink has shape mode excitations, then the endpoint of the decay will be a superposition of the ground state kink and the kink with an excited shape mode.}.  In more dimensions, the energy difference and even the tension difference is infinite, and so such a state does not exist as it has an infinite energy density \cite{erice}.

To see this from another perspective, it has long been claimed \cite{vinc72,cornwall74} that solitons correspond to coherent states, which are states obtained by acting on the vacuum with an operator of the form $e^A$ where $A$ is linear in the fields.  Our displacement operator is just of this form.  And so our trial state $\df|\Omega\rangle_0$ is indeed a coherent state.  

\section{One-Loop Quantum Kinks are Squeezed, Coherent States} \label{unosez}

In this section we will explicitly construct the quantum state corresponding to the ground state of the quantum kink.  It will have several differences with respect to the coherent state constructed above:
\begin{enumerate}
  \item \hypertarget{one}{The ground state will be a Hamiltonian eigenstate.}
  \item \hypertarget{two}{The ground state will have a lower energy than the coherent state.} 
  \item \hypertarget{three}{The ground state will be translation-invariant.}
  \item \hypertarget{four}{The ground state will be a squeezed, coherent state.}
  \item \hypertarget{five}{The ground state, being translation-invariant, will not be normalizable.}
\end{enumerate}
An analogy with quantum mechanics is as follows.  The coherent state will play the role of an eigenstate of the position operator $x$, corresponding to a fixed kink solution or equivalently a single point in the moduli space of center points $x_0$.  Such a localization costs energy, and in 2+1 or more dimensions the energy price will be infinite as it is in quantum mechanics.  The squeezed state will correspond to the unnormalizable state $\psi(x)=c$ for a constant $c$.

\subsection{The Kink Hamiltonian}

Let $|K\rangle$ be the kink ground state.  It will satisfy
\beq
H|K\rangle=Q|K\rangle \label{kse}
\eeq
where $Q$ is the energy of the kink state.  Upon subtracting the energy of the vacuum $|\Omega\rangle$ from $Q$ one obtains the kink mass.  It is often convenient to add a $c$-number counterterm to the Hamiltonian so that $H|\Omega\rangle=0$ in which case $Q$ is the kink mass.  As always, we may decompose $Q$ in powers of the coupling
\beq
Q=\sum_{i=0}^\infty Q_i\hsp Q_i\sim O(\lambda^{i-1})
\eeq
where $Q_0$ is the mass of the classical kink.

The fact that $Q_0$ is proportional to the inverse coupling indicates that $|K\rangle$ will be a very nonperturbative state, in the sense that it cannot be obtained by adding any perturbative corrections to $|\Omega\rangle$.  This motivates us to peel off a nonperturbative piece as we have done in quantum mechanics and with the coherent state, by defining
\beq
\vac=\df^\dag |K\rangle. \label{vacdef}
\eeq
Unlike our quantum mechanics construction of the shifted vacuum in Sec.~\ref{qmsez} and our construction of the coherent states in Sec.~\ref{cosez}, in the present case, the remaining part, $\vac$, will not be perturbative.  Instead, when we peel off the negative powers of $\lambda$ we will still be left with an $O(\lambda^0)$ correction which is nonperturbative and has no analogue in quantum mechanics.

Now instead of trying to construct $|K\rangle$ directly, we will try to construct $\vac$.  After all, once we know $\vac$, we can obtain $|K\rangle$ from (\ref{vacdef}), although we will rarely need to.

What is $\vac$?  Let us define, in analogy with the shifted Hamiltonian, the kink Hamiltonian
\beq
H\p=\df^\dag H\df. \label{hpdef}
\eeq
Then one can easily see that
\beq
H\p\vac=\df^\dag H\df\df^\dag|K\rangle=\df^\dag H|K\rangle=Q\df^\dag |K\rangle=Q\vac. \label{vse}
\eeq
In other words, while our desired ground state $|K\rangle$ is an eigenvector of $H$, the state $\vac$ is an eigenvector of $H\p$.  Note that the eigenvalue of $\vac$ under the action of $H\p$ is exactly equal to the eigenvalue of the ground state $|K\rangle$ under $H$.  This argument is trivially extended to all eigenstates of $H$.  This exact equality of the eigenvalues of $H$ and $H\p$, independent of any perturbative expansion or choice of renormalization condition, is one of the main strengths of our approach.

We have replaced one eigenvalue problem (\ref{kse}) with another (\ref{vse}).  Is this progress?  We will see that it is, because, now that we have peeled off the most nonperturbative part, we can solve (\ref{vse}) using, up to a squeeze, perturbation theory.

Note that our kink Hamiltonian $H\p$ is quite similar to the kink Hamiltonian often constructed in the literature, in which one expands $\phi(x)=f(x)+\eta(x)$ and rewrites the Hamiltonian in terms of $\eta(x)$. However, that procedure is performed on the Hamiltonian before regularization, leading to the regularization matching ambiguities described above, which ultimately lead $H$ and $H\p$ to have different spectra.  In our case, we regularize first, in fact in 1+1 dimensions it is sufficient to normal order but in higher dimensions the regularization will be less trivial, and then the definition (\ref{hpdef}) implies that, whatever regularization we use, $H\p$ and $H$ will have the same spectrum, and so either can be used to obtain the mass.   Said differently, most authors transform the classical Hamiltonian, together with all of its ordering ambiguities when one quantizes.  However our $H\p$ is defined by a unitary transformation of the full quantum Hamiltonian $H$.  As our $H$ and $H\p$ are guaranteed to have the same spectra for each value of the regulator, the eigenvalue of the vacuum with respect to $H$ can be subtracted from the eigenvalue of the kink ground state with respect to $H\p$ to obtain a kink mass which depends on only a single regulator, and so no regulator matching is needed. 

\subsection{Constructing the Kink Hamiltonian}

We have defined $H$ in Eq.~(\ref{hdef}) and $\df$ in Eq.~(\ref{dfdef}).  We can now plug those definitions into Eq.~(\ref{hpdef}) to calculate $H\p$.

To do this, let us note that for any operator $F[\phi(x),\pi(x)]$, the conjugation by $\df$ acts as
\beq
\df^\dag F[\phi(x),\pi(x)]\df=F[\phi(x)+f(x),\pi(x)].
\eeq

In particular, we may decompose $A^\ddag_p$ and $A_{-p}$ in terms of $\phi(x)$ and $\pi(x)$ as described in Eq.~(\ref{invdec}), and we learn that
\bea
\df^\dag A^\ddag_p\df &=&\int dx \left( \frac{\phi(x)+f(x)}{2}-\frac{i\pi(x)}{2\omega_p}
\right) e^{ipx}=A^\ddag_p+\frac{\tilde{f}_{-p}}{2}\\
\df^\dag A_{-p}\df&=&\int dx \left( \omega_p(\phi(x)+f(x))+i\pi(x)
\right) e^{ipx}=A_{-p}+\omega_p\tilde{f}_{-p}\nonumber
\eea
where we have defined the Fourier transform
\beq
\tilde{f}_p=\int dx f(x)e^{-ipx}.
\eeq

To conjugate $H$ by $\df$ we will need to understand how $\df$ acts on normal ordered products.  Now that we know how it acts of $A^\ddag_p$ and $A_{-p}$, this is straightforward.

Let us consider, for example, the operator
\beq
:A_{-p}A^\ddag_q: \ =A^\ddag_q A_{-p}.
\eeq
Conjugation gives
\bea
\df^\dag:A_{-p}A^\ddag_q:\df&=&\df^\dag A^\ddag_q\df\df^\dag A_{-p}\df=A^\ddag_qA_{-p}+\frac{\tilde{f}_{-q}}{2}A_{-p}+\omega_{p}\tilde{f}_{-p}A^\ddag_q+\frac{\omega_p}{2}\tilde{f}_{-q}\tilde{f}_{-p}\nonumber\\
&=&:\left(A_{-p}+\omega_p\tilde{f}_{-p}\right)\left(A^\ddag_q+\frac{\tilde{f}_{-q}}{2}\right):.
\eea
In other words, the correct procedure is to first evaluate the normal ordering and then do the conjugation, as is seen on the first line.  However this gives the same answer as one would obtain by first conjugating and then normal ordering.  

The reason for this is as follows.  Consider string $\co$ of $A$ and $A^\ddag$ operators.  Now $:\co:$ will be another such string containing the same operators but normal ordered.  Let us calculate $\df^\dag:\co:\df$.  We can insert $\df\df^\dag$ between each two operators in the string.  Then, they act by shifting the operator by a $c$-number while preserving the ordering of all other operators.  The $c$-number can be factored out, and the remaining string will be normal ordered.  Therefore, conjugation by $\df$ preserves normal ordering.  

What if we instead first applied the $\df$ and then normal ordered, corresponding to the expression $:\df^\dag\co\df:$?   The conjugation by $\df$ replaces each operator by itself plus a $c$-number, so that each term in $\df^\dag\co\df$ is equal, up to ordering of the operators, to a term in the series describing $\df^\dag:\co:\df$.  Now $:\df^\dag\co\df:$ will clearly be normal ordered, whereas we have already shown that $\df^\dag:\co:\df$ is normal ordered, and so these two series are in fact equal
\beq
\df^\dag:\mathcal{O}:\df= \ :\df^\dag\mathcal{O}\df:.
\eeq

With this identity in hand, we may obtain our kink Hamiltonian
\beq
H\p=\int d x: \mathcal{H\p}(x):, \quad \mathcal{H}\p(x)=\frac{\pi^2(x)}{2}+\frac{\left(\partial_x \left(\phi(x)+f(x)\right)\right)^2}{2}+\frac{V(\sqrt{\lambda} \left(\phi(x)+f(x)\right))}{\lambda}.  \label{hpesp}
\eeq
It will be convenient to decompose the kink Hamiltonian and its density into operators with fixed numbers $n$ of fields
\beq
H\p=\sum_{n=0}^\infty H\p_n\hsp \ch\p=\sum_{n=0}^\infty  \ch\p_n\hsp H\p_n\sim\ch\p_n\sim O(\lambda^{n/2-1}).
\eeq
Substituting (\ref{hpesp}) into this decomposition one finds that the $c$-number term
\beq
H\p_0=\frac{1}{2}\int dx \left[{f^{\prime 2}(x)}+\frac{V(\sqrt{\lambda} f(x))}{\lambda}\right]=Q_0
\eeq
is just equal to the classical kink mass $Q_0$.  At linear order, integrating by parts the $\phi$-linear terms in Eq.~(\ref{hpesp}), we find
\beq
H\p_1=\int dx\left[-f\pp(x)+\frac{\V1}{\sl}
\right]\phi(x)=0
\eeq
where the term in the square brackets vanishes using the classical equations of motion that define $f(x)$.  The term $H\p_1$ is the nonperturbative contribution to the tadpole, of order $O(1/\sl)$.  While there will be further contributions to the tadpole starting at $O(\sl)$, they can be handled by standard perturbative methods.  Unlike them, this term, had it not vanished, would have been a disaster.  Intuitively, if the kink is not a classical solution, then it will accelerate, a fact which is manifested in the quantum theory by a large tadpole.  This is the familiar fact that tadpoles correspond to forces in field space.

Our goal is to solve the $H\p$ eigenvalue equation.  We hope that this can be done perturbatively.  We have seen that the operator $H\p$ is not entirely perturbative, as the terms $H\p_n$ are of order $O(\lambda^{n/2-1})$ and so $H\p_{0}$, $H\p_1$ and $H\p_2$ are not suppressed by any powers of the coupling and must be diagonalized exactly before perturbation theory may begin.  However, we have now seen that $H\p_0$ is a $c$-number and so is diagonalized by any state, and $H\p_1$ vanishes.  Therefore, if we can diagonalize $H\p_2$, then the rest, the position-dependent $n$-point interactions
\beq
H\p_{n}=\int dx:\ch\p_{n}(x): \hsp\ch_{n\geq 3}\p(x)=\frac{\lambda^{n/2-1}}{n!}\V{n}\phi^n(x) \label{hnp}
\eeq
can be handled perturbatively.

The key step of the problem is then diagonalizing the free kink Hamiltonian $H\p_2$.  This is the integral of the normal-ordered Hamiltonian density $\ch\p_2$
\beq\label{h2eq}
H\p_2=\int dx:\ch\p_2(x): \hsp \ch\p_2=\frac{\pi^2(x)+\partial_x\phi(x)\partial_x\phi(x)+\V2 \phi^2(x)}{2}.
\eeq
Can we diagonalize it?  It looks like the free massive scalar theory, but the mass is position-dependent.  As is well-known, such a Hamiltonian can be diagonalized by a Bogoliubov transform.  This transform will occupy most of the rest of this section.

Note that the position-dependence $\V2$ of the mass squared is the new ingredient with respect to the quantum mechanical case.  In quantum mechanics, the conjugation with the displacement operator $\dx$ yielded the usual harmonic oscillator position term.  Had we also changed the frequency of the harmonic oscillator, then a squeeze would have been required in that case as well.

\subsection{Normal Modes}

If we simply substitute our decomposition (\ref{adec}) into the free kink Hamiltonian (\ref{h2eq}) we would find terms of the form $A^\ddag A^\ddag$ and $AA$, and so we would not be able to diagonalize $H\p_2$.  

The problem is as follows.  In the vacuum sector, it is reasonable to decompose the field $\phi(x)$ and its conjugate $\pi(x)$ into operators $A^\ddag_p$ and $A_{-p}$ that create and destroy plane waves, because plane waves (\ref{pw}) are the constant frequency perturbations of a vacuum state, in other words they solve the massive Klein-Gordon equation.

On the other hand, the constant frequency perturbations 
\beq
\phi(x,t)=f(x)+e^{-i\omega t}\g(x)
\eeq
of a kink solve a
Sturm-Liouville equation
\beq
\V{2}{\g}(x)=\omega^2{\g}(x)+{\g}^{\prime\prime}(x). \label{sl}
\eeq
Here the functions $\g(x)$ are called normal modes.  It is a theorem that the solutions of a Sturm-Liouville equation are a basis of functions on all intervals $(a,b)$ of the real numbers.  This generalizes the fact, in the case of the vacuum sector, the plane waves $e^{-ipx}$ provide a basis for all functions.

In the case of the plane waves, this property was essential for the following reason.  We defined the operators $A^\ddag_p$ and $A_{-p}$ in terms of the fields $\phi(x)$ and $\pi(x)$ in Eq.~(\ref{invdec}).  The coefficients in this construction were the plane waves.  Because the plane waves were a basis of functions, we were able to invert this construction, yielding the decomposition of $\phi(x)$ and $\pi(x)$ in terms of $A^\ddag_p$ and $A_{-p}$ in (\ref{adec}), in other words, using the Fourier transform. 

If, instead of (\ref{invdec}) we tried using an incomplete set of functions, for example if we excluded some momenta, then we would not have been able to invert the construction and so obtain a decomposition (\ref{adec}).

Similarly, the fact that the normal modes are a basis of functions on any interval means that any local operator $\co(x)$ can be decomposed in terms of the basis of functions times $x$-independent operators.  In particular, we may decompose the Schrodinger picture fields $\phi(x)$ and $\pi(x)$ in terms of the normal modes.  Note that this would not be possible in the interaction or Heisenberg picture because in that case the operators would also depend on the time $t$, which would need to somehow be matched with the normal modes.  In the Schrodinger picture, any decomposition in terms of a basis of functions is acceptable, because the operators are defined without reference to time and so the choice of Hamiltonian, which generates time evolution, does not affect the operators.  The decomposition is performed in space, not time.

So what are the normal modes?  The normal modes are classified by their frequency $\omega$.  We can take $\omega\geq 0$ because anyway only $\omega^2$ appears in the Sturm-Liouville equation (\ref{sl}).  Furthermore, we will be interested in classically stable solutions, and so $\omega$ will be real.  More precisely, there may be isolated quasinormal modes with complex frequencies, however they will only contribute to the kink mass via a deformation of the normal modes at nearby real frequencies, and so the quasinormal modes may be neglected in the computations below. 

The translation-invariance of our Hamiltonian implies that, for any $V$, there will be a zero mode
\beq
\g_B(x)=-\frac{f\p(x)}{\sqrt{Q_0}}.
\eeq
Here the constant factor $\sqrt{Q_0}$ is chosen because it leads to the convenient normalization
\beq
\int dx \g^2_B(x)=1 \label{norm1}
\eeq
where we have used the fact that kinks in such models are BPS and so their classical energy is just twice the kinetic energy, which is the integral of $f^{\prime 2}(x)/2$.

This zero-mode corresponds to an infinitesimal translation and so is related to the collective coordinate $x_0$
\beq
f(x-x_0)=f(x)+\sqrt{Q_0}\g_B(x) x_0+O(x_0^2).
\eeq
Intuitively, shifting the field by the zero mode times $\sqrt{Q_0}$ is, up to a nonlinear corrections, the same as shifting the collective coordinate $x_0$.  The linear correspondence is the origin of the name {\it{linearized}} soliton perturbation theory, in contrast with collective coordinate methods which treat $x_0$ exactly.  This relationship will become more precise later.

In the case of the Sine-Gordon model (\ref{sgpot}) the zero mode is
\beq
\g_{B}(x)=-\sqrt{\frac{m}{2}}\sech\left(mx\right) \label{sgz}
\eeq
while in the case of the $\phi^4$ double-well (\ref{kpot}) it is
\beq
\g_B(x)=-\sqrt{\frac{{3m}}{8}}\sech^2(m x/2). \label{p4z}
\eeq

For all frequencies $\omega>m$ there are also two continuum normal modes, with a single normal mode at $\omega=m$.  We will label these by the real numbers $k$ that satisfy
\beq
k^2=\omega^2-m^2.
\eeq
In the case of the Sine-Gordon model these are
\beq
\g_k(x)=\frac{e^{-ikx}}{\ok{}}\left[ik+ m\tanh(m x)\right] \label{sgm}
\eeq
while for the $\phi^4$ model they are
\beq
\g_k(x)=\frac{e^{-ikx}}{\ok{} \sqrt{m^2+4k^2}}\left[2k^2-m^2+(3/2)m^2\sech^2\left(\frac{mx}{2}\right)-3im k\tanh\left(\frac{mx}{2}\right)\right]. \label{p4m}
\eeq
The continuum modes are essentially plane waves plus the perturbations at small $m|x|$ caused by the kink.  Note that these plane waves asymptote to $e^{-ikx}$ at both $x\rightarrow +\infty$ and also $x\rightarrow-\infty$.  That is because these two potentials lead to reflectionless kinks.  For general potentials, the continuum modes will also contain reflected terms with opposite momenta.

Unlike the discrete modes, we assemble the continuum modes into complex combinations such that
\beq
\g^*_k(x)=\g_{-k}(x).
\eeq
We will always normalize the continuum modes so that
\beq
\int dx \g_{k_1}(x)\g^*_{k_2}(x)=2\pi\delta(k_1-k_2). \label{norm2}
\eeq

Finally, it can happen that there are discrete modes besides the zero mode.  Such modes are called shape modes $\g_S(x)$ and their frequencies lie beneath the continuum threshold
\beq
0<\omega_S<m.
\eeq
We will take the shape modes to be real and normalized so that
\beq
\int dx \g_{S_1}(x)\g_{S_2}(x)=\delta_{S_1S_2} \label{norm3}
\eeq
where $S_1$ and $S_2$ are two shape modes and now $\delta$ is a Kronecker delta.  The Sine-Gordon kink has no shape modes, but the $\phi^4$ kink has one
\beq
\g_S(x)=\frac{\sqrt{3 m}}{2}\tanh\left(\frac{mx}{2}\right)\sech\left(\frac{mx}{2}\right)\hsp \omega_S=\frac{\sqrt{3}}{2}m. \label{p4s}
\eeq

We note that, as the normal modes are eigenvectors of the Sturm-Liouville operator with distinct eigenvalues except for $\g_{\pm k}$, the normal modes are mutually orthogonal
\beq
\int dx \g_{k_1}(x)\g^*_{k_2}(x)=0
\eeq
unless $k_1$ and $k_2$ are equal.  Here we have adopted a condensed notation in which $k_i$ also runs over discrete modes.  Therefore this observation applies when $\g_{k_i}$ are continuum modes, shape modes or zero modes.

\subsection{Decomposing Fields Using Normal Modes}

With the normal modes in hand, we can assemble a new basis of the operators
\bea
\phi_0&=&\int dx \phi(x)\g_B(x)\hsp \phi_S=\int dx \phi(x) \g_S(x)\hsp
\phi_k=\int dx \phi(x) \g^*_k(x)\label{bdec}\\
\pi_0&=&\int dx \pi(x)\g_B(x)\hsp \pi_S=\int dx \pi(x) \g_S(x)\hsp
\pi_k=\int dx \pi(x) \g^*_k(x).\nonumber
\eea
Here $S$ runs over all shape modes, if there are any.  We have introduced the operators $\{\phi_0,\phi_S,\phi_k,\pi_0,\pi_S,\pi_k\}$ whice are a new basis of the operators.  Any operator can be written in terms of $\phi(x)$ and $\pi(x)$, it can be written in terms of $A^\ddag_p$ and $A_p$ and it can also be written in this new basis.  

Note that an analogous decomposition of $\phi(x)$ and $\pi(x)$ suitable for the vacuum sector is quite simple, it is just a Fourier transform, but we skipped this step in the case of the vacuum sector because it was easier to decompose them directly in terms of creation and annihilation operators in Eq.~(\ref{adec}).  In the case at hand, not all terms can be decomposed into oscillators and so we provide this intermediate decomposition.  However, shortly we will find a generalization of creation and annihilation operators that is suitable for the kink sector.

Using the canonical commutation relations and the various normalization conditions above we find the algebra satisfied by this new basis of operators
\beq
[\phi_0,\pi_0]=i\hsp
[\phi_{S_1},\pi_{S_2}]=i\delta_{S_1S_2}\hsp 
[\phi_{k_1},\pi_{k_2}]=2\pi i\delta(k_1+k_2) \label{phik}
\eeq
with all other commutators vanishing.  Here $S_1$ and $S_2$ are two shape modes.

As the normal modes $\g(x)$ are a basis of functions, we may invert the construction (\ref{bdec}) to arrive at the decomposition~\cite{cahill76}
\bea
\phi(x)&=&\phi_0\g_B(x)+\sum_S \phi_S \g_S(x)+\pin{k}\phi_k\g_k(x)\label{decprin}\\
\pi(x)&=&\pi_0\g_B(x)+\sum_S \pi_S \g_S(x)+\pin{k}\pi_k\g_k(x).\nonumber
\eea
Again, this decomposition generalizes a Fourier decomposition (\ref{adec}) of $\phi(x)$ and $\pi(x)$ into plane wave modes of momentum $p$.  However, instead of decomposing in terms of plane waves, one decomposes in terms of normal modes, which are more suitable for kinks.

Although the shape modes $\g_S$ and zero mode $\g_B$ are both discrete, we will see below that the shape modes often appear in formulas together with the continuum modes $\g_k$.  Indeed, one often sums over all shape modes and integrates over all continuum modes.  Therefore it will be useful to introduce the notation
\beq
\ppin{k} F_k =\sum_S F_S + \pin{k} F_k.
\eeq
On the left hand side, $k$ runs over all shape modes $S$ as well as all real numbers, representing continuum modes.  $F_k$ is anything that depends on this index.  On the right hand side there is a sum over shape modes $S$ as well as an integral over all real numbers $k$.

In this condensed notation our decomposition becomes
\beq
\phi(x)=\phi_0\g_B(x)+\ppin{k}\phi_k\g_k(x)\hsp
\pi(x)=\pi_0\g_B(x)+\ppin{k}\pi_k\g_k(x).
\eeq

Let us substitute this decomposition into the terms in our free kink Hamiltonian $H\p_2$ (\ref{h2eq}).  Note that, for now, the normal ordering will play no role in these manipulations.  The first term is proportional to
\beq
\int dx :\pi^2(x): \ = \ :\pi_0^2:+\ppin{k}:\pi_k\pi_{-k}:
\eeq
where we have used the orthonormality of the normal modes to perform the $x$ integrations.  

The rest is proportional to
\bea
&&\int dx \left[:\partial_x \phi(x)\partial_x\phi(x):+\V2:\phi^2(x):\right]\\
&&\ \ \ \ \ = \ :\phi_0^2:\int dx \g_B(x)\left(-\partial_x^2\g_B(x)+\V2 \g_B(x)\right)\nonumber\\
&&\ \ \ \ \ \ \ +2\ppin{k}:\phi_0\phi_k:\int dx \g_B(x)\left(-\partial_x^2\g_k(x)+\V2 \g_k(x)\right)\nonumber\\
&&\ \ \ \ \ \ \ +\ppin{k_1}\ppin{k_2}:\phi_{k_1}\phi_{k_2}:\int dx \g_{k_1}(x)\left(-\partial_x^2\g_{k_2}(x)+\V2 \g_{k_2}(x)\right)\nonumber\\
&&\ \ \ \ \ = \ :\phi_0^2:\int dx \g_B(x)\left(0\right)+2\ppin{k}:\phi_0\phi_k:\int dx \g_B(x)\left(\ok{}^2\g_k(x)\right)\nonumber\\
&&\ \ \ \ \ \ \ +\ppin{k_1}\ppin{k_2}:\phi_{k_1}\phi_{k_2}:\int dx \g_{k_1}(x)\left(\ok{2}^2 \g_{k_2}(x)\right)\nonumber\\
&&\ \ \ \ \ =\ppin{k}\ok{}^2:\phi_{k}\phi_{-k}:\nonumber
\eea
where in the second equality we used the Sturm-Liouville equation (\ref{sl}) and in the last we used the orthonormality of the normal modes.

Assembling these two terms together we find that the free kink Hamiltonian is
\beq
H\p_2=\frac{:\pi_0^2:}{2}+\frac{1}{2}\ppin{k} \left[:\pi_k\pi_{-k}:+\ok{}^2:\phi_k\phi_{-k}:\right].
\eeq
In other words, up to normal ordering, it is a sum of commuting quantum harmonic oscillators, one at each $k$, plus a free particle quantum mechanics model for the zero mode.  As each term is a bilinear, the role of normal ordering is only to introduce a constant.  Therefore, $H\p_2$ can be diagonalized, one normal mode at a time, and so we will be able to construct our kink ground state.

\subsection{The Kink Ground State}

As the normal ordering only shifts the free kink Hamiltonian $H\p_2$ by a constant $Q_1$, it does not affect the eigenstates and so we can construct the ground state of $H\p_2$ as the product of quantum harmonic oscillator ground states.

Remember that we have an oscillator for each $k$ above, where $k$ in our combined sum-integral runs both over shape modes and continuum modes.  For each oscillator we can construct creation and annihilation operators as in the case of the quantum harmonic oscillators in the vacuum sector
\beq
B^\ddag_k=\frac{\phi_k}{2}-\frac{i\pi_k}{2\omega_k}
\hsp
B_{-k}= \omega_k\phi_k+i\pi_k
\label{binvdec}
\eeq
where, in the case of the shape modes, the indices $-S$ and $S$ are to be identified as the modes are real.  For completeness, we state the inverse
\beq
\phi_k=B^\ddag_k+\frac{B_{-k}}{2\ok{}}\hsp \pi_k=i\left(\ok{}B^\ddag_k-\frac{B_{-k}}{2}\right)
\eeq
so that, in all, our decomposition is
\bea
\phi(x) &=&\phi_0 \mathfrak{g}_B(x)+\ppin{k} \left(B_k^{\ddag}+\frac{B_{-k}}{2 \omega_k}\right) \mathfrak{g}_k(x) \label{dec}\\
\pi(x) &=&\pi_0 \mathfrak{g}_B(x)+i \ppin{k}\left(\omega_k B_k^{\ddag}-\frac{B_{-k}}{2}\right) \mathfrak{g}_k(x). \nonumber
\eea

Inserting the algebra (\ref{phik}) into our decomposition (\ref{binvdec}) we obtain the algebra of our new and most useful basis of operators
\beq
[\phi_0,\pi_0]=i\hsp [B_{S_1},B^\ddag_{S_2}]=\delta_{S_1S_2}\hsp [B_{k_1},B^\ddag_{k_2}]=2\pi \delta(k_1-k_2). \label{bccr}
\eeq
We stress that any operator in the theory may be expressed in this $\{\phi_0,\pi_0,B^\ddag_S,B_S,B^\ddag_k,B_k\}$ basis.  The operators $B^\ddag$ and $B$ are the creation and annihilation opertors for our series of quantum harmonic oscillators, while the operators $\phi_0$ and $\pi_0$ are the position and momentum for the free quantum mechanical theory seen in $H\p_2$.  We will see shortly that the oscillators describe the excitations of the normal modes and the free quantum mechanics describes the motion of the center of mass of the kink.

The normal-ordering yields a constant $Q_1$, and so we may write the free kink Hamiltonian as
\beq
H\p_2=Q_1+\frac{\pi_0^2}{2}+\ppin{k}\omega_k B^\ddag_k B_k. \label{h2peq}
\eeq
Now it is clear that the ground state of $H\p_2$ is the state $\vac_0$ which satisfies
\beq
\pi_0\vac_0=B_S\vac_0=B_k\vac_0=0
\eeq
where $S$ runs over all shape modes and $k$ runs over all continuum normal modes.  In other words, all normal modes are in their ground states and the center of mass has no momentum.

Our leading order kink ground state is therefore
\beq
\df\vac_0.
\eeq
The condition that $\pi_0$ annihilates $\vac_0$ is the leading order expression for translation-invariance of $\df\vac_0$.

To see this, let us define the momentum operator
\beq
P=\int dx \phi(x)\partial_x\pi(x). \label{pdef}
\eeq
Conjugated with the displacement operator it becomes
\bea
P\p&=&\df^\dag P\df=P+\int dx f(x)\partial_x\pi(x)=P-\int dx f\p(x)\pi(x)\\
&=&P+\sqrt{Q_0}\int dx \g_B(x)\pi(x)=P+\sqrt{Q_0}\pi_0.\nonumber
\eea
Intuitively, the first term is the momentum operator for the mesons and the second for the kink's center of mass.  Translation invariance of the state $\df\vac_0$ is the vanishing of
\beq
P\df\vac_0=\df P\p \vac_0=\df\left(P+\sqrt{Q_0}\pi_0\right)\vac_0.
\eeq
Now $\vac_0$ only contains terms of order $O(\lambda^0)$.  On the other hand, $Q_0$ is of order $O(1/\lambda)$ and so the second term in the parenthesis is of the leading order.  Translation-invariance at leading order is thus the condition that $\pi_0\vac_0=0$, which we have seen is already required for $\vac_0$ to be a minimum energy eigenstate of $H\p_2$.  The translation-invariance was the \hyperlink{three}{third} claim at the beginning of this section.  From this the \hyperlink{five}{fifth} claim, that $\df\vac_0$ is not normalizable, follows.

As $\vac_0$ satisfies the eigenvalue equation
\beq
H\p_2\vac_0=Q_1\vac_0
\eeq
our leading ground state $\df\vac_0$ has, at leading order, energy given by the eigenvalue in
\beq
H\df\vac_0=\df H\p\vac_0=\df(H\p_0+H\p_2)\vac_0=(Q_0+Q_1)\df\vac_0.
\eeq
That is, the energy is not that of the coherent state, $Q_0$, but there is a further correction $Q_1$.  This is the celebrated one-loop correction of Ref.~\cite{dhn2}, which we will calculate in Subsec.~\ref{q1sez}.

We have thus shown that, up to order $O(\lambda^0)$, $\df\vac_0$ is a Hamiltonian eigenstate, which was the \hyperlink{one}{first} claim at the beginning of this section.  We have also seen that the energy differs from that of a coherent state which is part of the \hyperlink{two}{second} claim, although we have not yet established the sign of $Q_1$.

We are left with the \hyperlink{four}{fourth} claim, that $\df\vac_0$ is a squeezed coherent state.  Since $\df$ is a displacement operator, it suffices to show that $\vac_0$ is a squeezed state.  A squeezed state is a state which is annihilated by a set of operators which are a Bogoliubov transform of the usual $A_p$ annihilation operators.  In other words, we need to find a Bogoliubov transformation between the $A_p$ operators that annihilate the tree-level vacuum $|\Omega\rangle_0$ and the $\{\pi_0,B_k\}$ operators that annihilate $\vac_0$.  Here the index $k$ runs over the shape modes and the continuum modes, but not the zero mode.  We will adopt this convention for the rest of this section except when otherwise specified.

The Bogoliubov transformation is obtained by substituting the decomposition (\ref{adec}) into (\ref{bdec}) and substituting this into (\ref{binvdec}), yielding
\bea
\phi_0&=&\int dx \g_B(x) \phi(x)=
\pin{p}\tilde{\g}_B(p)\left(A^\ddag_p+\frac{A_{-p}}{2\omega_p}\right)\label{bog}\\
\pi_0&=&\int dx \g_B(x) \pi(x)=i\pin{p}\tilde{\g}_B(p)\left(\omega_p A^\ddag_p-\frac{A_{-p}}{2}\right)\nonumber
\\
B^\ddag_k&=&\int dx \g_{-k}(x)\left(\frac{\phi(x)}{2}-i\frac{\pi(x)}{2\ok{}} 
\right)=\pin{p}\tilde{\g}_{-k}(p)\left( 
\frac{\omega_p+\omega_k}{2\omega_k}A^\ddag_p+(\ok{}-\omega_p)\frac{A_{-p}}{4\ok{}\omega_p}
\right)\nonumber\\
B_{-k}&=&\int dx \g_{-k}(x)\left( \omega_k\phi(x)+i\pi(x)\right)
=\pin{p}\tilde{\g}_{-k}(p)\left((\ok{}-\omega_p) A^\ddag_p+\frac{\ok{}+\omega_p}{2\omega_p}A_{-p} 
\right)
\nonumber
\eea
in terms of the Fourier transforms
\beq
\tilde{\g}(p)=\int dx \g(x)e^{-ipx}.
\eeq
Note that in the case of continuum modes, when $k\sim p$, this transform implies that $B_k\sim A_p$.
The inverse transformation is
\bea
A^\ddag_p&=&\int dx \left( \frac{\phi(x)}{2}-\frac{i\pi(x)}{2\omega_p}
\right) e^{ipx}\label{ibog}\\
&=&\tilde{\g}_B(-p)\left(\frac{\phi_0}{2}-\frac{i\pi_0}{2\omega_p}\right)+\ppin{k}\tilde{\g}_k(-p)\left( 
\frac{\ok{}+\omega_p}{2\omega_p}B^\ddag_k+\frac{\omega_p-\ok{}}{4\ok{}\omega_p}   B_{-k}\right)\nonumber\\
A_{-p}&=&\int dx \left( \omega_p\phi(x)+i\pi(x)
\right) e^{ipx} \nonumber\\
&=&\tilde{\g}_B(-p)\left(\omega_p\phi_0+i\pi_0\right)+\ppin{k}\tilde{\g}_k(-p)\left( 
(\omega_p-\ok{})B^\ddag_k+\frac{\omega_p+\ok{}}{2\ok{}}   B_{-k}\right).\nonumber
\eea

\subsection{Interpretations and Excited States} \label{excsez}

First, in Subsec.~\ref{DfOmega0}, we constructed a coherent state $\df|\Omega\rangle_0$, corresponding to a fixed kink solution $f(x)$, and so broke translation invariance.  Then we found $\df\vac_0$ which is a Hamiltonian eigenstate, at least at leading order.   This corresponds to a superposition of solutions $f(x-x_0)$ for all $x_0$.  Now that we have restored translation-invariance, how can we discuss the location of the kink?

To do this, let us decompose our state $\vac_0$ into a superposition of position eigenstates $|y\rangle$ defined by
\beq
\phi_0|y\rangle=y|y\rangle\hsp B_S|y\rangle=B_k|y\rangle=0\hsp \langle y|y\rangle=1.
\eeq
This is a decomposition of $\vac_0$ because, up to a normalization constant
\beq
\vac_0=\int dy |y\rangle. \label{wfun}
\eeq
To see this, notice that, for any wavefunction $F(y)$
\beq
\pi_0\int dy F(y)|y\rangle=-i\int dy F\p(y)|y\rangle.
\eeq
Now let $F(y)$ be the wavefunction for $\vac_0$
\beq
\vac_0=\int dy F(y)|y\rangle. \label{wfuna}
\eeq
Then, using $\pi_0\vac_0=0$ we find
\beq
0=\pi_0\vac_0=-i\int dy F\p(y)|y\rangle.
\eeq
As the states $|y\rangle$ are linearly independent, the only solution to this equation is $F\p(y)=0$.  Substituting this into (\ref{wfuna}), up to a $y$-independent constant $F(y)$, it reduces to (\ref{wfun}).

We will now show that our new states $\df|y\rangle$ correspond to kink solutions $f(x-y)$.  As they are eigenstates of position, albeit for an extended object, they will have infinite energies.  Indeed, the $\pi_0^2$ term in $H\p_2$ has an infinite expectation value for such states.  But we will press on.

Let us compute the expectation value of the field in such a state
\bea
\langle y|\df^\dag \phi(x)\df|y\rangle&=&\langle y|\left(\phi(x)+f(x)\right)|y\rangle=\langle y|\left(\phi_0\g_B(x)+f(x)\right)|y\rangle
=f(x)+y\g_B(x)\nonumber\\
&=&f(x)-y\frac{f\p(x)}{\sqrt{Q_0}}=
f\left(x-\frac{y}{\sqrt{Q_0}}\right)+O(y^2).
\eea
We have learned that the collective coordinate $x_0$ is related to $y$ by
\beq
x_0=\frac{y}{\sqrt{Q_0}}+O(y^2).
\eeq
This is a relationship between eigenvalues, but it implies a similar relationship between the corresponding operators.  In other words, the operator $\phi_0/\sqrt{Q_0}$ is a linear approximation to the collective coordinate operator $x_0$.  This is the reason that our formalism is called {\it{linearized}} soliton perturbation theory, because we deal with $\phi_0$ instead of $x_0$.  The advantage of $\phi_0$ is its simple commutation relations (\ref{bccr}).

We have interpreted $\phi_0$ as the center of mass position of the kink, times $\sqrt{Q_0}$.  This suggests that $\sqrt{Q_0}\pi_0$ should be the momentum of the center of mass of the kink.  As the kink mass $Q$ is dominated by $Q_0$ which is of order $O(1/\lambda)$, our semiclassical expansion in $\lambda\hbar$ only applies to nonrelativistic kinks, and so we expect a potential energy of
\beq
\frac{(\sqrt{Q_0}\pi_0)^2}{2Q_0}=\frac{\pi_0^2}{2}.
\eeq
This is precisely the kinetic term that we found in $H\p_2$ in Eq.~(\ref{h2peq}).

What about $B^\ddag_S$?  Let us define the state
\beq
|S\rangle_0=B^\ddag_S\vac_0.
\eeq
Then, using the quantum harmonic oscillator terms in Eq.~(\ref{h2peq}), we may compute
\beq
(H\p_0+H\p_1+H\p_2)|S\rangle_0=(Q_0+Q_1+\omega_S)|S\rangle_0
\eeq
where we have also used the fact that $H\p_0=Q_0$ while $H\p_1$ vanishes.  We have learned that, $\df|S\rangle_0$ is, up to corrections of order $O(\sl)$, a Hamiltonian eigenstate whose mass is $\omega_S$ above the ground state.  Following Ref.~\cite{dhn2}, we identify this state not as the ground state kink, but as the kink with a single shape mode excited.  The same calculation shows that
\beq
\df|k\rangle_0=\df B^\ddag_k\vac_0
\eeq
is a kink plus a meson which, far from the kink, has momentum $k$.  Near the kink, the momenta of the meson and kink are not conserved separately.  Actually, as our state is annihilated by $\pi_0$, we are working in the center of mass and so the kink and meson have opposite momenta.  However the kink is nonperturbatively massive, and so its velocity cannot be detected at leading order.  The state $\df|k\rangle_0$ consists of a kink and a meson which is distributed throughout space.  The meson wavefunction is therefore nonvanishing both near the kink and also far.  However, only a finite amount of its normalization lies close to the kink while the remaining infinite part lies far away.  Therefore, the probability for a single meson in this state to lie close to the kink is zero, and so the kink cannot contribute to the meson's self-energy.

More generally we will define states
\beq
|k_1\cdots k_{\nn}\rangle_0=B^\ddag_{k_1}\cdots B^\ddag_{k_{\nn}}\vac_0
\eeq
where each $k_i$ can be a real number or a discrete index representing a shape mode.  These, after multiplication by $\df$, correspond to the kink excitations with excitation energies of $\sum_{i=1}^{\nn}\ok{i}$.  One can also make the center of mass move by multiplying by an operator of the form $e^{i\alpha\phi_0}$, which is useful for calculating form factors but we will not discuss it further.

\subsection{One-Loop Mass Correction}
\label{q1sez}

Now that we understand the physics behind the kink, we can turn to the calculation of the one-loop correction to its energy using Eq.~(\ref{h2peq})
\beq
Q_1= \frac{:\pi_0^2:-\pi_0^2}{2}+\ppin{k}\omega_k \left(:B^\ddag_k B_k:-B^\ddag_k B_k\right). \label{q1def}
\eeq

Recall that our normal ordering is defined in terms of the operators $A^\ddag_p$ and $A_{-p}$ but these expressions are written in the normal mode basis.  Therefore, to evaluate the normal ordering, one must first use the Bologiubov transformation. 

To evaluate, for example, $:\pi_0^2:-\pi_0^2$, we use the Bogoliubov transform (\ref{bog}) to write the expression as a bilinear in $A^\ddag_p$ and $A_{-p}$.  Normal ordering only affects the $A_{-p}A^\ddag_{-p}$ terms, and so the other terms will cancel between the $:\pi_0^2:$ and the $\pi_0^2$.  We conclude
\bea
:\pi_0^2:-\pi_0^2&=&-\pin{p_1}\pin{p_2}\tilde{\g}_B(p_1)\tilde{\g}_B(p_2)\left(-\frac{i}{2}\right)\left(i\omega_{p_2}\right)[A_{-p_1},A^\ddag_{p_2}]\\
&=&-\frac{1}{2}\pin{p}|\tilde{\g}_B(p)|^2\omega_p\nonumber
\eea
where we have used
\beq
\tilde{\g}_B(-p)=\tilde{\g}^*_B(p).
\eeq

Similarly, the contribution from the $k$th oscillator is proportional to
\bea
:B^\ddag_k B_k:-B^\ddag_k B_k&=-&\pin{p_1}\pin{p_2}\tilde{\g}_{-k}(p_1)\tilde{\g}_{k}(p_2)\left(\frac{\ok{}-\omega_{p_1}}{4\ok{}\omega_{p_1}}\right)\left(\ok{}-\omega_{p_2}\right)[A_{-p_1},A^\ddag_{p_2}]\nonumber\\
&=&-\frac{1}{4\ok{}}\pin{p}|\tilde{\g}_k(p)|^2\frac{(\ok{}-\omega_p)^2}{\omega_p}
\eea
where we have used
\beq
\tilde{\g}_{-k}(p)=\tilde{\g}^*_k(-p)
\eeq
and then changed the sign of the dummy variable $p$.

Assembling these contributions into Eq.~(\ref{q1def}) we find that the one-loop contribution to the kink mass is \cite{cahill76}
\beq
Q_1=-\frac{1}{4}\pin{p}|\tilde{\g}_B(p)|^2\omega_p-\frac{1}{4}\ppin{k}\pin{p}|\tilde{\g}_k(p)|^2\frac{(\ok{}-\omega_p)^2}{\omega_p} \label{ccg}
\eeq
which is manifestly negative, verifying our claim that the squeezed coherent states that we are constructing have a lower energy than the coherent states from the previous section.  The difference is finite, and so a coherent state will decay to a squeezed coherent state, perhaps with an excited shape mode, after emitting a meson or two. In higher dimensions the kink becomes a domain wall, our derivation above carries through with few modifications and one finds the same expression with a higher-dimensional integral.  In three or more spatial dimensions this integral is divergent, and so we learn that, in those cases, the squeeze is necessary to obtain a finite energy density.
Recall that we have defined $\dint$ such that shape modes are summed but not zero modes.  However, we note that, if we did include the zero mode in the sum, then the first term on the right hand side would be just this zero mode contribution because $(\omega_B-\omega_p)^2/\omega_p=\omega_p$.

In summary, we now understand the kink at one-loop.  This is one-loop in the sense of our perturbative expansion for the ground state $\vac$ of $H\p$, and its excited states $|k_1\cdots k_{\nn}\rangle$.  These lead to the states $\df\vac$ and $\df|k_1\cdots k_{\nn}\rangle$ which are very nonperturbative, because of the exponentiated inverse coupling in $\df$.  So far, however, we have constructed neither $\vac$ nor $|k_1\cdots k_{\nn}\rangle$ but merely their leading approximations $\vac_0$ and $|k_1\cdots k_{\nn}\rangle_0$.  We will turn to higher order corrections in the next section. 

\section{Higher Loop Corrections: Translation Invariance}

The real strength of LSPT is that it can be applied beyond one loop, unlike many of the more efficient competing formalisms such as spectral methods \cite{gw22} and the classical-quantum correspondence \cite{cq1,cq2}.  To go beyond one loop, we need two ingredients.  First, we need to understand the space of states in the kink sector, where we are working.  Next, we need to understand how the kink Hamiltonian acts on those states.

Once we have these two ingredients, our strategy will be to apply them to arrive at higher order corrections to our Hamiltonian eigenstates.  These higher order corrections may then be applied to directly determine mass corrections or may be assembled into wave packets to understand scattering.  

At leading order we can also construct unstable states and measure their lifetimes, however at higher orders these lifetimes are sensitive to choices in the constructions of unstable states, which are somewhat arbitrary as they are not Hamiltonian eigenstates.  Nonetheless, the lifetime may be defined as the inverse width of a resonance in which the unstable state is created.  In the future, we intend to calculate subleading corrections to lifetimes by calculating the scattering amplitudes of stable states and extracting the inverse widths of the corresponding resonances~\cite{bilguun}.  This reduces the problem of understanding unstable states to that of understanding wavepackets of stable states.

\subsection{The Kink Sector States}

We now have introduced many bases for our operators.  We started with the $\{\phi(x),\pi(x)\}$ basis.  Then we introduced the operators $\{A^\ddag_p,A_{-p}\}$ which create and destroy plane waves.  We folded the normal modes into the fields, generalizing the Fourier transform by using normal modes instead of plane waves, to arrive at the basis $\{\phi_0,\pi_0,\phi_S,\pi_S,\phi_k,\pi_k\}$ where $S$ runs over any discrete shape modes and $k$ is a real number labeling the continuum modes.  We recall that $\phi_0$ is a linear approximation to the kink's center of mass position and $\pi_0$ to the center of mass momentum.  Finally, we decomposed this into $\{\phi_0,\pi_0,B^\ddag_S,B_S,B^\ddag_k,B_{-k}\}$ where $B^\ddag_S$ and $B_S$ excite and de-excite shape modes and $B^\ddag_k$ and $B_{-k}$ are creation and annihilation operators for mesons. 

It will be most convenient to construct the kink sector using this last basis.  As $\phi_0$ and $\pi_0$ describe the quantum mechanics of a free particle, in this case the kink's center of mass, they play the same role as $x$ and $p$ in quantum mechanics.  Therefore a state is entirely characterized by a wavefunction $\psi(\phi_0)$ for each $|k_1\cdots k_{\nn}\rangle_0$.  For example, one state is
\beq
\psi_1(\phi_0)|3,S\rangle_0+\psi_2(\phi_0)|4,2\rangle_0.
\eeq
This state is a linear combination of two states.  The first has a meson with momentum $3$, a kink whose shape mode is excited and a position wavefunction $\psi_1$ while the second has two mesons with momenta $4$ and $2$, no shape mode excitations and a position wavefunction $\psi_2$.

Now we will make a rather strange choice.  We will decompose the wavefunctions $\psi(\phi_0)$ into monomials $\phi_0^{\mm}$.  We will choose the basis of states
\beq
\phi_0^{\mm}|k_1\cdots k_{\nn}\rangle_0.
\eeq
In other words, all of our wavefunctions will be polynomials.  This may seem absurd, as polynomials will never be normalizable, and in fact when we want to treat wave packets of kink-meson systems with different momenta we will need to assemble these states into wave packets by multiplying them by normalizable functions, such as Gaussians in $\phi_0$. 

Our main interest, however, will be translation-invariant states.  These are states which are invariant under a simultaneous translation of the kink and the mesons, which fixes the kink-meson separation.  Such states are sufficient for calculating processes such as kink-meson scattering.  These states are not normalizable, and, for the part of the wavefunction where $\phi_0$ is small, such polynomials provide a good approximation to the state.

Our choice is sensible for the following reason.  We will find that at each order in the coupling $\sl$, only a finite number of powers of $\phi_0$ appear in perturbation theory\footnote{At order $O(\lambda^j)$, a Hamiltonian eigenstate will have powers $\phi_0^\mm$ with $\mm\leq 4j$.}.  Therefore, our expansion in $\mm$ is consistent with the perturbative expansion.  How do we obtain translation-invariance?  We will impose it by demanding that $P$ annihilates the state, again order by order in the coupling.  This condition will turn out to make life much simpler.

Summarizing, we will write a general state in the kink sector as $\df|\Psi\rangle$ where
\beq
|\Psi\rangle=\sum_{\mm\nn}\ppink{\nn} \gamma^{\mm\nn}(k_1\cdots k_{\nn})
\phi_0^{\mm}|k_1\cdots k_{\nn}\rangle_0
\eeq
and $\gamma^{\mm\nn}(k_1\cdots k_{\nn})$ are complex coefficients.

In the case of a kink Hamiltonian eigenstate $|\Psi\rangle$, it is often convenient to decompose the state in powers of the coupling
\beq
|\Psi\rangle=\sum_{i=0}^\infty |\Psi\rangle_i\hsp |\Psi\rangle_i\sim O(\lambda^{i/2}). \label{psidec}
\eeq
Here we will refer to $|\Psi\rangle_i$ as the $(i/2)+1$ loop correction, and in fact a diagrammatic interpretation of the components will not be difficult to see\footnote{As is obvious from the fact that $i$ can be odd, in the presence of external legs, the number of loops in these diagrams will, unfortunately, not always equal $(i/2)+1$.}.   In this case, the coefficients at each order also factorize, and so one may write
\beq
|\Psi\rangle_i=\sum_{\mm\nn}\ppink{\nn} \gamma_{i}^{\mm\nn}(k_1\cdots k_{\nn})
\phi_0^{\mm}|k_1\cdots k_{\nn}\rangle_0 \label{gdec}
\eeq
where $\gamma_i^{\mm\nn}$ is of order $O(\lambda^{i/2})$.  The goal of the present section is to explain how the coefficients $\gamma_{i}$ may be calculated at each order, and to present $\gamma_1$ and $\gamma_2$ in the case of the kink's ground state $\df|\Psi\rangle=\df\vac$.  

We have already found the ground state at one loop, it is $\df\vac_0$ and it corresponds to
\beq
\gamma_0^{00}=1.
\eeq

We stress that the displacement operator $\df$ itself plays no role in the calculation of the coefficients $\gamma_i^{\mm\nn}$, once it has been used to calculate $H\p$.  Only after one calculates $|\Psi\rangle$, one may act on it with $\df$ to obtain a state.  However, to merely calculate matrix elements such as amplitudes, the $\df$ in the ket will cancel the $\df^\dag$ in the bra and so this step can be skipped.  Therefore in what follows we will not discuss $\df$ further, and sometimes we will loosely refer to $|\Psi\rangle$ as our state.  

More precisely, we will introduce frames as we did in quantum mechanics in Sec.~\ref{qmsez}.  The identification of quantum field theory kets and states used above, in which $H$ generates time evolution, will be referred to as the defining frame.  We will define the kink frame as follows.  If in the defining frame a state corresponds to the ket $\df|\Psi\rangle$ then in the kink frame that same state corresponds to the ket $|\Psi\rangle$.  In other words, we formalize dropping the $\df$ before each kink sector state, or more precisely multiplying each ket by $\df^\dag$ while not changing the physical state, by saying that we are working in the kink frame and not the defining frame.  In the kink frame, time translations are generated by $H\p$ and spatial translations by $P\p$.  We remind the reader that the choice of frame is a matter of convenience.  One could always work in the defining frame, and the price would merely be that one would need to write a $\df$ before each ket corresponding to a kink sector state.

\subsection{Translation Invariance}

LSPT can be applied to models without translation invariance.  For example, kink-impurity scattering and spectral walls \cite{swall} were considered in Ref.~\cite{chris}.  However, in these lecture notes we consider potentials $V$ with no explicit $x$-dependence.  As a result our models are translation-invariant and so the Hamiltonian $H$ commutes with the momentum $P$.  Therefore, each Hamiltonian eigenspace can be decomposed into momentum eigenspaces.  

Consider a state which in the kink frame is represented by $|\psi\rangle$ and so in the defining frame is represented by $\df|\psi\rangle$.  We will be interested in Hamiltonian eigenspaces that contain a state with no total momentum.  After all, a state with momentum can be obtained by a boost.  Therefore, we will be interested in states $\df|\Psi\rangle$ such that
\beq
P\df|\Psi\rangle=0
\eeq
where $P$ was defined in Eq.~(\ref{pdef}).  There we saw that this condition is equivalent to
\beq
P\p|\Psi\rangle=0\hsp P\p=\df^\dag P\df=P+\sqrt{Q_0}\pi_0 \label{ppinv}
\eeq
where $P$ is the momentum carried by the mesons and $\sqrt{Q_0}\pi_0$ is that of the kink's center of mass.  In other words, in the defining frame, the kets corresponding to translation-invariant states are those annihilated by $P$ while in the kink frame, where we will work from now on, the kets corresponding to the same states are in the kernel of $P\p$.  In the present subsection, we will solve this condition order by order.

Let us start by decomposing the state $|\Psi\rangle$ in powers of the coupling.  Recalling that $Q_0$ is of order $O(1/\lambda)$, substituting this decomposition (\ref{psidec})  into the translation-invariance condition (\ref{ppinv}) one finds
\beq
\pi_0|\Psi\rangle_0=0
\eeq
at leading order and, at higher orders, the recursion relation
\beq
\pi_0|\Psi\rangle_{i+1}=-\frac{P}{\sqrt{Q_0}}|\Psi\rangle_i\hsp i\geq 0. \label{rra}
\eeq
The left hand side is a translation of the kink's center of mass to the right, while the right hand side is a translation of the mesons to the left.  Invariance under a rigid translation of the entire kink meson system implies that these two operations are equivalent.
If $\pi_0$ were invertible, this recursion relation would yield every order $|\Psi\rangle_i$ in terms of the leading order $|\Psi\rangle_0$, which, for the Hamiltonian eigenstates, we have already found in Subsec.~\ref{excsez}.  In other words, translation-invariance alone would determine all multi-loop corrections to our Hamiltonian eigenstates.

However, $\pi_0$ is not invertible.  Recall that our basis of kink-sector states is, with the $\df$ peeled-off as always\footnote{Recall that $P$ is the translation operator on the ket expressed in the defining frame $\df|\Psi\rangle$ while $P\p$ is the translation operator in the kink frame, where the same state is denoted by the ket $|\Psi\rangle$.  We are trying to calculate the kink frame kets $|\Psi\rangle$ such that the corresponding state is translation-invariant.}, $\phi_0^{\mm}|k_1\cdots k_{\nn}\rangle_0$.  $\pi_0$ acts on this basis as
\beq
\pi_0 \phi_0^{\mm}|k_1\cdots k_{\nn}\rangle_0=-i\mm\phi_0^{\mm-1}|k_1\cdots k_{\nn}\rangle_0.  \label{piaz}
\eeq
Therefore the kernel of $\pi_0$ consists of all states with $\mm=0$, in other words with no powers of $\phi_0$.

We will refer to the states spanned by $|k_1\cdots k_\nn\rangle_0$ as primary states and those generated by $\phi_0^\mm|k_1\cdots k_\nn\rangle_0$ with $\mm>1$ as descendants.  Therefore the kernel of $\pi_0$ consists of primary states.  Given $|\Psi\rangle_i$, the recursion relation (\ref{rra}) fixes only the descendant part of $|\Psi\rangle_{i+1}$, but does not constrain the primary part.  

It is convenient to use our decomposition (\ref{gdec}) in terms of the matrices $\gamma_i^{\mm\nn}$.  Then the recursion relation fixes $\gamma_{i+1}^{\mm\nn}$ with $\mm\geq 1$ in terms of all $\gamma_i^{\mm\p\nn\p}$.  

To write this relation explicitly, we will need to decompose $P$ in our normal mode basis.  More precisely, if we define vectors $\vec{\phi}=(\phi_0,\phi_k)$ and $\vec{\pi}=(\pi_0,\pi_k)$ where $k$ runs over the shape modes and also all real numbers corresponding to continuum modes, then\footnote{This matrix notation statement is rewritten in integral notation in the first line of Eq.~(\ref{pdeca}).} $P=\vec{\phi}^\perp \Delta\vec{\pi}$.  Here the matrix $\Delta$ is\footnote{The matrices $\Delta$ in the case of the Sine-Gordon and $\phi^4$ models are given in Appendix~\ref{sapp}.}
\beq
\Delta_{k_1k_2}=\int dx \g_{k_1}(x)\partial_x\g_{k_2}(x) \label{deldef}
\eeq
where $k_1$ and $k_2$ run over $B$, $S$ and also all real numbers corresponding to continuum modes.  Note that
\beq
\Delta_{k_1k_2}+\Delta_{k_2k_1}=\g_{k_1}(\infty)\g_{k_2}(\infty)-\g_{k_1}(-\infty)\g_{k_2}(-\infty).
\eeq
When $k_1$ and $k_2$ is a discrete mode, then the corresponding normal mode $\g$ vanishes at $\pm\infty$ and so this vanishes.  When both are continuum modes, so long as $k_1\neq -k_2$, this difference oscillates quickly at large $|x|$ and so, as a distribution, vanishes by the Lebesgue Lemma.  If $k_1=-k_2$ then the terms at $\pm\infty$ are equal and so cancel.  Needless to say, we do not attempt to treat these distributions rigorously, but at this intuitive level, we conclude that
\beq
\Delta_{k_1k_2}=-\Delta_{k_2k_1}. \label{asyeq}
\eeq

We will now return to our usual convention in which $k$ runs over continuum and shape modes but not the zero mode.  Using the antisymmetry (\ref{asyeq}), we decompose $P$ as
\bea
P&=&\int dx \phi(x)\partial_x\pi(x)= \ppin{k} \Delta_{Bk} (\phi_0\pi_k-\phi_k\pi_0)+\ppin{k_1}\ppin{k_2}\Delta_{k_1k_2}\phi_{k_1}\pi_{k_2}\label{pdeca}\\
&=&\ppin{k} \Delta_{kB} 
\left[ 
B^\ddag_k\left( 
\pi_0-i\ok{}\phi_0
\right)
+B_{-k}\left(
\frac{\pi_0}{2\ok{}}+\frac{i}{2}\phi_0
\right)
\right]
\nonumber\\
&&\hspace{-.5cm}+i\ppin{k_1}\ppin{k_2}\Delta_{k_1k_2}
\left[\frac{\ok{2}-\ok{1}}{2}B^\ddag_{k_1}B^\ddag_{k_2}-\frac{\ok{1}+\ok{2}}{2\ok{2}}B^\ddag_{k_1}B_{-k_2}+\frac{\ok{1}-\ok{2}}{8\ok{1}\ok{2}}B_{-k_1}B_{-k_2}
\right].
\nonumber
\eea

The action of $P$ on our basis elements is then
\bea
P\phi_0^\mm|k_1\cdots k_\nn\rangle_0&=&-i\ppin{k\p}\Delta_{k\p B}\left(\mm \phi_0^{\mm-1}
+\okp{}\phi_0^{\mm+1}\right)|k_1\cdots k_\nn,k\p\rangle_0\label{paz}\\
&&
+i\sum_{j=1}^\nn\Delta_{-k_j B}\left(-\frac{\mm}{2\ok{j}}\phi_0^{\mm-1} +\frac{1}{2}\phi_0^{\mm+1}\right)|k_1\cdots\hat{k}_j\cdots k_{\nn}\rangle_0\nonumber\\
&&+i\ppin{k\p_1}\ppin{k\p_2}\Delta_{k\p_1k\p_2}\frac{\okp{2}-\okp{1}}{2}\phi_0^\mm|k_1\cdots k_{\nn},k\p_1,k\p_2\rangle_0\nonumber\\
&&-i\sum_{j=1}^\nn\ppin{k\p}\Delta_{k\p,-k_j}\frac{\okp{}+\ok{j}}{2\ok{j}}\phi_0^\mm|k_1\cdots \hat{k}_j\cdots k_{\nn},k\p\rangle_0\nonumber\\
&&+i\sum_{j_1=1}^{\nn-1}\ \sum_{j_2=j_1+1}^{\nn}\Delta_{-k_{j_1},-k_{j_2}}\frac{\ok{j_1}-\ok{j_2}}{4\ok{j_1}\ok{j_2}}\phi_0^\mm|k_1\cdots \hat{k}_{j_1}\cdots \hat{k}_{j_2}\cdots k_{\nn}\rangle_0\nonumber
\eea
where the $\hat{k}$ means that a given $k$ has been omitted.

Note that our decomposition (\ref{gdec}) is ambiguous, in that the each $\gamma(k_1\cdots k_\nn)$ is contracted with the state $|k_1\cdots k_\nn\rangle_0$ which is symmetric under permutations of the $k_i$.  When convenient, we will fix this ambiguity by demanding that each $\gamma$ is symmetric under permutations of its arguments.


Inserting Eq.~(\ref{paz}) into our decomposition (\ref{gdec}) and imposing the convention that $\gamma_i^{\mm\nn}(k_1\cdots k_\nn)$ is symmetric in its arguments, one obtains
\bea
P|\Psi\rangle_i&=&\sum_{\mm\nn}\ppink{\nn} \gamma_{i}^{\mm\nn}(k_1\cdots k_{\nn})
P\phi_0^{\mm}|k_1\cdots k_{\nn}\rangle_0 \label{ppsi}\\
&=&i\sum_{\mm\nn}\left[ -\ppink{\nn}
\Delta_{k_\nn B}\left[\mm  \gamma_{i}^{\mm,\nn-1}(k_1\cdots k_{\nn-1})+\ok{\nn} \gamma_{i}^{\mm-2,\nn-1}(k_1\cdots k_{\nn-1})\right]\right.
\nonumber\\
&&+\ppink{\nn+1}  
(\nn+1)\Delta_{-k_{\nn+1} B}\left[-\frac{\mm}{2\ok{\nn+1}}\gamma_{i}^{\mm,\nn+1}(k_1\cdots k_{\nn+1})  +\frac{1}{2}\gamma_{i}^{\mm-2,\nn+1}(k_1\cdots k_{\nn+1})
\right]
\nonumber\\
&&+\ppink{\nn} 
\Delta_{k_{\nn-1}k_{\nn}}\frac{(\ok{\nn}-\ok{\nn-1})}{2}\gamma_{i}^{\mm-1,\nn-2}(k_1\cdots k_{\nn-2})
\nonumber\\
&&-\ppink{\nn+1} 
\nn\Delta_{k_\nn,-k_{\nn+1}}\frac{(\ok{\nn}+\ok{\nn+1})}{2\ok{\nn+1}}\gamma_{i}^{\mm-1,\nn}(k_1\cdots k_{\nn-1},k_{\nn+1})
\nonumber\\
&&\left.\hspace{-2cm}+\ppink{\nn+2} 
\frac{(\nn+2)(\nn+1)}{8}\Delta_{-k_{\nn+2},-k_{\nn+1}}\frac{(\ok{\nn+2}-\ok{\nn+1})}{\ok{\nn+2}\ok{\nn+1}}\gamma_{i}^{\mm-1,\nn+2}(k_1\cdots k_{\nn+2}) \right]\phi_0^{\mm-1}|k_1\cdots k_{\nn}\rangle_0
\nonumber
\eea
for any state $|\Psi\rangle$ such that $\df|\Psi\rangle$ is in the kink sector. Here $\sum_{\mm\nn}$ is to be understood as $\sum_{\mm=0}\sum_{\nn=0}$ and we adopt the convention that $\gamma_i^{mn}$ vanishes whenever $m<0$ or $n<0$. The recursion relation (\ref{rra}) identifies (\ref{ppsi}) with
\beq
-\sqrt{Q_0}\pi_0|\Psi\rangle_{i+1}=i\sqrt{Q_0}\sum_{\mm\nn}\mm\ppink{\nn} \gamma_{i+1}^{\mm\nn}(k_1\cdots k_{\nn})
\phi_0^{\mm-1}|k_1\cdots k_{\nn}\rangle_0.
\eeq

Matching the coefficients of each $\phi_0^{\mm-1}|k_1\cdots k_\nn\rangle_0$ we arrive at the recursion relation \cite{me2loop}
\bea
\sqrt{Q_0}\gamma_{i+1}^{\mm\nn}(k_1\cdots k_\nn)&=&\left.-\Delta_{k_\nn B}\left(\gamma_i^{\mm,\nn-1}(k_1\cdots k_{\nn-1})+\frac{\omega_{k_\nn}}{\mm}\gamma_i^{\mm-2,\nn-1}(k_1\cdots k_{\nn-1})\right)
\right. \label{rra2}\\
&&+\frac{\nn+1}{2}\ppin{k\p}\Delta_{-k\p B}\left(-\frac{\gamma_i^{\mm,\nn+1}(k_1\cdots k_\nn,k\p)}{\omega_{k\p}}
+\frac{\gamma_i^{\mm-2,\nn+1}(k_1\cdots k_\nn,k\p)}{\mm}\right)\nonumber\\
&&-\frac{\omega_{k_{\nn-1}}\Delta_{k_{\nn-1}k_\nn}}{\mm}\gamma_i^{\mm-1,\nn-2}(k_1\cdots k_{\nn-2})\nonumber\\
&&-\frac{\nn}{2\mm}\ppin{k\p}\Delta_{k_\nn,-k\p}\left(1+\frac{\omega_{k_\nn}}{\omega_{k\p}}\right)\gamma^{\mm-1,\nn}_i(k_1\cdots k_{\nn-1},k\p)
\nonumber\\
&&\left.+\frac{(\nn+2)(\nn+1)}{4\mm}\ppinkp{2}\frac{\Delta_{-k\p_1,-k\p_2}}{\omega_{k\p_2}} \gamma_i^{\mm-1,\nn+2}(k_1\cdots k_{\nn},k\p_1,k\p_2).
\right.
\nonumber
\eea
Recall that on the right hand side, it is assumed that the coefficients $\gamma_i$ are already symmetrized in their arguments $k_j$.  The recursion relation gives $\gamma_{i+1}$ with arguments that are not symmetrized, but must be symmetrized before one substitutes them into the right hand side to find $\gamma_{i+2}$.  The fact that this recursion relation may only be applied to the coefficients with $\mm\geq 1$ is clear, as the right hand side contains $\mm$ in the denominators.

In summary, we have arrived at a powerful result (\ref{rra2}).  It states that any translation-invariant state may be decomposed into symmetric functions $\gamma$ as in (\ref{gdec}) and that the $\gamma_{i+1}$ at order $O(\lambda^{(i+1)/2})$ may be determined from those of $\gamma_i$ at order $O(\lambda^{i/2})$ except for the terms $\gamma^{0\nn}_{i+1}$ which contain no zero modes.  Said differently, the zero mode coefficients are determined entirely from translation invariance.  This observation will greatly simplify the task of finding Hamiltonian eigenstates, as it remains only to find the coefficients $\gamma^{0\nn}_i$.

The fact that the translation symmetry fixes all but a relatively small subset of the coefficient leads one to wonder which coefficients are fixed and left unfixed by other symmetries.  In particular, in the future we would like to perform a similar decomposition of the Hilbert space of a supersymmetric theory, to understand to what extent the states are fixed by supersymmetry and to interpret supersymmetric nonrenormalization theorems from this point of view.  In particular, in the case of $\mathcal{N}=2$ superQCD, where supersymmetry implies that the monopole mass only receives instanton corrections, we would like to see just how much of the monopole state is fixed by the supersymmetry.  Is it enough to find the monopole state at strong coupling?  Or does the state factorize into a part that can be found at strong coupling and a part that cannot but however does not affect the mass?

\subsection{Example: The Ground State}

As an example of the use of this recursion relation, let us consider the kink ground state $\df\vac$.  The goal is to find $\vac_i$.  The leading term of course is $\vac_0$ and so
\beq
\gamma^{\mm\nn}_0=\delta_{\mm 0}\delta_{\nn 0}.
\eeq

A single application of the recursion relation then yields
\beq\label{gamma121}
\gamma_1^{21}(k)=-\frac{\Delta_{kB}\ok{}}{2\sqrt{Q_0}}\hsp \gamma_1^{12}(k_1,k_2)=-\frac{\Delta_{k_1k_2}\ok{1}}{\sqrt{Q_0}}
\eeq
and $\gamma_1^{\mm\nn}=0$ for all other combinations of $\mm\neq 0$ and $\nn$.  We remind the reader that only the part of the coefficients $\gamma_i^{\mm\nn}(k_1\cdots k_{\nn})$ which is completely symmetric in the $k_i$ arguments will contribute to the state $\vac_i$, and so one can freely modify that coefficients so long as the completely symmetric part remains unchanged.  

For use in the next recursion relation, to obtain $\gamma_2$, we will need to symmetrize the $\gamma_1^{12}$
\beq
\gamma_1^{12}(k_1,k_2)=\frac{\Delta_{k_1k_2}(\ok{2}-\ok{1})}{2\sqrt{Q_0}}.
\eeq
The primary coefficients $\gamma^{0\nn}_1$ are not constrained by translation invariance, and so to find them we will need to solve the Hamiltonian eigenvalue problem.

We will see later, when treating the Hamiltonian eigenvalue problem, that the only nonvanishing primary coefficients at order $O(\sl)$ are $\gamma_1^{01}$ and $\gamma_1^{03}$.  The recursion relations fix the descendants $\gamma_2^{\mm\nn}$ in terms of these, although there remains usual ambiguity of a shift which vanishes when fully symmetrized.  Those nonvanishing coefficients with a single zero mode $\phi_0$ are
\bea
\sqrt{Q_0}\gamma_2^{11}(k)&=&-\frac{1}{2}\pin{k\p}\Delta_{k,-k\p}\left(1+\frac{\ok{}}{\okp{}}\right)\gamma_1^{01}(k\p)+\frac{3}{2}\ppinkp{2}\frac{\Delta_{-k\p_1,-k\p_2}}{\omega_{k\p_2}} \gamma_1^{03}(k,k\p_1,k\p_2)\nonumber\\
&&-\ppin{k\p}\Delta_{-k\p B}\frac{\gamma_1^{12}(k,k\p)}{\okp{}}\nonumber\\
\sqrt{Q_0}\gamma_2^{13}(k_1,k_2,k_3)&=&
-{\omega_{k_{2}}\Delta_{k_{2}k_3}}{}\gamma_1^{01}(k_1)
-\frac{3}{2}\ppin{k\p}\Delta_{k_3,-k\p}\left(1+\frac{\omega_{k_3}}{\omega_{k\p}}\right)\gamma^{03}_1(k_1,k_2,k\p)\nonumber\\
&&-\Delta_{k_3B}\gamma_1^{12}(k_1,k_2)\nonumber\\
\sqrt{Q_0}\gamma_2^{15}(k_1\cdots k_5)&=&
-\omega_{k_{4}\Delta_{k_{4}k_5}}\gamma_1^{03}(k_1,k_2,k_3).
\eea
Those nonvanishing coefficients with two zero modes, or more precisely $\phi_0^2$, are
\bea
\sqrt{Q_0}\gamma_2^{20}&=&
\ppin{k\p}\Delta_{-k\p B}\left(-\frac{\gamma_1^{21}(k\p)}{2\omega_{k\p}}
+\frac{\gamma_1^{01}(k\p)}{4}\right)
+\frac{1}{4}\ppinkp{2}\frac{\Delta_{-k\p_1,-k\p_2}}{\omega_{k\p_2}} \gamma_1^{12}(k\p_1,k\p_2)\nonumber\\
\sqrt{Q_0}\gamma_2^{22}(k_1,k_2)&=&
-\Delta_{k_2 B}\left(\gamma_1^{21}(k_1)+\frac{\omega_{k_2}}{2}\gamma_1^{01}(k_1)\right)
-\frac{1}{2}\ppin{k\p}\Delta_{k_2,-k\p}\left(1+\frac{\omega_{k_2}}{\omega_{k\p}}\right)\gamma^{12}_1(k_1,k\p)\nonumber\\
&&+\frac{3}{4}\ppin{k\p}\Delta_{-k\p B}\gamma_1^{03}(k_1,k_2,k\p)\nonumber\\
\sqrt{Q_0}\gamma_2^{24}(k_1\cdots k_4)&=&
-\frac{\ok 4}{2}\left(\Delta_{k_4 B}\gamma_1^{03}(k_1,k_2,k_3)+\Delta_{k_4k_3}\gamma_1^{12}(k_1,k_2)\right).\label{duez}
\eea
Finally, the nonvanishing coefficients with three or four zero modes are
\bea
\sqrt{Q_0}\gamma_2^{31}(k_1)&=&
\frac{1}{3}\ppin{k\p}\Delta_{-k\p B}{\gamma_1^{12}(k_1,k\p)}
-\frac{1}{6}\ppin{k\p}\Delta_{k_1,-k\p}\left(1+\frac{\omega_{k_1}}{\omega_{k\p}}\right)\gamma^{21}_1(k\p)\nonumber\\
\sqrt{Q_0}\gamma_2^{33}(k_1,k_2,k_3)&=&
-\frac{\omega_{k_3}}{3}\left(\Delta_{k_3 B}\gamma_1^{12}(k_1,k_2)+\Delta_{k_{3}k_2}\gamma_1^{21}(k_1)\right)\nonumber\\
\sqrt{Q_0}\gamma_2^{40}&=&
\frac{1}{8}\ppin{k\p}\Delta_{-k\p B}{\gamma_1^{21}(k\p)}\nonumber\\
\sqrt{Q_0}\gamma_2^{42}(k_1,k_2)&=&-\frac{\omega_{k_2}\Delta_{k_2 B}}{4}\gamma_1^{21}(k_1).
\eea

We note that so far we have not used the form of the potential, and so the results above are valid in a model with any potential.  The potential will only affect the terms above via the value of the primary coefficients $\gamma_i^{0\nn}$.

\section{Higher Loop Corrections: Hamiltonian Eigenstates} \label{twosez}
\subsection{Wick's Theorem} \label{wsez}

Our states are written in terms of the action of the basis $\{\phi_0,\pi_0,B^\ddag_S,B_S,B^\ddag_k,B_{-k}\}$ on the leading order ground state $\vac_0$.  To find eigenstates of the kink Hamiltonian $H\p$, it will therefore be convenient to rewrite $H\p$ in this basis.

We have already written $H\p_2$ in this basis in Eq.~(\ref{h2peq}).  The higher-order terms in the kink Hamiltonian were given in Eq.~(\ref{hnp}).  These are normal-ordered in terms of the $\{A^\ddag_p,A_{-p}\}$ basis, which is not the basis that we are using.  In the case of $H\p_2$ this mismatch merely led to a constant shift $Q_1$.  However, in the case of an $n$-point interaction like that in $H\p_n$, the change in normal-ordering leads to a change in the $(n-2)$-point coupling, the $(n-4)$-point coupling and so on.  Therefore, before we may proceed, we need to make the normal ordering compatible with our basis.

With this goal in mind, we will introduce a new normal ordering $::_b$ called normal mode normal ordering.  Given any operator $\co$, we define the operator $:\co:_b$ as follows.  One first expresses $\co$ in terms of the $\{\phi_0,\pi_0,B^\ddag_S,B_S,B^\ddag_k,B_{-k}\}$ basis.  One then moves all $\pi_0,$\ $B_S$\ and $B_k$ to the end.  Remember, as in the case of ordinary normal ordering, when one moves the annihilation operators to the right one should not add the commutators.  That would result in the same operator, whereas normal ordering should change the operator.  For example
\beq
:B_{k_1}\pi_0 B^\ddag_{k_2}\phi_0:_b=B^\ddag_{k_2}\phi_0B_{k_1}\pi_0.
\eeq

The interactions in our kink Hamiltonian~(\ref{hnp}) consist of weighted sums of plane wave normal ordered $n$-point functions
\beq
:\phi^n(x):.
\eeq
To act as simply as possible on our basis of states, we would like to rewrite them in terms of normal mode normal ordered operators.  

How can we do this?  This question was answered in Ref.~\cite{mewick}.  Let us start with the case $j=2$.  In this case, in Appendix~\ref{capp} it is shown that
\beq
:\phi^2(x): \ = \ :\phi^2(x):_b+\I(x) \label{idef}
\eeq
where we have defined the contraction factor $\I(x)$ to be
\beq
\I(x)=\pin{k}\frac{\left|\g_{k}(x)\right|^2-1}{2\omega_k}+\sum_S\frac{\left|{\g}_{S}(x)\right|^2}{2\omega_S}. \label{ival}
\eeq
What about other values of $j$?  In Ref.~\cite{mekink} we showed that
\beq
:\phi^j(x): \ =\sum_{m=0}^{\lfloor{\frac{j}{2}}\rfloor}\frac{j!}{2^m m!(j-2m)!}\I^m(x):\phi^{j-2m}(x):_b\label{wick}
\eeq
and we referred to this result as Wick's theorem.  The combinatoric factor simply states that there is a factor of $\I(x)$ for each possible contraction of two fields at the same $j$-point vertex.

Intuitively, $\I(x)$ corresponds to a contraction between the two copies of the field in $:\phi^2:$ which computes the difference between the commutator of the $A^\ddag_p$ and $A_p$ components and the $B^\ddag_k$ and $B_k$ components.  The $A$ basis and $B$ basis commutators are both divergent, which is the reason that normal-ordering eliminates ultraviolet divergences in our theories.  However the difference $\I(x)$ is bounded and well-behaved, at least in 1+1 dimensions.  It does, however, depend on the position.  It vanishes far from the kink, where the plane waves and normal modes become equal, up to a phase shift.  

In the vacuum sector, normal ordering eliminates diagrams in which a loop begins and ends at the same vertex.  In the kink sector, normal mode normal ordering does the same thing.  However, as the original Hamiltonian was plane-wave normal ordered, there will be $\I(x)$ terms once we normal mode normal order.  And these $\I(x)$ terms can be depicted precisely as such loops, beginning and ending at the same vertex.  Therefore, in this sense, the loops are back in the kink sector.  However, since $\I(x)$ is not divergent, these loops are not divergent.

In summary, one has two ways to diagrammatically represent the interactions of $H\p$.  First, one may first express everything in terms of normal mode normal ordered operators.  In this case, there are many interactions, with $\I(x)$-dependent coefficients, but there are no loops beginning and ending at the same vertex.  Alternately one may write the diagrams before changing the normal ordering.  In this case no interactions depend on $\I(x)$ and there are not so many interactions, but one must remember to sum over diagrams including loops starting and ending at the same vertex, with a factor of $\I(x)$ for each loop.

Going up to two loops, in other words calculating the leading and subleading corrections $|\Psi\rangle_1$ and $|\Psi\rangle_2$ to states, we will only need cubic and quartic interactions.  In these cases, Wick's theorem is
\beq
:\phi^3(x): \ = \ :\phi^3(x):_b+3\I(x)\phi(x)\hsp :\phi^4(x): \ = \ :\phi^4(x):_b+6\I(x):\phi^2(x):_b+3\I^2(x).
\eeq
Inserting these into Eq.~(\ref{hnp}) we find that the leading interactions in the Hamiltonian density are
\bea
:\ch_{3}\p(x):&=&\sl\V{3}\left(\frac{:\phi^3(x):_b}{6}+\frac{\I(x)\phi(x)}{2}\right)\label{hnno}\\
:\ch_{4}\p(x):&=&\lambda\V{4}\left(\frac{:\phi^4(x):_b}{24}+\frac{\I(x):\phi^2(x):_b}{4}+\frac{\I^2(x)}{8}\right).\nonumber
\eea

\subsection{Interaction Vertices}
Let us consider the leading interaction $\ch\p_3$.  The corresponding term in the kink Hamiltonian is
\beq
H\p_3=\int dx :\ch\p_3(x): \ =\sl\int dx\left[ \V{3}\left(\frac{:\phi^3(x):_b}{6}+\frac{\I(x)\phi(x)}{2}\right)\right].
\eeq
In the vacuum sector, instead of $\V{3}$ we would have a position-independent coupling.  We could expand $\phi(x)$ in the momentum basis in terms of plane waves, and then perform the $x$ integration.  The $x$ integration would give a momentum-conserving Dirac $\delta$ and we would be left with a coupling constant that could be inserted into Feynman rules.  

The situation here in a kink sector is a bit different.  For one, the coupling $\V3$ depends on the position.  Second, if we decompose $\phi(x)$ in a normal mode basis, the coefficients are our normal modes $\g$ instead of plane waves.  However we may still perform the $x$-integration, and whatever we find will be our $k$-dependent couplings which can again be used to draw diagrams if desired.

Consider, for example, a tadpole interaction
\beq
\int dx \V3\I(x)\phi(x).
\eeq
Expanding $\phi(x)$ in terms of normal modes (\ref{dec}), this is
\bea
\int dx \V3\I(x)\phi(x)\,\,\,&=&\,\,\,\int dx \V3\I(x)\left[ 
\phi_0 \mathfrak{g}_B(x)+\ppin{k} \left(B_k^{\ddag}+\frac{B_{-k}}{2 \omega_k}\right) \mathfrak{g}_k(x) 
\right]\nonumber\\
&=&V_{\I B}\phi_0+\ppin{k}V_{\I k}\left(B_k^{\ddag}+\frac{B_{-k}}{2 \omega_k}\right)
\eea
where we have defined the shorthand notation
\beq
V_{\I B}=\int dx \V3 \I(x) \g_B(x)\hsp
V_{\I k}=\int dx \V3 \I(x) \g_k(x).
\eeq
These play the role of our coupling constants.  Each factor of $\I(x)$ we recall corresponded to a commutator in our Wick's theorem, and so can be seen as a loop beginning and ending at the same point.  It begins and ends at the same point because it involves two factors of $\phi(x)$ at the same interaction, the two which are commuted.  Then $V_{\I B}$ is a vertex factor for a tadpole consisting of a vertex at which there is a loop and also a zero mode is created.  Similarly, $V_{\I k}$ corresponds to tadpole consisting of a vertex with a single loop and a meson which is either created, corresponding to $B^\ddag_k$, or destroyed, corresponding to $B_{-k}$.  Recall that in our condensed notation, $k$ may also be a shape mode $S$, and so our notation includes $V_{\I S}$ at which a shape mode is excited or de-excited.  

The short-hand notation is easily generalized to\footnote{The three-point functions $V$ in the case of the Sine-Gordon and $\phi^4$ models are given in Appendix~\ref{sapp}.}
\beq
V_{\I^m k_1\cdots k_n}=\int dx \V{2m+n} \I^m(x) \g_{k_1}(x)\cdots\g_{k_n}(x). \label{vdef}
\eeq
This is the coupling constant corresponding to a vertex hosting $m$ loops and also creating or destroying $n$ shape modes and mesons.  Clearly, one may also replace any number $j$ of these indices $k$ with $B$ indices, in which case it corresponds to a vertex at which $j$ zero modes are created.

With this notation in hand, we may rewrite $H\p_3$ in $k$-space as
\bea
H\p_3\,\,\,&=&\,\,\,\sl\left[ 
\ppink{3}V_{k_1k_2k_3}\left(\frac{\Bd1\Bd2\Bd3}{6}+\frac{\Bd1\Bd2 B_{-k_3}}{4\ok 3}+\frac{\Bd1 B_{-k_2} B_{-k_3}}{8\ok 2\ok 3}+\frac{B_{-k_1}B_{-k_2}B_{-k_3}}{48\ok 1\ok 2\ok 3}
\right)\right.\nonumber\\
&&+\phi_0\ppink{2} V_{Bk_1k_2}\left(\frac{\Bd1\Bd2}{2}+\frac{\Bd1 B_{-k_2}}{2\ok 2}+\frac{B_{-k_1}B_{-k_2}}{8\ok 1\ok 2}
\right)\nonumber\\
&&+\left.\ppin{k} \left(\phi^2_0 V_{BBk}+V_{\I k}\right)\left(\frac{\Bd{}}{2}+\frac{B_{-k}}{4\ok{}}\right)+\phi_0\frac{V_{\I B}}{2}\right]. \label{h3k}
\eea
Note that there is no $V_{BBB}$ term.  This is because $V_{BBB}$ vanishes as the result of a Ward Identity for the translation symmetry, which we will study in Subsec.~\ref{wardsez}.  Were there a $V_{BBB}$ term, then,  in the case of the kink ground state, $\gamma^{30}_1$ would have been nonzero, but we have seen above that translation symmetry allows only the descendants $\gamma^{12}_1$ and $\gamma^{21}_1$ at order $O(\sl)$.

At order $O(\lambda)$ one finds the interaction
\bea
H\p_4\,\,\,&=&\,\,\,\int dx:\ch_{4}\p(x): \ =\,\,\int dx \lambda\V{4}\left(\frac{:\phi^4(x):_b}{24}+\frac{\I(x):\phi^2(x):_b}{4}+\frac{\I^2(x)}{8}\right)\label{h4dec}\\
&=&\lambda\left[ 
\ppink{4}V_{k_1k_2k_3k_4}\left(\frac{\Bd1\Bd2\Bd3\Bd4}{24}+\frac{\Bd1\Bd2\Bd3 B_{-k_4}}{12\ok 4}+\frac{\Bd1\Bd2 B_{-k_3} B_{-k_4}}{16\ok 3\ok 4}\right.\right.\nonumber\\
&&\left.+\frac{\Bd1 B_{-k_2}B_{-k_3}B_{-k_4}}{48\ok 2\ok 3\ok 4}+\frac{B_{-k_1}B_{-k_2}B_{-k_3}B_{-k_4}}{384\ok 1\ok 2\ok 3\ok 4}
\right)\nonumber\\
&&+\phi_0\ppink{3}V_{Bk_1k_2k_3}\left(\frac{\Bd1\Bd2\Bd3}{6}+\frac{\Bd1\Bd2 B_{-k_3}}{4\ok 3}+\frac{\Bd1 B_{-k_2} B_{-k_3}}{8\ok 2\ok 3}+\frac{B_{-k_1}B_{-k_2}B_{-k_3}}{48\ok 1\ok 2\ok 3}
\right)\nonumber\\
&&+\ppink{2} \left(\phi^2_0V_{BBk_1k_2}+V_{\I k_1k_2}\right)\left(\frac{\Bd1\Bd2}{4}+\frac{\Bd1 B_{-k_2}}{4\ok 2}+\frac{B_{-k_1}B_{-k_2}}{16\ok 1\ok 2}
\right)\nonumber\\
&&\left.+\ppin{k}\left(\phi_0^3V_{BBBk}+3\phi_0 V_{\I B k}\right)\left(\frac{\Bd{}}{6}+\frac{B_{-k}}{12\ok{}}\right)+\frac{\phi_0^4 V_{BBBB}+6\phi_0^2 V_{\I BB}+3 V_{\I^2}}{24}\right].\nonumber
\eea

\subsection{The Hamiltonian Eigenvalue Problem}

To compute the primary coefficients, we need to use the kink Hamiltonian.  It will be most convenient to do this in the normal mode normal-ordered form.  

Let us ignore the $c$-number terms $H\p_0=Q_0$ and also the $Q_1$ in $H\p_2$, as these merely contribute constants $Q_0$ and $Q_1$ to the energies of kink sector states, that can easily be inserted later.  They contribute to the eigenvalues of $H\p$ but not to its eigenvectors.

Let us expand the eigenvalue equation (\ref{vse}) for $\vac$ order by order in the coupling.  At leading order we found $H\p_2\vac_0=Q_1\vac_0$ which defined $\vac_0$ and so is already solved.  At order $O(\sl)$ one finds
\beq
H\p_2\vac_1=-H\p_3\vac_0. \label{se1}
\eeq
There is no energy correction, which will be clear momentarily as we will be able to solve this equation with no energy correction.

Using $H\p_2$ from Eq.~(\ref{h2peq}), without the $Q_1$, the left hand side is
\bea
H\p_2\vac_1&=&\left(\frac{\pi_0^2}{2}+\ppin{k}\omega_k B^\ddag_k B_k\right)\sum_{\mm\nn}\ppink{\nn} \gamma_1^{\mm\nn}(k_1\cdots k_{\nn})
\phi_0^{\mm}|k_1\cdots k_{\nn}\rangle_0\label{sel}\\
&=&\sum_{\mm\nn}\ppink{\nn} \gamma_1^{\mm\nn}(k_1\cdots k_{\nn})
\left[-\frac{\mm(\mm-1)}{2}\phi_0^{\mm-2}+\phi_0^{\mm}\sum_{j=1}^\nn \ok{j}
\right]|k_1\cdots k_{\nn}\rangle_0\nonumber\\
&&\hspace{-2cm}=\sum_{\mm\nn}\ppink{\nn} 
\left[-\frac{(\mm+2)(\mm+1)}{2}\gamma_1^{\mm+2,\nn}(k_1\cdots k_{\nn})+\gamma_1^{\mm\nn}(k_1\cdots k_{\nn})\sum_{j=1}^\nn \ok{j}
\right]\phi_0^{\mm}|k_1\cdots k_{\nn}\rangle_0.\nonumber
\eea

To evaluate the right hand side of Eq.~(\ref{se1}), we will use $H\p_3$ from Eq.~(\ref{h3k})
\bea
H\p_3\vac_0&=&\sl\left[ 
\ppink{3}\frac{V_{k_1k_2k_3}}{6}{|k_1k_2k_3\rangle_0}+\ppink{2} \frac{V_{Bk_1k_2}}{2}{\phi_0|k_1k_2\rangle_0}\right. \label{ser}\\
&&+\left.\ppin{k} \left( \frac{V_{BBk}}{2}\phi_0^2|k\rangle_0+\frac{V_{\I k}}{2}|k\rangle_0\right)+\frac{V_{\I B}}{2}\phi_0\vac_0\right].\nonumber
\eea
Our eigenvalue equation implies that, term by term, Eqs.~(\ref{sel}) and (\ref{ser}) differ by a sign.

Let us first match the coefficients of $|k\rangle_0$.  On the left hand side this corresponds to $\mm=0$ and $\nn=1$.  We then find
\beq
-\gamma_1^{21}(k)+\gamma_1^{01}(k)\ok{}=-\frac{\sl V_{\I k}}{2}.
\eeq
As the descendant $\gamma_1^{21}(k)$ was already determined in Eq.~(\ref{gamma121}) using translation invariance, we can fix the primary coefficient
\beq
\gamma_1^{01}(k)=\frac{2\gamma_1^{21}(k)-\sl V_{\I k}}{2\ok{}}=-\frac{\Delta_{kB}}{2\sqrt{Q_0}}-\frac{\sl V_{\I k}}{2\ok{}}.
\eeq
Similarly, matching the coefficients of $|k_1k_2k_3\rangle_0$ with a sign flip we find the other primary
\beq
\gamma_1^{03}(k_1,k_2,k_3)=-\frac{\sl V_{k_1k_2k_3}}{6(\ok1+\ok2+\ok3)}
\eeq
where we have used the fact that $\gamma_1^{23}(k_1,k_2,k_3)$ vanishes as a result of the recursion relation~(\ref{rra2}).

\subsection{The Two-Loop Kink Mass}

To calculate the two-loop correction $Q_2$ to the mass of a ground state kink\footnote{Actually, $Q_2$ is the energy of the state with the kink.  To arrive at the kink mass, one needs to subtract the energy of the vacuum.  Both are in general infrared divergent.  In the case of the $\phi^4$ model this divergence arises already at two loops.  However the divergences are easily canceled by adding a $c$-number counterterm to the Hamiltonian which is fixed by hand to make the vacuum energy vanish.  This was done for the $\phi^4$ model in Ref.~\cite{phi42loop}.}, we need only solve the eigenvalue equation at order $O(\lambda)$
\beq
H\p_4\vac_0+H\p_3\vac_1+H\p_2\vac_2=Q_2\vac_0 \label{se2}
\eeq
where for brevity we have again dropped the $c$-number $Q_1$ contribution to $H_2\p$.

Let us expand the $i$th order correction to the ground state $\vac_i$ as
\beq
\vac_i=\sum_{\mm\nn}\vac_i^{\mm\nn}\hsp
\vac_i^{\mm\nn}=\ppink{\nn} \gamma_i^{\mm\nn}(k_1\cdots k_{\nn})
\phi_0^{\mm}|k_1\cdots k_{\nn}\rangle_0.
\eeq
We will match the coefficient of $\vac_0$ on both sides of Eq.~(\ref{se2}).  We will use the notation $\supset$ to denote the projection of a state onto $\vac_0$.

\begin{figure} 
\begin{center}
\includegraphics[width=2.8in,height=1.7in]{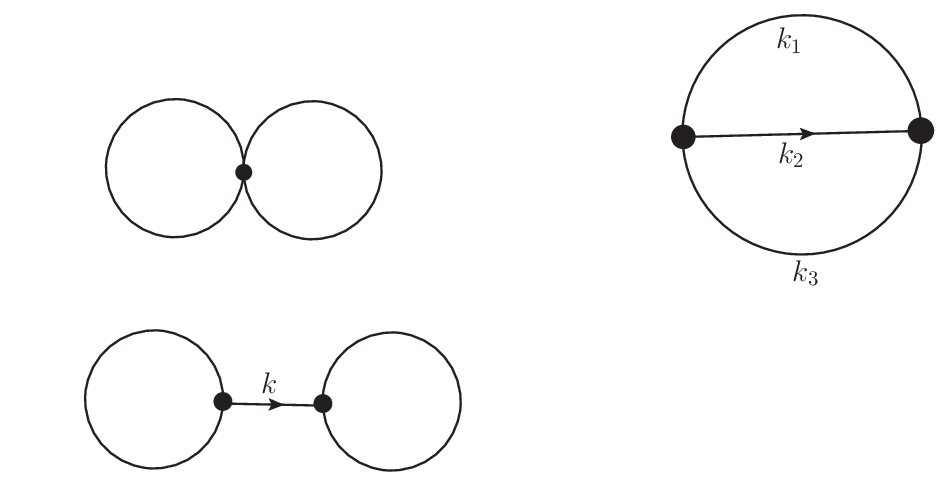}
\caption{The first three terms in $Q_2$ can be represented abstractly as two-loop diagrams.  Each factor of $\I(x)$ corresponds to a loop at the vertex $x$ while each $\g_k(x)$ is an internal line corresponding to a meson with momentum $k$ or a shape mode.}
\label{q2afig}
\end{center}
\end{figure}

Our goal is to evaluate each term in Eq.~(\ref{se2}), projected on to the $\vac_0$ component.  The first term in Eq.~(\ref{se2}), after this projection, only receives a contribution from the $c$-number part of $H\p_4$, since any other term will either annihilate $\vac_0$ or else produce mesons, shape modes or zero modes.  There is only one $c$-number term in $H\p_4$, it is the last term in Eq.~(\ref{h4dec}).  It leads to
\beq
H\p_4\vac_0\supset \lambda \frac{V_{\I^2}}{8}\vac_0.
\eeq
Although we have not defined diagrams at this point, it intuitively corresponds to a diagram consisting of a single vertex with two loops starting and ending at the vertex.  It is the first diagram in Fig.~\ref{q2afig}.

The second contribution in Eq.~(\ref{se2}) is the most complicated, consisting of the other two diagrams in Fig.~\ref{q2afig} with two three-point interaction vertices, two loops and no external legs.  In Fig.~\ref{q2bfig} one can see the contribution with a $\Delta$ factor, which has zero modes that are annihilated by the kinetic term $\pi_0^2/2$, as will be clear later when we turn to Ward Identities.  Using the expansion of $H\p_3$ in Eq.~(\ref{h3k}), the contributions to this term are
\bea
H\p_3\vac_1&\,\,\supset\,\,&
\sl\ppink{3}V_{k_1k_2k_3}\frac{B_{-k_1}B_{-k_2}B_{-k_3}}{48\ok 1\ok 2\ok 3}\vac_1^{03}+\sl\ppin{k}V_{\I k}\frac{B_{-k}}{4\ok{}}\vac_1^{01}\\
&&\hspace{-.9cm}=\,-\lambda \ppink{3}\frac{|V_{k_1k_2k_3}|^2\vac_0}{48\ok 1\ok 2\ok 3(\ok 1+\ok 2+\ok 3)}-\,\lambda \ppin{k}\left[ \frac{|V_{\I k}|^2}{8\ok{}^2}
+\frac{1}{\sqrt{\lambda Q_0}}\frac{V_{\I k}\Delta_{-k,B}}{8\ok{}}
\right]\vac_0
\nonumber
\eea
where we have used the identity
\beq
V^*_{\I^m k_1...k_n}=V_{\I^m,-k_1,...,-k_n}
\eeq
which follows from $\g^*_k=\g_{-k}$.  The analogous identity also applies to the $\Delta$ symbol.

The last contribution on the left hand side of Eq.~(\ref{se2}) can either arise from the kinetic term $\pi_0^2/2$ of $H\p_2$ acting on $\vac_2^{20}$ or else the oscillator terms in $H\p_2$ acting on $\vac_2^{00}$.  However, the oscillator terms annihilate $\vac_2^{00}$ and more generally $\vac_2^{00}$ is unconstrained.  This is because it may be absorbed into the overall normalization of $\vac$.  On the other hand, using Eq.~(\ref{duez}) we find
\bea
\gamma_2^{20}&=&
\frac{1}{\sqrt{Q_0}}\ppin{k\p}\Delta_{-k\p B}\left(-\frac{\gamma_1^{21}(k\p)}{2\omega_{k\p}}
+\frac{\gamma_1^{01}(k\p)}{4}\right)
+\frac{1}{4\sqrt{Q_0}}\ppinkp{2}\frac{\Delta_{-k\p_1,-k\p_2}}{\omega_{k\p_2}} \gamma_1^{12}(k\p_1,k\p_2)\nonumber\\
&=&\,\,\,
\frac{1}{\sqrt{Q_0}}\ppin{k\p}\Delta_{-k\p B}\left(\frac{\Delta_{k\p B}}{4\sqrt{Q_0}}
-\frac{\Delta_{k\p B}}{8\sqrt{Q_0}}-\frac{\sl V_{\I k\p}}{8\okp{}}\right)
+\frac{1}{4\sqrt{Q_0}}\ppinkp{2}\frac{|\Delta_{k\p_1k\p_2}|^2(\okp2-\okp1)}{2\sqrt{Q_0}\okp 2}\nonumber\\
&=&\,\,\,
\ppin{k\p}\left(\frac{|\Delta_{k\p B}|^2}{8Q_0}
-\frac{\sl \Delta_{-k\p B}V_{\I k\p}}{8\sqrt{Q_0}\okp{}}\right)
-\ppinkp{2}\frac{|\Delta_{k\p_1k\p_2}|^2(\okp2-\okp1)^2}{16Q_0\okp 1\okp 2}.\nonumber
\eea
We conclude that the third term in the eigenvalue equation (\ref{se2}) is
\bea
H\p_2\vac_2&\supset&\frac{\pi_0^2}{2}\vac_2^{20}=-\gamma_2^{20}\vac_0\\
&=&
\left[ \ppin{k\p}\left(-\frac{|\Delta_{k\p B}|^2}{8Q_0}
+\frac{\sl \Delta_{-k\p B}V_{\I k\p}}{8\sqrt{Q_0}\okp{}}\right)
+\ppinkp{2}\frac{|\Delta_{k\p_1k\p_2}|^2(\okp2-\okp1)^2}{16Q_0\okp 1\okp 2}\right] \vac_0.\nonumber
\eea
Adding these three contributions and, following Eq.~(\ref{se2}), identifying the coefficient with the two-loop mass correction $Q_2$, we obtain our master formula for the two-loop correction to a kink mass
\bea
Q_2&=&\lambda \frac{V_{\I^2}}{8}
-\lambda \ppink{3}\frac{|V_{k_1k_2k_3}|^2 }{48\ok 1\ok 2\ok 3(\ok 1+\ok 2+\ok 3)}-\,\lambda \ppin{k} \frac{|V_{\I k}|^2}{8\ok{}^2}\label{q2pad}\\
&&-\ppin{k}\frac{|\Delta_{k B}|^2}{8Q_0}
+\ppink{2}\frac{|\Delta_{k_1k_2}|^2(\ok2-\ok1)^2}{16Q_0\ok 1\ok 2}.
\nonumber
\eea
The three terms on the first line have simple diagrammatic interpretations, shown in Fig.~\ref{q2afig}, as two-loop diagrams with one four-point vertex in the first term and in the other two terms, two three-point vertices connected by three and one internal meson lines respectively.  The terms on the second line, shown in Fig.~\ref{q2bfig}, are related to translation-invariance, as $\Delta$ consists of matrix elements of the momentum operator.  To see their connection to zero modes, we will next turn to the Ward Identities.

\subsection{Ward Identities} \label{wardsez}

Whenever an $n$-point vertex $V$ contains a zero mode index $B$, translation-invariance allows it to be decomposed into more primitive vertices.  This is because each $B$ index corresponds to a zero mode related to the kink momentum, which is just minus the meson momentum for a translation-invariant state.

To see that the various $n$-point vertices are related, it suffices to take the $x$-derivative
\beq
\partial_x \V{n}=\V{n+1}\partial_x(\sl f(x))=-\sqrt{\lambda Q_0}\V{n+1}\g_B(x)
\eeq
where we remind the reader that the notation $V^{(n)}$ denotes the $n$th derivative of $V$ with respect to its argument.  Applying this to the three-point function $V_{BBB}$, one finds
\bea
V_{BBB}&=&\int dx \V3 \g^3_B(x)=-\frac{1}{\sqrt{\lambda Q_0}}\int dx \left(\partial_x\V2\right)\g_B^2(x)\\
&=&\frac{2}{\sqrt{\lambda Q_0}}\int dx \V2 \g_B(x)\g\p_B(x)=\frac{2}{\sqrt{\lambda Q_0}}\int dx \g\pp_B(x)\g\p_B(x)\nonumber\\
&=&\frac{1}{\sqrt{\lambda Q_0}}\int dx \partial_x(\g_B^{\prime 2}(x))=0\nonumber
\eea
where in the second line we used the Sturm-Liouville equation (\ref{sl}) satisfied by $\g_B(x)$.

\begin{figure} 
\begin{center}
\includegraphics[width=2.8in,height=1.7in]{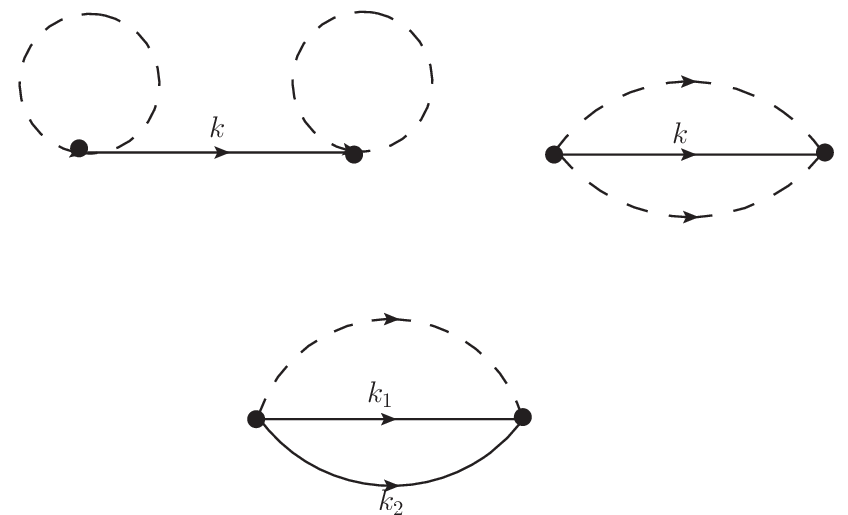}
\caption{The last three terms in $Q_2$ can be represented abstractly as two-loop diagrams involving zero modes, which are depicted as dashed lines.}
\label{q2bfig}
\end{center}
\end{figure}

Next let us replace a zero mode with a meson
\bea
V_{BBk}&=&\int dx \V3 \g^2_B(x)\g_k(x)\\
&=&-\frac{1}{\sqrt{\lambda Q_0}}\int dx \left(\partial_x\V2\right)\g_B(x)\g_k(x)\nonumber\\
&=&\frac{1}{\sqrt{\lambda Q_0}}\int dx \V2 \left(\g_B(x)\g\p_k(x)+\g_k(x)\g\p_B(x)\right)\nonumber\\
&=&\frac{1}{\sqrt{\lambda Q_0}}\int dx \left( \g\pp_B(x)\g\p_k(x)+\g\pp_k(x)\g\p_B(x)+\ok{}^2\g_k(x)\g\p_B(x)
\right) \nonumber\\
&=&\frac{1}{\sqrt{\lambda Q_0}}\int dx  \partial_x(\g_B^{\prime}(x)\g\p_k(x))+\frac{\ok{}^2}{\sqrt{\lambda Q_0}}\int dx \g_k(x)\g\p_B(x)\nonumber\\
&=&\frac{\ok{}^2}{\sqrt{\lambda Q_0}}\Delta_{kB}.
\eea
This means that our momentum matrix element $\Delta_{kB}$ is proportional to the vertex factor for a three-point vertex which creates two zero modes, corresponding to the operator $\phi_0^2$ and also creates or annihilates a meson, corresponding to the operators $B^\ddag_k$ and $B_{-k}$.  For example, the $\Delta_{kB}$ term in $Q_2$ in Eq.~(\ref{q2pad}) corresponds to the sum of the two diagrams on the top line of Fig.~\ref{q2bfig}.

What about the vertex with two mesons and a single zero mode?  In that case
\bea
V_{Bk_1k_2}&=&\int dx \V3 \g_B(x)\g_{k_1}(x)\g_{k_2}(x)\\
&=&-\frac{1}{\sqrt{\lambda Q_0}}\int dx \left(\partial_x\V2\right)\g_{k_1}(x)\g_{k_2}(x)\nonumber\\
&=&\frac{1}{\sqrt{\lambda Q_0}}\int dx \V2 \left(\g_{k_1}(x)\g\p_{k_2}(x)+\g_{k_2}(x)\g\p_{k_1}(x)\right)\nonumber\\
&=&\frac{1}{\sqrt{\lambda Q_0}}\int dx \left[\partial_x\left( \g\p_{k_1}(x)\g\p_{k_2}(x)\right)+\ok{1}^2\g_{k_1}(x)\g\p_{k_2}(x)+\ok{2}^2\g_{k_2}(x)\g\p_{k_1}(x)\right] \nonumber\\
&=&\frac{(\ok{1}^2-\ok{2}^2)}{\sqrt{\lambda Q_0}}\Delta_{k_1 k_2}.\nonumber
\eea
We learn that the momentum matrix element $\Delta_{k_1k_2}$ is proportional to the three-point vertex with two mesons and a single zero mode.  For example, the last term in $Q_2$ in Eq.~(\ref{q2pad}) is shown on the bottom line of Fig.~\ref{q2bfig}.

In summary, we have learned that three point vertices sourcing zero modes are proportional to our momentum matrix elements $\Delta$.  What about higher interactions $\V{n}$?  Here the same procedure does not apply as we do not have an equation satisfied by $\V{n-1}\g$.  Also, we do not obtain another $V_{k_1\cdots k_n}$ interaction as, after integration by parts, will have some $\g\p$.  Our strategy will be to separate out the $\g\p$ factors using the completeness relation
\beq
\g_B(x)\g_B(y)+\ppin{k}\g_{k}(x)\g^*_k(y)=\delta(x-y) \label{cr}
\eeq
which is obtained by inverting the normalization conditions (\ref{norm1}), (\ref{norm2}) and (\ref{norm3}).  The normalization matrix is invertible because the normal modes are a basis of the space of bounded functions, which is a general property of solutions of the Sturm-Liouville equation.  The completeness relation will allow us to insert a factor of unity in the form
\beq
G(x)=\int dy \delta(x-y)G(y)=\int dy\left[ 
\g_B(x)\g_B(y)+\ppin{k}\g_{k}(x)\g^*_k(y)
\right]G(y).
\eeq
For example, we may use it to simplify
\bea
\int dx \V3 \g_{k_1}(x)\g_{k_2}(x)\g\p_{k_3}(x)&=&\int dx\int dy \V3 \g_{k_1}(x)\g_{k_2}(x)\nonumber\\
&&\hspace{-3cm}\ \ \times\left[ 
\g_B(x)\g_B(y)+\ppin{k\p}\g_{k\p}(x)\g^*_{k\p}(y)
\right] \g\p_{k_3}(y)\nonumber\\
&&\hspace{-3cm}=V_{Bk_1k_2}\Delta_{Bk_3}+\ppin{k\p}V_{k_1k_2k\p}\Delta_{-k\p,k_3}.
\eea
Recalling that $\Delta$ consists of matrix elements of the momentum operator $P$, we conclude that this is an action of the momentum operator on $V$, as one might expect from an $x$ derivative.

Consider for example $V_{Bk_1k_2k_3}$.  This can be decomposed as follows
\bea
V_{Bk_1k_2k_3}&=&\int dx \V4 \g_B(x)\g_{k_1}(x)\g_{k_2}(x)\g_{k_3}(x)\label{vbkkk}\\
&=&-\frac{1}{\sqrt{\lambda Q_0}}\int dx \left(\partial_x\V3\right)\g_{k_1}(x)\g_{k_2}(x)\g_{k_3}(x)\nonumber\\
&=&\frac{1}{\sqrt{\lambda Q_0}}\int dx \V3 \left[\g_{k_1}(x)\g_{k_2}(x)\g\p_{k_3}(x)+\g_{k_1}(x)\g_{k_3}(x)\g\p_{k_2}(x)\right.\nonumber\\
&&\left.\ \ \ +\g_{k_2}(x)\g_{k_3}(x)\g\p_{k_1}(x)\right]\nonumber\\
&=&\frac{1}{\sqrt{\lambda Q_0}}\left[
V_{Bk_1k_2}\Delta_{Bk_3}+V_{Bk_2k_3}\Delta_{Bk_1}+V_{Bk_3k_1}\Delta_{Bk_2}\right.\nonumber\\
&&\left. +\ppin{k\p}\left( 
V_{k\p k_1k_2}\Delta_{-k\p, k_3}+V_{k\p k_2k_3}\Delta_{-k\p, k_1}+V_{k\p k_3k_1}\Delta_{-k\p, k_2}
\right)
\right]\nonumber\\
&=&\frac{(\ok{1}^2-\ok{2}^2)\Delta_{k_1k_2}\Delta_{Bk_3}+(\ok{2}^2-\ok{3}^2)\Delta_{k_2k_3}\Delta_{Bk_1}+(\ok{3}^2-\ok{1}^2)\Delta_{k_3k_1}\Delta_{Bk_2}}{\lambda Q_0}\nonumber\\
&&+\frac{1}{\sqrt{\lambda Q_0}}\ppin{k\p}\left( 
V_{k\p k_1k_2}\Delta_{-k\p, k_3}+V_{k\p k_2k_3}\Delta_{-k\p, k_1}+V_{k\p k_3k_1}\Delta_{-k\p, k_2}
\right).\nonumber
\eea
Note that this result also follows if any of the $k_i$ are replaced by $B$, although in that case one will arrive at $V_{Bkk}$ terms in the last line which may be rewritten as $\Delta$ symbols.  

The generalization to higher-point functions is straightforward \cite{elaslungo}. An $n$-point function containing a single index $B$ can be rewritten as a sum on $(n-1)$-point functions, where the $B$ index and another index $k$ were removed, integrated over an index $k\p$ which may be $B$ and contracted with $\Delta_{-k\p, k}$.  One then sums over which index $k$ was removed.  In the case of the terms in which $k\p$ is $B$, one may again apply the Ward identity to reduce to $(n-2)$-point functions, corresponding to the second to last line in Eq.~(\ref{vbkkk}), and so on.

In Ref.~\cite{me2loop}, Ward Identities were also derived for interaction vertices including the contraction factor $\I$, using a derivation similar to those above.  For example, in Eq.~(A.9) of that reference it was shown that
\beq
V_{\I B}=0.
\eeq

\subsection{Example: One Kink and One Meson}

Let us consider a kink Hamiltonian eigenstate $|\kt\rangle$ consisting of one ground state kink and one meson with asymptotic momentum $\kt$.  At leading order this state is
\beq
|\kt\rangle_0=B^\ddag_\kt\vac_0.
\eeq
Note that the current discussion can be applied equally to a kink with an excited shape mode and no mesons $|S\rangle_0=B^\ddag_S\vac_0$, by simply replacing $\kt$ with the abstract index $S$.

The quantum corrections to this state can be found by the same methods as were applied to the ground state $\vac$ of a lone kink above.  First one writes the coefficient $\gamma$ at leading order
\beq
\gamma_{0\kt}^{01}(k)=2\pi\delta(k-\kt)
\eeq
where the $0$ in the subscript indicates the leading order and the $\kt$ indicates that we are now studying the one kink one meson state $|\kt\rangle$.
Next, at order $O(\sl)$, one uses the recursion relation to compute the descendants
\bea
\gamma_{1\kt}^{11}(k_1)&=&\frac{\Delta_{-\kt k_1}}{2\sqrt{Q_0}}\left(1+\frac{\ok1}{\omega_{\kt}}\right)\hsp
\gamma_{1\kt}^{13}(k_1,k_2,k_3)=\frac{\ok3\Delta_{k_2k_3}}{\sqrt{Q_0}}2\pi\delta(k_1-\kt)\nonumber\\
\gamma_{1\kt}^{22}(k_1,k_2)&=&-\frac{\ok2 \Delta_{k_2 B}}{2\sqrt{Q_0}}2\pi\delta(k_1-\kt)\hsp
\gamma_{1\kt}^{20}=\frac{\Delta_{-\kt B}}{4\sqrt{Q_0}}.\label{gammakt}
\eea
Here we have not bothered to symmetrize over the momenta since we do not intend to substitute this into the recursion relation again to obtain the order $i=2$ corrections $\gamma_{2\kt}^{\mm\nn}$.

Solving the Hamiltonian eigenvalue problem yields the primaries
\bea
\gamma_{1\kt}^{00}&=& \frac{\sl V_{\I, -\kt}}{4\omega^2_{\kt}}-\frac{\Delta_{-\kt B}}{4\sqrt{Q_0}\ \omega_{\kt}}\label{kt}\\
\gamma_{1\kt}^{02}(k_1,k_2)&=& -\frac{\sl V_{\I  k_1}}{2\ok1}2\pi\delta(k_2-\kt) -\frac{\Delta_{k_2 B}}{2\sqrt{Q_0}}2\pi\delta(k_1-\kt)+\frac{\sl V_{-\kt k_1 k_2}}{4\omega_{\kt}\left(\omega_{\kt}-\ok1-\ok2\right)}\nonumber\\
\gamma_{1\kt}^{04}(k_1\cdots k_4)&=& -\frac{\sl V_{k_1 k_2 k_3}}{6\sum_{j=1}^3 \ok{j}}2\pi\delta(k_4-\kt)
.\nonumber
\eea
Note that $\gamma_{1\kt}^{02}$ has a pole at $\omega_\kt=\ok1+\ok2$, just where the process
\beq
{\rm{kink\ +\ meson}}\rightarrow {\rm{kink\ +\ 2\ mesons}} \label{mmt}
\eeq
conserves energy.  There are many prescriptions for defining this pole.  The different prescriptions correspond to different definitions of the state $|\kt\rangle$ and they all satisfy the Hamiltonian eigenvalue equation.  These different definitions all correspond to Hamiltonian eigenstates, all with the same eigenvalue.  Which eigenstate one should choose, and so which prescription should be used to define the pole, depends on the state which is desired for a given application.  When we turn to scattering in Sec.~\ref{scatsez} we will see, for example, that the $+i\epsilon$ prescription yields a suitable {\it{in}} state for scattering.

\subsection{Summary}

In Sec.~\ref{unosez}, working at one-loop, we found the spectrum of the kink sector.  More precisely we found the spectrum of $H\p_2$.  In the center of mass frame, it consists of the kink ground state $\vac_0$ plus arbitrary excitations of shape modes, achieved with $B^\ddag_S$, and mesons, which are created with $B^\ddag_k$.  

In the present section, we described how to lift an eigenstate of $H\p_2$ to an eigenstate of the full $H\p$, order by order in the semiclassical expansion.   We worked out in detail the case of the kink ground state at two loops, however any other choice of $H\p_2$ eigenstate can be treated similarly.  One need merely choose $\gamma_0^{0\nn}$ to include a total of $\nn$ mesons and shape modes, and then use the recursion relation to get the descendants at each order and the Hamiltonian eigenvalue problem to arrive at the primaries.   For example, the two-loop state corresponding to a $\phi^4$ kink with an excited shape mode was explicitly constructed in Ref.~\cite{unbind}.


\section{Translation-Invariant Matrix Elements from the Reduced Inner Product}

\subsection{The Initial Value Problem}

Given a theory, there are two basic problems in physics.  The first is finding the time-independent states.  In a quantum theory, these are the Hamiltonian eigenstates.  Above we described how this problem can be solved perturbatively for states consisting of a kink plus bound excitations and unbound mesons.

The second problem is the initial value problem.  One begins with a state $|t=0\rangle$ at time $0$ and asks the state at time $t$.  In a quantum theory this is
\beq
e^{-iHt}|t=0\rangle.
\eeq
This final state is a vector in an abstract Hilbert space.  To convert it into a probability that a given outcome, corresponding to the state $|f\rangle$ will be measured, one need only take the norm-squared of the normalized amplitude
\beq
\frac{\langle f|e^{-iHt}|t=0\rangle }{\sqrt{\langle f|f\rangle\langle t=0|t=0\rangle}}\label{ampa}.
\eeq

\subsection{LSPT's Weakness}

The Hamiltonian eigenstates in their center of mass frames, as usual, are translation-invariant.  This means that if space is infinite, then they will be not be normalizable.  As a result, the inner product (\ref{ampa}) does not exist.

The usual solution to this problem is to assemble the Hamiltonian eigenstates into normalizable wave packets.  In other words, instead of states which are annihilated by the momentum operator $P$, one considers superpositions of $P$-eigenstates with various small eigenvalues and adds them together, weighted for example by a narrow Gaussian, to arrive at a nearly monochromatic yet normalizable state.

Such a wave packet corresponds to a sum of states such that the center of mass of the kink-meson system is at some collective coordinate $x_0$, and the states with different values of $x_0$ are summed with some weight.  Now recall that $x_0$, when it is small, is proportional to the eigenvalue of our operator $\phi_0$.  Therefore, we sum over states with various eigenvalues of $\phi_0$.  If the wave packet is nearly monochromatic, then these eigenvalues cover a wide range and so some are large.

This is a problem for the perturbative approach to finding states in Sec.~\ref{twosez}.  Recall that for each $\nn$-tuple of momenta $|k_1\cdots k_\nn\rangle_0$ a state is characterized by a wavefunction $\psi(\phi_0)$.  More precisely, each state is a sum of terms of the form
\beq
\psi(\phi_0)|k_1\cdots k_\nn\rangle_0.
\eeq
We found these states perturbatively by expanding $\psi(\phi_0)$ in a power series in $\phi_0$.  This power series expansion turned out to be justified because, when we calculated the states, we found that each power of $\phi_0^2$ in a state is accompanied by at least a power of $\sl$.  This meant that our expressions for the states converge in the sense of an asymptotic series so long as the eigenvalue of $\phi_0$ is smaller than order $O(\lambda^{-1/4})$, corresponding to an eigenvalue of $x_0\sim \phi_0/\sqrt{Q_0}$ of order $O(\lambda^{1/4})$.  As a result, convergence of the $\phi_0$ series expansion requires that the position spreads of our wave packets must be smaller than $\lambda^{1/4}/m^{3/2}$.  

This maximum distance scale is much larger than the de Broglie wavelength of the kink, $\lambda/m^3$.  That is fortunate, as the semiclassical expansion of solitons is only reasonable when they are much larger than their de Broglie wavelength.  Otherwise the uncertainty principle will lead to an energy uncertainty of order their classical mass.  Nonetheless, an upper bound on the position-smearing is a lower bound on the momentum smearing and so the energy resolution.  In particular, it means that we cannot achieve arbitrary precision.

There are several ways out of this problem.

First, one could achieve arbitrary precision if one chooses to simply not expand $\psi(\phi_0)$ in a power series.  In this case, the Hamiltonian eigenvalue condition can be interpreted as a constraint on $\psi(\phi_0)$ which needs to be solved.

Second, one can deal with relatively narrow wave packets, and accept the resulting broadening.  Indeed, in some physical situations, one is interested in relatively narrow wave packets of solitons, for example for pinned Abrikosov vortices or solitonic dark matter halos.  This approach to LSPT was used to calculate form factors in Refs.~\cite{kinkff,hengyuanff}.

In the remainder of these lectures, we will consider a third approach.  We will continue to restrict our attention to states which are annihilated by the momentum operator $P$.  These are states which are invariant under a simultaneous translation of the kink together with all bound modes and unbound mesons.  In short, to avoid considering states with a large $x_0$ smearing, we will consider states with an infinite $x_0$ smearing.  

Does this make matters even worse?  Amazingly, it does not.  The reason for this is that once we have restricted our initial state to be translation-invariant, then a translation-invariant Hamiltonian will guarantee a translation-invariant final state.  This means that, just like in a gauge theory where gauge orbits have infinite volumes, the divergences in the numerator and denominator of the amplitude (\ref{ampa}) factorize out, and cancel each other.  As a result, the amplitude (\ref{ampa}) may be evaluated fixing the translation symmetry in the numerator and denominator, rendering them finite, so long as one multiplies by the appropriate Jacobian factor corresponding to the determinant of the action of the translation symmetry. More concretely, we will find that amplitudes for the evolution of translation-invariant states depend only on the primary coefficients in the evolved state, and so there is no $\phi_0$ expansion to be summed and thus no divergence.

We will refer to the translation-fixed inner product, including the Jacobian factor, as the reduced inner product.  It was computed in Ref.~\cite{menorm}, and it turns out to be even simpler than the usual inner product.  It is just the usual inner product on only the primary part of the state, which we remind the reader is the part with no factors of $\phi_0$, plus a single correction suppressed by a power of $\sl$ resulting from the determinant.  This correction, besides being subleading, also vanishes for inner products in which all mesons are localized in wave packets far from the kink, as they are in the far past and future of a kink-meson scattering experiment.

Thus in these cases the amplitude (\ref{ampa}) can be calculated by simply ignoring all terms $\gamma^{\mm\nn}(k_1\cdots k_\nn)$ with $\mm>0$ in the final state $e^{-iHt}|t=0\rangle$.

\subsection{A Very Simple Example of a Reduced Inner Product}

Consider the quantum mechanics of a particle with two spin states $\up$ and $\dn$ with inner products
\beq
\langle\uparrow\,\up=\langle\downarrow\,\dn=1\hsp \langle\uparrow\,\dn=0.
\eeq
Now consider the unnormalized state
\beq
|\psi\rangle=\up+\dn.
\eeq
If one wants to know the probability that an up-down measurement reveals that it is in the $\up$ state, one first computes the inner product of the actual state with the $\up$ state, remembering to normalize by the norm squared of both
\beq
\mathcal{A}=\frac{\langle\uparrow\,|\psi\rangle}{\sqrt{\langle\uparrow\up}\sqrt{\langle\psi|\psi\rangle}}=\frac{1}{\sqrt{1}\sqrt{2}}=\frac{1}{\sqrt{2}}.
\eeq
The probability is just
\beq
P_\uparrow=|\mathcal{A}|^2=\frac{1}{2}.
\eeq

Now, imagine that the particle lives on a circle with coordinate $x$ valued on the interval $[-\pi R,\pi R)$ for some $R$.  Then each state needs an additional quantum number $x\in [-\pi R,\pi R)$ specifying the position of the particle.  Such a basis of states is given by
\beq
\upx{x},\ \dnx{x}
\eeq
with inner products
\beq
\langle\uparrow ,x\upx{y}=\langle\downarrow ,x\dnx{y}=\delta(x-y)\hsp \langle\uparrow,x\dnx{y}=0. \label{ipa}
\eeq
Now imagine that we are given a translation-invariant state
\beq
|\psi\rangle=\int_{-\pi R}^{\pi R} dx \left(\upx{x}+\dnx{x}\right)
\eeq
and we are asked the probability that the measured spin will be up.

Working in position space, this is straightforward.  First we can compute the matrix element corresponding to a spin up and a position at some arbitrary but fixed $x$
\beq
\mathcal{A}_x=\frac{\langle\uparrow\,, x|\psi\rangle}{\sqrt{\langle\uparrow,x \upx{x}}\sqrt{\langle\psi|\psi\rangle}}=\frac{1}{\sqrt{\langle\uparrow,x \upx{x}}\sqrt{4\pi R}}.
\eeq
This squares to the probability that the particle is spin up and at $x$, which later should be summed over all values of $x$.  However, the inner product in the denominator is infinite, and so this probability is zero.  Of course, this is to be expected for the probability of the particle lying at a point.

To deal with the divergence, we should be more careful.  We can consider an amplitude for the particle to lie within a small interval $[x-\epsilon/2,x+\epsilon/2]$.  For this, one introduces a test state
\beq
|\phi\rangle_x=\int_{x-\frac{\epsilon}{2}}^{x+\frac{\epsilon}{2}}dx\p \upx{x\p}\hsp 
{}_x\langle\phi|\phi\rangle_x=\epsilon.
\eeq
Then the matrix element corresponding to this interval is
\beq
\mathcal{A}_{[x-\frac{\epsilon}{2},x+\frac{\epsilon}{2}]}=\frac{{}_x\langle\phi|\psi\rangle}{\sqrt{{}_x\langle\phi|\phi\rangle_x}\sqrt{\langle\psi|\psi\rangle}}=\frac{\epsilon}{\sqrt{4\pi R\epsilon}}.
\eeq
The probability that the particle is spin up and lies in the interval is then
\beq
P_{\uparrow,[x-\frac{\epsilon}{2},x+\frac{\epsilon}{2}]}=\left|\mathcal{A}_{[x-\frac{\epsilon}{2},x+\frac{\epsilon}{2}]}\right|^2=\frac{\epsilon}{4\pi R}.
\eeq
The total probability that the particle is spin up is just the sum over all $2\pi R/\epsilon$ such intervals, which is
\beq
P_{\uparrow}=\frac{2\pi R}{\epsilon}P_{\uparrow,[x-\frac{\epsilon}{2},x+\frac{\epsilon}{2}]}=\frac{2\pi R}{\epsilon}\frac{\epsilon}{4\pi R}=\frac{1}{2}.
\eeq

What if $R$ is infinite and $\epsilon$ is fixed?  The probability does not depend on $R$ and clearly the probability that the spin is up should not depend on $R$.  This is rather obvious as the probability that the spin is up is independent of the position $x$.  However, in this case the probability in each finite interval vanishes, as there are now infinitely many intervals.

Imagine that we know, as we do know in this example, that our state $|\psi\rangle$ is translation-invariant.  In this case, the probability that the particle is spin up is equal to the probability that the particle is spin up given that the particle is in a certain finite interval.  Concretely, if we divide space into intervals $A_i$ which are related by a translation, then
\beq
P_\uparrow=\sum_{i}P_{\uparrow,x\in A_i} =\sum_i P_{\uparrow|x\in A_i}P_{x\in A_i}=P_{\uparrow|x\in A_1}\sum _i P_{x\in A_i}=P_{\uparrow|x\in A_1}
\eeq
where the comma denotes joint probability and the vertical line denotes conditional probability.  $P_{x\in A_i}$ is the probability that $x$ is in the interval $A_i$, and so the sum over these probabilities is unity because we have precisely one particle.  Here $A_1$ is any interval, it does not matter which interval is chosen since they are related by a translation symmetry.

We have therefore reduced our problem of computing the total probability that the spin is up to calculating the probability that the spin is up assuming that $x\in A_1$ where $A_1$ is any interval.  Of course this is just our previous problem where $\epsilon$ is the length of $A_1$.

Formally we may compute this conditional probability by replacing the inner products in the definition of the matrix element with the inner products with $x$ restricted to lie in our interval.  We will refer to such inner products as {\it{reduced inner products}}.  Let us see how these work in our example.

Let us choose the interval
\beq
A_1=[x-\frac{\epsilon}{2},x+\frac{\epsilon}{2}].
\eeq
Now the inner products (\ref{ipa}) imply that the usual inner product between any two states is
\bea
|\psi_1\rangle&=&\int_{-\infty}^\infty dx \left[a_\uparrow(x) \upx{x}+a_\downarrow(x) \dnx{x}\right]\hsp |\psi_2\rangle=\int_{-\infty}^\infty dx \left[b_\uparrow (x)\upx{x}+b_\downarrow (x)\dnx{x}\right]\nonumber\\
\langle \psi_1|\psi_2\rangle&=&\int_{-\infty}^\infty dx \left[a^*_\uparrow (x) b_\uparrow(x)+a^*_\downarrow (x) b_\downarrow(x)
\right].
\eea
Our reduced inner product will be defined by restricting the integrals to $A_1$ and, for later convenience, we will divide by the length of $A_1$, which in the present case is $\epsilon$.  This division will make the reduced inner product independent of the length of $A_1$.  Our definition of the reduced inner product is then
\beq
\langle \psi_1|\psi_2\rangle_{\rm{red}}= \frac{\int_{A_1} dx \left[a^*_\uparrow (x) b_\uparrow(x)+a^*_\downarrow (x) b_\downarrow(x)\right]}{\epsilon}.
\eeq
Note that in the present case, since our variable $x$ is simply shifted by the translation symmetry, the reduced inner product is simply equal to $a^*_\uparrow (x) b_\uparrow(x)+a^*_\downarrow (x) b_\downarrow(x)$ for any $x$.

How do we use this to get the conditional probability?  Defining
\beq\label{ket_phi_qm}
|\phi\rangle=\int_{-\infty}^\infty  dx\upx{x}
\eeq
so that
\beq
\langle\phi|\phi\rangle_{\rm red}=\langle\phi|\psi\rangle_{\rm red}=1\hsp
\langle\psi|\psi\rangle_{\rm red}=2
\eeq
the probability is simply
\beq
P_\uparrow=P_{\uparrow|x\in A_1}=\frac{\left|\langle\phi|\psi\rangle_{\rm red}\right|^2}{{\langle\phi|\phi\rangle_{\rm{red}}}{\langle\psi|\psi\rangle}_{\rm red}}=\frac{1}{1\times 2}=\frac{1}{2}.
\eeq
Note that this may be written as the square of a reduced amplitude
\beq
\mathcal{A}_{\rm red}=\frac{\langle\phi|\psi\rangle_{\rm red}}{\sqrt{\langle\phi|\phi\rangle_{\rm{red}}}\sqrt{\langle\psi|\psi\rangle}_{\rm red}}.
\eeq
In conclusion, we see that if we replace all inner products by reduced inner products, we arrive at the conditional probability which, in the case of a translation-invariant state, is equal to the total probability.

Let us review the key steps of this argument.  First, we used translation invariance to show that the probability $P_\uparrow$ that the particle is spin up is equal to the conditional probability $P_{\uparrow|x\in A_i}$ that the spin is up given that the particle is in a region $A_i$.  Using the $\delta$ function normalization of $\upx{x}$ to turn the incoherent sum over values of $x$ into an an integral, the original probability was
\beq
P_\uparrow=\int_{-\infty}^\infty dx \frac{\left|\langle\uparrow, x|\psi\rangle\right|^2}{{\langle\psi|\psi\rangle}}
\eeq
which is ill-defined as $\langle\psi|\psi\rangle$ is infinite and also the integrand is $x$-independent implying that the integration gives an infinite factor.  However, in the case of the conditional probability, we only need to integrate over some fixed $A_1$, let us choose it to be the interval $[-\epsilon/2,\epsilon/2]$, and the norm $\langle\psi|\psi\rangle$ is replaced with the norm restricted to $A_1$
\bea
|\psi\rangle&=&\int_{-\infty}^\infty dx \left(\upx{x}+\dnx{x}\right)\hsp |\psi\rangle_{\left[-\frac{\epsilon}{2},\frac{\epsilon}{2}\right]}=\int_{-\frac{\epsilon}{2}}^{\frac{\epsilon}{2}} dx \left(\upx{x}+\dnx{x}\right)\nonumber\\
&&{}_{\left[-\frac{\epsilon}{2},\frac{\epsilon}{2}\right]}\langle\psi|\psi\rangle_{\left[-\frac{\epsilon}{2},\frac{\epsilon}{2}\right]}=2\epsilon.
\eea
The conditional probability is therefore
\beq
P_{\uparrow|x\in A_1}=\int_{-\frac{\epsilon}{2}}^{\frac{\epsilon}{2}} dx \frac{\left|\langle\uparrow, x|\psi\rangle\right|^2}{ {}_{\left[-\frac{\epsilon}{2},\frac{\epsilon}{2}\right]}\langle\psi|\psi\rangle_{\left[-\frac{\epsilon}{2},\frac{\epsilon}{2}\right]}}
=\int_{-\frac{\epsilon}{2}}^{\frac{\epsilon}{2}} dx \frac{1}{2\epsilon}=\frac{1}{2}.
\eeq
Identifying this with the total probability $P_\uparrow$ we find the same answer at finite and infinite $R$, as can be expected. 

The key step is that we used the translation invariance to simultaneously reduce the infinite integration to a finite region and also reduce the normalization factor in the denominator.  In other words, we quotiented the numerator and the denominator by the translation symmetry, taking the ill-defined $P_\uparrow$ to the conditional probability.  The translation symmetry acted by a simple shift of the coordinate $x$ and so this quotient was quite simple.

\subsection{The Reduced Inner Product with Linearized Coordinates}

What if, instead of the collective coordinate basis $\{\upx{x},\dnx{x}\}$, we work in another basis $\{\upx{y}_y,\dnx{y}_y\}$ where the basis transformation mixes not only the coordinate but also the arrows so that $\upx{y}$ is not necessarily spin up
\bea
\upx{y}_y&=&\int dx\left[ b^\uu(y,x)\upx{x}+ b^\ud(y,x) \dnx{x}\right]\\
\dnx{y}_y&=&\int dx\left[ b^\du(y,x)\upx{x}+ b^\dd(y,x) \dnx{x}\right].\nonumber
\eea
Let us introduce the translation operator $p$ defined on an arbitrary state by
\beq
p\int dx\left[\psi^\ua(x) \upx{x}+ \psi^\da(x) \dnx{x}
\right]=-i\int dx\left[\psi^{\ua\prime}(x) \upx{x}+ \psi^{\da\prime}(x) \dnx{x}
\right].
\eeq
The translation operator acts on our new basis as
\bea
p\upx{y}_y&=&-i\int dx\left[ \left(\partial_x b^\uu(y,x)\right)\upx{x}+ \left(\partial_x b^\ud(y,x)\right) \dnx{x}\right]\\
p\dnx{y}_y&=&-i\int dx\left[ \left(\partial_x b^\du(y,x)\right)\upx{x}+ \left(\partial_x b^\dd(y,x)\right) \dnx{x}\right].\nonumber
\eea

Now let us, for the moment, consider the simple case in which the matrix $b(y,x)$ is chosen so that this same operator $p$ acts on the $y$-basis as
\bea
p\int dy \phi^\ua(y)\upx{y}_y&=&-i \int dy  \left(\partial_y \phi^\ua(y)\right)\left[ \ga^\uu \upx{y}_y+\ga^\ud\dnx{y}_y\right]\\
p\int dy \phi^\da(y)\dnx{y}_y&=&-i \int dy  \left(\partial_y \phi^\da(y)\right)\left[ \ga^\du \upx{y}_y+\ga^\dd\dnx{y}_y\right].\nonumber
\eea
Let us consider first the top line, expressing it in the $x$ basis using, alternately, the action of $p$ in the coordinate $x$ basis and that in the new $y$ basis
\bea
p\int dy \phi^\ua(y)\upx{y}_y&=&-i\int dx\int dy \left(\partial_y \phi^\ua(y)\right)\left[\left(\ga^\uu b^\uu(y,x)+\ga^\ud b^\du(y,x)\right)\upx{x}\right.\nonumber\\
&&\left.+
\left(\ga^\ud b^\dd(y,x)+\ga^\uu b^\ud(y,x)\right)\dnx{x}
\right]\nonumber\\
&=&\int dy \phi^\ua(y) p\upx{y}_y\nonumber\\
&=&-i\int dy \phi^\ua(y) \int dx  \left[ 
\left(\partial_x b^\uu(y,x)\right)\upx{x}+ \left(\partial_x b^\ud(y,x)\right) \dnx{x}
\right]\nonumber.
\eea
Identifying the coefficients, for example, of $\upx{x}$ and integrating $y$ by parts in the first expressions one finds
\beq
i\int dy \phi^\ua(y)\left(\ga^\uu \partial_y b^\uu(y,x)+\ga^\ud \partial_y b^\du(y,x)\right)=-i\int dy \phi^\ua (y)\partial_x b^\uu(y,x).
\eeq
As the function $\phi^\ua(y)$ is arbitrary, we may identify its coefficients
\beq
\ga^\uu \partial_y b^\uu(y,x)+\ga^\ud \partial_y b^\du(y,x)=-\partial_x b^\uu(y,x).
\eeq

Choosing other spins, one finds three similar relations which can be combined in matrix form as
\beq
\ga \partial_y b(y,x)=-\partial_x b(y,x). \label{geqa}
\eeq
What does this mean?  For a moment let us pretend that there is only one spin, so $\ga$ is a single number.  In this case, Eq.~(\ref{geqa}) would be solved by
\beq
b(y,x)=b(x-y/\ga).
\eeq
Now, we want $y$ to be a {\it{linearized}} coordinate, so that $\upx{0}_y$ and $\dnx{0}_y$ only depend on $\upx{0}$ and $\dnx{0}$.  This implies that $b(y,0)$ vanishes unless $y=0$ and so
\beq
b(x-y/\ga)=B \delta(x-y/\ga)
\eeq
for some constant $B$.  In other words, in the absence of internal degrees of freedom, the $x$ and $y$ coordinates are proportional and $A$ is their constant of proportionality.

Of course we have two spins, not one.  So imagine that we rotate our spin axis in the $x$ basis so that $\ga$ is diagonal.  In that case our matrix equation~(\ref{geqa}) becomes
\bea
\ga^\uu \partial_y b^\uu(y,x)&=&-\partial_x b^\uu(y,x)\hsp
\ga^\uu \partial_y b^\ud(y,x)=-\partial_x b^\ud(y,x)\\
\ga^\dd \partial_y b^\du(y,x)&=&-\partial_x b^\du(y,x)\hsp
\ga^\dd \partial_y b^\dd(y,x)=-\partial_x b^\dd(y,x)
\nonumber
\eea
and so we conclude
\bea
b^\uu(y,x)&=&b^\uu(x-y/\ga^\uu)=B^\uu\delta(x-y/\ga^\uu)\hsp
b^\ud(y,x)=b^\ud(x-y/\ga^\uu)=B^\ud\delta(x-y/\ga^\uu)\nonumber\\
b^\du(y,x)&=&b^\du(x-y/\ga^\dd)=B^\du\delta(x-y/\ga^\dd)\hsp
b^\dd(y,x)=b^\dd(x-y/\ga^\dd)=B^\dd\delta(x-y/\ga^\dd ).\nonumber
\eea
Intuitively, with this simple form of $p$, an $x$-translation and a shift of the variable $y$ are equivalent up to multiplication by the matrix $-\ga$ in spin-space.

How can we use this fact to evaluate reduced inner products, so that we may calculate probabilities?  Let us consider a general translation-invariant state written in the $y$ basis
\bea
|\psi\rangle&=&\int dy \left[\gamma^\ua (y)\upx{y}_y+\gamma^\da (y)\dnx{y}_y 
\right]\\
&=&\,\,\,\int dx dy \left[\left( \gamma^\ua(y) b^\uu (y,x) +\gamma^\da (y) b^\du (y,x)\right)\upx{x} 
+\left( \gamma^\ua(y) b^\ud (y,x) +\gamma^\da (y) b^\dd (y,x)\right)\dnx{x} 
\right]\nonumber\\
&=&\,\,\,\int\, dx \left[\left( B^\uu |\ga^\uu| \gamma^\ua(\ga^\uu x)  +B^\du |\ga^\dd| \gamma^\da (\ga^\dd x) \right)\upx{x}\right.\nonumber\\
&&\qquad\left.\,+\,\left(  B^\ud |\ga^\uu| \gamma^\ua(\ga^\uu x) +B^\dd |\ga^\dd| \gamma^\da (\ga^\dd x) \right)\dnx{x} 
\right].\nonumber
\eea

Now let us find the spin bases at $x=y=0$ so that $B$ is the identity matrix.  Then
\beq
|\psi\rangle=\int dx \left[|\ga^\uu| \gamma^\ua(\ga^\uu x)\upx{x}+|\ga^\dd| \gamma^\da(\ga^\dd x)\dnx{x} \right].
\eeq
Defining $|\phi\rangle$ as in (\ref{ket_phi_qm}) the reduced inner products are easily calculated
\bea
\langle \phi|\psi\rangle_{\rm{red}}&=&\frac{1}{\epsilon}\int_{-\epsilon/2}^{\epsilon/2}dx |\ga^\uu| \gamma^\ua(\ga^\uu x)=|\ga^\uu| \gamma^\ua(0) +O(\epsilon)\\
\langle \psi|\psi\rangle_{\rm{red}}&=&|\ga^\uu \gamma^\ua(0) |^2+|\ga^\dd \gamma^\da(0) |^2+O(\epsilon)\nonumber
\eea
leading, in the $\epsilon\rightarrow 0$ limit, to a probability of
\beq
P_\ua=\frac{|\ga^\uu \gamma^\ua(0) |^2}{\langle \psi|\psi\rangle_{\rm{red}}}. \label{pqm}
\eeq

In other words, we learn that the probability is equal to the product of the coefficient $\gamma$ evaluated at $y=0$ times the coefficient $A$ of $\partial_y$ in an expansion of $p$.  As is shown in Ref.~\cite{menorm}, this result holds true even if $p$ contains other terms, as only those proportional to $\partial_y$ contribute to the determinant factor.  Such terms in fact are necessary to avoid translation-invariance overconstraining our states.

Notably, the numerator has two factors.  The first is the coefficient $\gamma^\ua$ that can be read off of the state, evaluated at $x=0$.  The second is $\ga^\uu$, which is the coefficient of $-i\partial_y$ in the translation operator $p$.  The coefficient $\gamma(x)$ at other values of $x$ does not affect the probability, nor do the other terms in $p$.



\subsection{Passing to Quantum Field Theory}

Now let us return to the case of kink sector quantum field theory.  The variables above in this section now have other names.  $x$ is our collective coordinate $x_0$ which is the position of the kink's center of mass, and so
\beq
p\rightarrow P\p\hsp y\rightarrow \phi_0\hsp \partial_y\rightarrow i\pi_0\hsp \gamma(0)\rightarrow \gamma^{0\nn}.
\eeq
Instead of two spins $\ua$ and $\da$, now one has the infinitely many quantum numbers associated with the mesons and shape modes.

So how do we write the probability (\ref{pqm}) in kink sector quantum field theory?  The coefficient $\gamma^\ua$ at collective coordinate $x=0$ is just our wavefunction $\psi(\phi_0)$ evaluated at $\phi_0=0$.  This is none other than the $\mm=0$ term in our coefficient functions $\gamma^{\mm\nn}$.  The descendants are not needed.  This is because, as we have seen, they are fixed by translation-invariance.  This will make computing probabilities involving translation-invariant states much easier, one need only compute the inner products of the $\mm=0$ terms.  

What about the $\ga^\uu$?  This was a term in the decomposition of the translation operator $p$.  In the kink frame, our translation operator can be written in the form
\beq
P\p=A\pi_0+B+C\phi_0.
\eeq
Here it is $\pi_0$ that plays the role of $-i\partial_y$ and so $A$ that plays the role of $\ga$.  $A$ in turn can be read off of Eqs.~(\ref{ppinv}) and (\ref{pdeca}) to be
\beq
A=\sqrt{Q_0}+\ppin{k}\Delta_{kB}\left(B^\ddag_k+\frac{B_{-k}}{2\ok{}}\right).
\eeq
The reduced inner product of two states characterized by coefficients $\gamma_\phi$ and $\gamma_\psi$ is the matrix element of $A$.  Explicitly our master formula is \cite{menorm}
\bea
{}_{\rm{}}{}\langle \phi|\psi\rangle_{\rm{red}}&=&
\sum_\nn \nn! \ppink{\nn}\frac{\gamma_\phi^{0\nn*}(k_1\cdots k_\nn)}{\prod_{i=1}^\nn(2\ok{i})}
\left[ \sqrt{Q_0}\gamma_\psi^{0\nn}(k_1\cdots k_\nn)
\right.\nonumber\\
&&\left.+
\gamma_\psi^{0,\nn-1}(k_1\cdots k_{\nn-1})\Delta_{k_\nn B}
+(\nn+1)\ppin{k_{\nn+1}}\gamma_\psi^{0,\nn+1}(k_1\cdots k_{\nn+1})\frac{\Delta_{-k_{\nn+1} B}}{2\ok{\nn+1}}
\right].
\nonumber
\eea
Finite probabilities may be derived by replacing all inner products with reduced inner products.

One sees that the leading term, of order $O(1/\sl)$, is on the first line.  It is simply the inner product restricted to the primaries $\mm=0$, with a factor of $\sqrt{Q_0}$ that we have seen relates $\phi_0$ to the position $x_0$.  The determinant factors are on the second line.  These are subleading by one power of $\sl$.  They are proportional to $\Delta_{kB}$ where $k$ labels the momenta of various mesons or shape modes.  In particular, if one calculates the reduced inner products when the mesons are far from a ground state kink, for example long after a scattering event, then the corresponding terms vanish.  Therefore, in scattering calculations computing probabilities of various well-separately final states, these corrections do not arise from the mesons.  They may arise, however, from shape modes in the case of an excited kink because $\Delta_{SB}$ is in general nonzero.

We note that there is only one correction, suppressed by a power of $\sl$, and no further corrections.  This is a result of the fact that our $P\p$ only contains a single power of $\pi_0$ due to the simple form of the displacement operator.  

Our derivation of these corrections has been rather abstract.  Are the corrections really there?  In Ref.~\cite{menorm} we have shown that the corrections are necessary in order for the kink sector Hamiltonian eigenstates to be mutually orthogonal.  

In Ref.~\cite{elaslungo} we considered elastic kink-meson scattering at order $O(\lambda)$.  Here, there is a contribution from a process in which a single virtual shape mode is nucleated, leading to a meson and a shape mode, which have a nonvanishing inner product with the two-meson subdominant correction to the one-meson final state $\gamma_1^{02}(S,k)$.  This final state correction is precisely canceled by the subleading term in the reduced inner product with the final state of the Stokes scattering process, in which, at $O(\sl)$, an incident meson excites a kink's shape mode.  We concluded that the reduced inner product conspires to avoid final state corrections arising from quantum corrections to the asymptotic states.  Seen differently, the subleading correction to the reduced inner product not only guarantees that distinct Hamiltonian eigenstates corresponding to various excited kink-meson systems are mutually orthogonal, but also that the end states of distinct processes, such as Stokes scattering and elastic kink-meson scattering, are orthogonal. 

In fact, so far in our studies of kink-meson scattering, corrections to scattering probabilities arising from the quantum corrections to initial and final states always seem to vanish.  This phenomenon is well understood in the vacuum sector, where it arises from LSZ reduction \cite{lsz}.  In that case it is known that wavefunction renormalization is the only effect that quantum corrections to initial and final states may have on the $S$-matrix.  However, a corresponding result is lacking in soliton sectors.  The degeneracy of states resulting from the zero modes obstructs a naive generalization of the usual derivations of the LSZ reduction formula to the kink sector. 

\section{Kink-Meson Scattering} \label{scatsez}

LSPT efficiently describes the kink sector, which consists of states describing a single kink with arbitrary numbers of internal excitations and mesons.  It also efficiently describes processes that occur in this sector, such as scattering processes of kinks and mesons that may or may not (de)excite the kink.  There are several distinct ways that LSPT has been used to calculate the probabilities of such scattering processes.  In this section, we will describe what is perhaps the most efficient, which is based on the Lippmann-Schwinger equations.  Our presentation will follow Ref.~\cite{elas}.

\subsection{Degenerate Eigenstates}

As described in Sec.~\ref{twosez}, higher order $O(\lambda^{i/2})$ corrections $|\psi\rangle_i$ to Hamiltonian eigenstates $|\psi\rangle$ can be found by solving the Hamiltonian eigenvalue equation
\beq
(H\p_2-E_1)|\psi\rangle_i=-\sum_{j=3}^{i+2}H\p_j|\psi\rangle_{i+2-j}+\sum_{j=2}^{\lfloor{\frac{i+2}{2}}\rfloor}E_j|\psi\rangle_{i+2-2j}. \label{sed}
\eeq
In the center of mass frame, a basis of kink sector states is
\beq
|\psi\rangle=|k_1\cdots k_n\rangle
\eeq
where $k_i$ are either shape modes $S$ or the momenta of unbound mesons.  For these we have already computed the one-loop energy
\beq
H\p_2|k_1\cdots k_n\rangle_0=E_1|k_1\cdots k_n\rangle_0\hsp E_1=Q_1+\sum_{j=1}^n \ok{j}.
\eeq

We will use the decomposition of the $i$th order state
\beq
|\psi\rangle_i=\sum_{\mm\nn}|\psi\rangle_i^{\mm\nn}
\eeq
where $\mm$ is the number of zero modes $\phi_0$ and $\nn$ is the number of shape modes and mesons.  Let us restrict our attention to primary terms $\mm=0$ in (\ref{sed}) , as only these contribute to the reduced inner product.  Let us consider a term in the $i$th order correction with $\nn\p$ mesons and shape modes with momentum or shape mode labels $k\p_i$.  Then there will be four contributions to the $\mm=0$ terms on the left hand side of (\ref{sed}).  The first arises from the $\pi_0^2/2$ in $H\p_2$ striking $|\psi\rangle_i^{2\nn\p}$
\beq
\frac{\pi_0^2}{2}\gamma_i^{2\nn\p}(k\p_1\cdots k\p_{\nn\p})\phi_0^2|k\p_1\cdots k\p_{\nn\p}\rangle_0=-\gamma_i^{2\nn\p}(k\p_1\cdots k\p_{\nn\p})|k\p_1\cdots k\p_{\nn\p}\rangle_0.
\eeq
The second arises from the oscillator terms in $H\p_2$ striking $|\psi\rangle_i^{0\nn\p}$ 
\beq
\ppin{k}\ok{}\Bd{}B_{k}\gamma_i^{0\nn\p}(k\p_1\cdots k\p_{\nn\p})|k\p_1\cdots k\p_{\nn\p}\rangle_0=\left( 
\sum_{j=1}^{\nn\p}\okp{j}
\right)\gamma_i^{0\nn\p}(k\p_1\cdots k\p_{\nn\p})|k\p_1\cdots k\p_{\nn\p}\rangle_0.
\eeq
The third arises from the $Q_1$ term in the $H\p_2$.  The fourth is the $-E_1$ term which also strikes $|\psi\rangle_i^{0\nn\p}$.  Altogether, on the left hand side of (\ref{sed}) the contribution proportional to $|k\p_1\cdots k\p_{\nn\p}\rangle_0$ is
\beq
\left[ 
-\gamma_i^{2\nn\p}(k\p_1\cdots k\p_{\nn\p})+\left( 
-E_1+Q_1+\sum_{j=1}^{\nn\p}\okp{j}\right)\gamma_i^{0\nn\p}(k\p_1\cdots k\p_{\nn\p})
\right]|k\p_1\cdots k\p_{\nn\p}\rangle_0.
\eeq
The right hand side consists of terms with coefficients $\gamma$ at orders less than $i$, and so using the usual inductive logic let us say that it is known.  The $\gamma_i^{2\nn\p}$ term on the left hand side is therefore also known, using the recursion relation, as it is a descendant.  

Matching the coefficient of $|k\p_1\cdots k\p_{\nn\p}\rangle_0$ on the right and left hand sides we therefore arrive at an equation of the form
\beq
\left( 
-E_1+Q_1+\sum_{j=1}^{\nn\p}\okp{j}\right)\gamma_i^{0\nn\p}(k\p_1\cdots k\p_{\nn\p})={\rm{something\ known}}. \label{nonin}
\eeq
If we can solve this equation for the primary coefficient $\gamma^{0\nn\p}_i$ then we have found the state at $i$th order.  To do this, we need to divide by $\left( 
-E_1+Q_1+\sum_{j=1}^{\nn\p}\okp{j}\right)$.  Therefore, we have a unique solution except when
\beq
Q_1+\sum_{j=1}^{n}\ok{j}=E_1=Q_1+\sum_{j=1}^{\nn\p}\okp{j}.
\eeq
This happens whenever $H\p_2$ has degenerate eigenvectors $|k_1\cdots k_n\rangle_0$ and $|k\p_1\cdots k\p_
{\nn\p}\rangle_0$, in other words whenever a scattering process is allowed by energy conservation.

\subsection{Elastic Kink-Meson Scattering}
Whenever there are degenerate eigenstates, one must be chosen.  In a kink-meson scattering experiment in 1+1 dimensions, the eigenstate can be chosen so that the initial configuration has the desired momenta $\{k_1\cdots k_n\}$.

How do we do this?  For concreteness, let us consider elastic scattering of a single meson and a single kink.  The generalization to arbitrary processes will be straightforward.  Since we start and end with one meson
\beq
\nn=\nn\p=1.
\eeq
Let the incoming meson momentum be $k$, so the outgoing momentum of interest $k\p$ will be near $-k$, perhaps with some arbitrarily small smearing due to wave packets and the uncertainty principle at finite times.  Now our equation (\ref{nonin}) for the primary is
\beq
(\okp{}-\ok{})\gamma_i^{01}(k\p)=\frac{k}{\ok{}}R(k\p) \label{peq}
\eeq
where $R(k\p)$ is something known from the computation at lower orders and the recursion relation, and the $k/\ok{}$ factor is included for convenience.

Let us write the solution as\footnote{It is easily shown that $\gamma_i^{0\nn}$ admits the same decomposition.  We will use this fact in Eq.~(\ref{mmm}).}
\beq
\gamma_i^{01}(k\p)=F(k,k\p)+\frac{R(k\p)}{k+k\p}
\eeq
where $F(k,k\p)$ and $R(k\p)$ are continuous at $k=-k\p$.  At $k=-k\p$ the denominator vanishes and we have not yet specified how this divergence is to be treated.  The corresponding term in our state is then
\beq
|k\rangle_i^{01}=\pin{k\p}\left(F(k,k\p)+\frac{R(k\p)}{k+k\p}\right)|k\p\rangle_0. \label{k01}
\eeq

Now let us search for the initial conditions for our scattering experiment, which will be at a time $t=0$.  

In order to localize the incoming particle on one side of the kink, we will need to construct a wave packet.  We will begin at time $t=0$ with the initial state
\beq
|t=0\rangle=\pin{k} e^{-\sigma^2(k-k_0)^2-i(k-k_0)x_0}|k\rangle\hsp
-x_0\gg \sigma\gg 1/k_0>0 \label{t0}
\eeq
describing a nearly monochromatic particle\footnote{Here we have restricted our attention to reflectionless kinks, corresponding to special choices of the potential which however include the popular Sine-Gordon and $\phi^4$ double-well models.  More generally, one should add a $e^{ikx_0}$ component for the reflected wave.  This does not lead to a significant complication of the scattering probabilities and is described in Ref.~\cite{memult}.} localized far to the left of the kink, near $x=x_0$, in a wave packet whose width is $\sigma$.  As $|k\rangle$ is an eigenstate of the Hamiltonian with eigenvalue $E_k$, we know just how it evolves in time
\beq
|t\rangle=\pin{k} e^{-\sigma^2(k-k_0)^2-i(k-k_0)x_0-i E_k t}|k\rangle.\label{t}
\eeq

Now we will use the fact that $\lambda\ll m^2$ to replace $E_k$ with $\ok{}$.  This will change the phase significantly as $t$ can be quite large, but only the derivative of the phase with respect to $k$ will be important below and so this approximation will be inconsequential.  Similarly, as $m\sigma\gg 1$ we will expand the frequency
\beq
\ok{}=\ok{0}+\frac{\partial \ok{0}}{\partial k_0}(k-k_0)+O\left((k-k_0)^2\right)=\ok{0}+\frac{k_0}{\ok 0}(k-k_0)+O\left((k-k_0)^2\right)
\eeq
and drop the $O((k-k_0)^2)$ term, which is equal to $\ok 0$ times a factor of order $O(1/(m\sigma)^2)$.  In dropping this term, we are neglecting the spread of the wave packet, which will not affect the time-integrated scattering probability.

If we define the position
\beq
x_t=x_0+\frac{k_0}{\ok 0} t
\eeq
of the center of the wave packet at time $t$, then
one can rewrite (\ref{t}) as
\beq
|t\rangle=e^{-i\ok 0 t}\pin{k} e^{-\sigma^2(k-k_0)^2-i(k-k_0)x_t}|k\rangle. \label{t2}
\eeq
In other words, as the wave packet evolves, the position $x_0$ changes to $x_t$.  

Thus far $|k\rangle$ is an all-orders Hamiltonian eigenstate.  At various orders it contains various numbers of mesons, corresponding to the clouds of virtual mesons about the incoming meson and about the kink.  Let us restrict our attention to the one-meson, no zero mode sector at order $i$.  Then Eq.~(\ref{k01}) leads to
\bea
|t\rangle_i^{01}&=&e^{-i\ok 0 t}\pin{k} e^{-\sigma^2(k-k_0)^2-i(k-k_0)x_t}|k\rangle_i^{01} \label{t3}\\
&=&e^{-i\ok 0 t}\pin{k} e^{-\sigma^2(k-k_0)^2-i(k-k_0)x_t}\pin{k\p}\left(F(k,k\p)+\frac{R(k\p)}{k+k\p}\right)|k\p\rangle_0.\nonumber
\eea

The $|k\p\rangle_0$ states are orthogonal and the coefficient of each is
\beq
e^{-i\ok 0 t}\pin{k} e^{-\sigma^2(k-k_0)^2-i(k-k_0)x_t}\left(F(k,k\p)+\frac{R(k\p)}{k+k\p}\right).
\eeq
Now let us consider the contribution to the state from $k\sim -k\p$, corresponding to elastic scattering.  Consider first the $F$ term.  If $|x_t|$ is taken to be very large, corresponding to times well before or after the collision, then the phase oscillates so quickly that the integral vanishes as $e^{-x_t^2}$, as one would obtain by treating $F$ as a constant and performing the Gaussian integration.  
Therefore, to calculate the probability of elastic scattering we may ignore the $F$ term.

Let us consider the initial condition at $t=0$ so that $x_{t=0}=x_0\ll 0$.  In this case, we will close the contour of the $k$ integration at positive imaginary $k$, as a large semicircle there will not affect the integral.  Now recall that this integral is measuring the amplitude for the initial particle to have momentum $-k_0$.  We are interested in an initial condition in which the initial momentum is $k_0$, and so we want this integral to vanish.  This choice of initial condition dictates how we should define the pole.  

Such a contour leads to a nonzero integral unless we move the pole so that it has a small, negative imaginary part.  In other words, we define the pole so that the coefficient is
\beq
e^{-i\ok 0 t}\pin{k} e^{-\sigma^2(k-k_0)^2-i(k-k_0)x_t}\left(F(k,k\p)+\frac{R(k\p)}{k+k\p+i\epsilon}\right).
\eeq
With this choice, the initial condition does not already have a reflected meson.  

Are we allowed to change the coefficient?  By changing the pole, we have only shifted the integral at $k=-k\p$, exactly where the eigenvalue equation Eq.~(\ref{peq}) did not constrain the state.  This corresponds to another Hamiltonian eigenstate with the same energy.  Therefore, we have simply chosen another of the degenerate Hamiltonian eigenstates.  We are allowed to choose whichever Hamiltonian eigenstate we like, but the initial condition that there is no reflected meson before the scattering implies that we should use the initial condition with the $+i\epsilon$ prescription.  Of course, this statement is well-known in the context of the Lippmann-Schwinger equation, the $+i\epsilon$ convention corresponds to the {\it{in}} state.

Now that we have defined our initial state, to find the final state one need only choose a $t$ so that $x_t\gg 0$.  In this case, the integral should be evaluated with the semicircle closing in the negative complex plane.  In this case the coefficient is, up to a phase
\beq
e^{-\sigma^2(k\p+k_0)^2}R(k\p)
\eeq
and so our $i$th order primary state, at $t\gg -\ok{0}x_0/k_0$, in the one-meson Fock space, is proportional to
\beq
|t\rangle_i^{01}=\pin{k\p}e^{-\sigma^2(k\p+k_0)^2} R(k\p)|k\p\rangle_0=R(-k_0)\pin{k\p}e^{-\sigma^2(k\p+k_0)^2}|k\p\rangle_0.
\eeq

The probability of elastic scattering is then
\beq
P(k_0)=\frac{{}^{01}_{\ i}\langle t|t\rangle_{i\rm{\ red}}^{01}}{\langle t=0|t=0\rangle_{\rm red}}=|R(-k_0)|^2
\frac{\sqrt{Q_0}/(2\ok 0 2\sqrt{2\pi}\sigma)}{\sqrt{ Q_0}/(2\ok 02\sqrt{2\pi}\sigma)}=|R(-k_0)|^2
\eeq
where we have taken the leading terms in the reduced inner products.  More precisely, this is the $O(\lambda^i)$ contribution that comes from squaring the $O(\lambda^{i/2})$ amplitude.  If the lowest order contribution arises at that order, this is the whole $O(\lambda^i)$ probability.  More generally, if there are contributions to the amplitude at lower orders than $O(\lambda^{i/2})$, then in addition to the subleading term in the reduced inner product, there will also be contributions to the probability from cross terms of amplitudes at different orders. 

We conclude that the leading order probability of elastic scattering depends only on the initial momentum $k_0$ and also the function $R$ which is determined by the perturbative expansion of the states.

\subsection{Meson Multiplication}

Let us now consider meson multiplication, which is the process (\ref{mmt}) in which a meson, colliding with a kink, turns into two mesons.  The final state has two mesons and so is described by the coefficient $\gamma^{02}$ of the decomposition of the one-meson state $|\kt\rangle$.  

The first nontrivial contribution to this coefficient is at order $O(\sl)$ and is given in Eq.~(\ref{kt}).  The pole arises from the term
\beq
\frac{\sl V_{-\kt k_1 k_2}}{4\omega_{\kt}\left(\omega_{\kt}-\ok1-\ok2\right)}.
\eeq
Separating out the pole contribution and using the fact that for our nearly monochromatic wave packet $\kt>0$, we can write (\ref{kt}) as
\bea
\gamma_{1\kt}^{02}(k_1,k_2)&=&F(k_1,k_2,\kt)+\frac{R(k_1,k_2)}{\kt-k_I}
\label{mmm}\\
R(k_1,k_2)&=&\frac{\sqrt{\lambda}V_{-k_I k_1 k_2}}{4 k_I}\hsp k_I=\sqrt{(\ok 1 + \ok 2)^2-m^2}.\nonumber
\eea
Here $k_I$ is not the momentum of a particle, but is the rather chosen so that $\ok{I}=\ok1+\ok2$.  

Let us begin in the same initial state (\ref{t0}) as before.  Again at time $t$ this becomes Eq.~(\ref{t2}).  Inserting (\ref{mmm}) and following (\ref{t3}) this state can be written
\bea
|t\rangle_1^{02}&=&e^{-i\ok 0 t}\pin{k} e^{-\sigma^2(\kt-k_0)^2-i(\kt-k_0)x_t}|\kt\rangle_1^{02}\\
&=&\,\,e^{-i\ok 0 t}\,\pin{\kt}\, e^{-\sigma^2(\kt-k_0)^2-i(\kt-k_0)x_t}\,\pink{2}\,\left(F(k_1,k_2,\kt)+\frac{R(k_1,k_2)}{\kt-k_I+i\epsilon}
\right)|k_1,k_2\rangle_0.\nonumber
\eea
After the interaction $x_t>0$, meaning that the initial particle would have passed the kink to the right had it not interacted.  Therefore we close the $\kt$ integration contour on the $-i$ part of the complex plane.  Then, up to an irrelevant phase
\beq
|t\rangle_1^{02}=\pink{2}  e^{-\sigma^2(k_I-k_0)^2} R(k_1,k_2) |k_1,k_2\rangle_0.
\eeq
The  probability of meson multiplication is then
\bea
P&=&\frac{{}^{02}_{\ 1}\langle t|e^{-iH\p t}|t=0\rangle_{\rm{\ red}}}{\langle t=0|t=0\rangle_{\rm red}}=\frac{{}^{02}_{\ 1}\langle t|t\rangle_{1\rm{\ red}}^{02}}{\langle t=0|t=0\rangle_{\rm red}}\nonumber\\
&=&\pink{2}e^{-2\sigma^2(k_I-k_0)^2} |R(k_1,k_2)|^2
\frac{2\sqrt{ Q_0}/(4\ok 1\ok 2 )}{\sqrt{Q_0}/(2\ok 02\sqrt{2\pi}\sigma)}\nonumber\\
&=&\pink{2}e^{-2\sigma^2(k_I-k_0)^2}\frac{\sqrt{\pi}\sigma \ok 0}{4\sqrt{2} k_0^2 \ok 1\ok 2}\lambda |V_{-k_0k_1k_2}|^2
\eea
in agreement with Eq.~(3.47) of Ref.~\cite{memult} in the case of a reflectionless kink.  Note that while the meson multiplication probability apparently depends on the wave packet width $\sigma$, in our monochromatic limit the Gaussian factor can be replaced with a Dirac delta function and the $\sigma$ dependence disappears.

\section{Domain Walls} \label{dimsez}

Essentially every quantum field theory contains ultraviolet divergences.  Yet they have not so far featured in our lectures.  This is because we have focused on (1+1)-dimensional scalar field theories with canonical kinetic terms, and in such theories the only ultraviolet divergences that appear are those resulting from contractions of creation and annihilation operators in the same interaction.  Diagrammatically these correspond to loops that begin and end at the same point.  As we have normal ordered our Hamiltonian $H$, each interaction is already normal ordered and so there is no such contraction.

In the case of the kink sector, interactions correspond to terms in the kink Hamiltonian $H\p$.  This is normal ordered in terms of the operators $A$ that create and destroy plane waves, not the operators $B$ that create and destroy normal modes.  Therefore there are loop diagrams at a point.  However the mismatch between these normal orderings is proportional to the loop factor $\I(x)$ which, in 1+1 dimensions and likely even in 2+1 dimensions, is finite.  Therefore, again we have avoided ultraviolet divergences. 

\subsection{More Dimensions}

What happens if we move beyond 1+1 dimensions?  The simplest generalization of our Hamiltonian (\ref{hdef}) to d+1 dimensions is
\beq
H=\int d^{d}\vx :\ch(\vx):\hsp
\ch(\vx)=\frac{\pi^2(\vx)+\nabla\phi(\vx)\cdot \nabla\phi(\vx)}{2}+\frac{V(\sl\phi(\vx))}{\lambda}.
\eeq
Again we may choose a time-indepedent solution of the classical equations of motion
\beq
\phi(\vx,t)=f(\vx)
\eeq
and use it to construct a displacement operator
\beq
\df=\exp{-i\int d^d\vx f(\vx) \pi(\vx)}
\eeq
and so a soliton Hamiltonian $H\p=\df^\dag H\df$.  

Again $H\p_0$ is the classical energy, $H\p_1$ vanishes if $f(\vx)$ solves the equations of motion and, decomposing into normal modes, the free soliton Hamiltonian is
\beq
H\p_2=\frac{1}{2}\kinv{d}{k}\left(:\pi_\vk\pi_{-\vk}:+\omega_{\vk}^2:\phi_\vk\phi_{-\vk}:\right). 
\eeq
The normal ordering again yields
\beq
H\p_2=Q_1+\frac{\pi_0^2}{2}+\kinv{d}{k}\omega_{\vk}B^\ddag_{\vk}B_{\vk} \label{h2f}
\eeq
where the one-loop mass correction is
\beq
Q_1=-\frac{1}{4}\kinv{d}{k}\pinv{d}{p} \gt_{-\vk}(\vp)\gt_\vk(-\vp)\frac{(\omega_\vk-\omega_\vp)^2}{\omega_\vp}. 
\eeq

So far the situation seems like that for the kink.  However, there are two problems.  First, $Q_1$ is ultraviolet-divergent in 3+1 dimensions or more.  Second, Derrick's theorem \cite{derrick} implies that there are no such solutions if we impose that $f(\vx)$ tends a minimum of the potential $V$ when $|\vx|\rightarrow\infty$.

How do we solve these problems?  For the first problem, recall that already at one loop even the vacuum sector already enjoys ultraviolet divergences.  These divergences necessitate adding counterterms to $H$, which invariably affect $H\p$.  One must check to see whether these divergences cancel that revealed above in $Q_1$.  That will be the goal of the present section.

What about the second problem?  In 1+1 dimensions, if $|x|\rightarrow\infty$ then either $x\rightarrow \infty$ or $x\rightarrow -\infty$.  In more dimensions, there are many more directions in which $|\vx|\rightarrow\infty$.  If the field configuration does not tend to a vacuum in all of these directions then the classical energy will indeed be infinite, but that does not mean that the corresponding classical configuration is physically irrelevant.

Let us briefly describe two examples of such interesting yet infinite-energy configurations.  The first is the domain wall soliton.  This arises when the potential enjoys two global minima so that there is a kink solution $f(x)$ of the corresponding (1+1)-dimensional theory with $f(\pm\infty)$ equal to the field values at the two global minima.  In this case
\beq
F(x_1\cdots x_d)=f(x_1)
\eeq
is a domain wall solution of the (d+1)-dimensional theory.  The classical tension $\rho_0$ of this domain wall solution in d+1 dimensions is equal to the mass $Q_0$ of the kink in 1+1 dimensions.  How can two quantities with different dimensions be equal?  Note that both are of order $m^3/\lambda$ which has dimensions of [mass${}^d$], which is a mass in $1+1$ dimensions and a tension in higher dimensions.

The energy is infinite but the classical tension is constant, the infinite energy only arises from the fact that the string or domain wall is infinitely extended.  Such long strings and domain walls, that may extend beyond the horizon and so be effectively infinite, arise in many interesting cosmological models.  We will focus on these domain wall solitons below.  Note that $Q_1$ has an infrared divergence in such cases, as it receives a uniform contribution along the domain wall, yet we will see below that the correction to the tension is finite.

Now we will briefly mention a different kind of soliton that evades Derrick's theorem.  For concreteness let us restrict our attention to 2+1 dimensions, although the generalization to more dimensions is trivial.  Let us consider a model with two scalar fields that again has a potential with two global minima.  These can be connected by kinks that follow a path in field space from one minimum to the other.  In some (1+1)-dimensional models, such as the Montonen–Sarker–Trullinger–Bishop (MSTB) model \cite{mstb1,mstb2,mstb3,albertomstb} there are two such kinks $f_\pm(x)$ which, since they have the same boundary conditions, are in the same topological sector.  Each such kink can be lifted to a domain wall string $F_\pm(x,y)=f_\pm(x)$ in 2+1 dimensions.  Now the string is infinitely extended in the $y$ direction, and so again $Q_1$ is infinite.  However, at some point in the $y$ direction, a string of type $f_+$ may switch to a string of type $f_-$.  Such solutions were constructed perturbatively and also numerically in Ref.~\cite{mstbinst} in a slightly different context.  This kink-in-a-kink is localized, and the additional energy with respect to a domain wall string is finite.  As a result, the contribution to $Q_1$ from the kink-in-a-kink is also finite, once one subtracts the background energy from the domain wall string itself.  Indeed, all kink masses are only infrared finite once one has subtracted the energy of the configuration with no kink, and so the need to subtract the domain wall string mass is of no surprise.

\subsection{The Domain Wall Soliton}

The one-loop mass correction $Q_1$ is infinite for an infinitely-extended object.  Consider, for example, a domain wall which is extended in the $\vec{y}$ directions.  One may choose to derive $Q_1$ by performing the $\vec{y}$ integral at the end, in which case
\beq
Q_1=\int d^{d-1}\vec{y}\  \rho_1(\vec{y})
\eeq
where $\rho_1(\vec{y})$ is the one-loop correction to the tension \cite{noi3d}
\beq
\rho_1=\rho_1(\vec{y})=-\frac{1}{4}\ppin{k_x}\int\frac{dp_x d^{d-1}\vp_y}{(2\pi)^d}\gt_{-k_x}(p_x)\gt_{k_x}(-p_x)
\frac{(\sqrt{\ok{x}^2+\sum_i^{d-1}p_{y_i}^2}-\omega_{\vec{p}})^2}{\omega_{\vec{p}}} \label{teq}
\eeq
and we have defined the frequency
\beq
\omega_\vp=\sqrt{m^2+p_x^2+\sum_{i=1}^{d-1} p^2_{y_i}}.
\eeq
Here, unlike our convention in the $d=1$ case of kinks (\ref{ccg}), $\dint$ sums over all modes, including the kink zero mode $k_x=B$.  $\omega_{k_x}$ is defined to be $\sqrt{m^2+k_x^2}$ for continuum modes, $\omega_S$ for shape modes $k_x=S$ and $0$ for the zero mode $k_x=B$.  Note that the zero mode of the full domain wall corresponds to not just $k_x=B$ but to $\vk=(B,\vec{0})$.  This distinction leads to some important physics, for example when $d\geq 3$ the ground states are not translation invariant \cite{mezero} in line with the theorem of Ref.~\cite{sydzero}.   In string theory, D-branes are domain wall solitons and the corresponding analysis was done in Refs.~\cite{oleg1,oleg2}

The Fourier transformed normal modes $\gt$ are simply those of the (1+1)-dimensional kink solution.  Again the squared term in the numerator is the difference between the energies of the plane waves and normal modes.

In the case of the domain wall string, $d=2$, and so there is a single $y$ direction along the string.  In this case $\rho_1$ is finite.  However in the case of a more general domain wall, when $d\geq 3$, then $\rho_1$ enjoys an ultraviolet divergence.


\subsection{The Domain Wall Membrane}

Let us restrict our attention to the $\phi^4$ double-well model in 3+1 dimensions.  In more dimensions it is nonrenormalizable.  In less dimensions, it is finite at one loop.  In $3+1$ dimensions, the one-loop tension $\rho_1$ above is ultraviolet divergent.  For example, if $\Lambda$ is an ultraviolet cutoff, then in the large $\Lambda$ limit
\beq
\rho_1=-\frac{3}{16\pi^2}{\rm{ln}}\left(\Lambda\right)+{\rm{finite}}. \label{r1}
\eeq
However, the $\phi^4$ theory has other ultraviolet divergences in 3+1 dimensions, necessitating both mass and also coupling constant renormalization.  Therefore $\rho_1$ gives a divergence to the tension written in terms of the bare mass $m_0$ and coupling $\lambda_0$, but this does not imply that the tension is infinite when expressed in terms of the renormalized mass $m$ and coupling $\lambda$.

More concretely, we begin with the bare Hamiltonian
\beq
H=\int d^{d}\vx :\ch(\vx):\hsp
\ch(\vx)=\frac{\pi^2(\vx)+\nabla\phi(\vx)\cdot \nabla\phi(\vx)}{2}+\frac{V(\sqrt{\lambda_0}\phi(\vx))}{\lambda_0}
\eeq
where our potential is
\beq
\frac{V(\sqrt{\lambda_0}\phi(\vx))}{\lambda_0}=\frac{\lambda_0}{4}\left(\phi^2-\frac{m_0^2}{2\lambda_0}\right)^2.
\eeq
Then we define the renormalized mass and coupling via
\beq
m^2=m_0^2+\delta m^2\hsp \sl=\sqrt{\lambda_0}+\delta\sl
\eeq
where $\delta m^2$ and $\delta\sl$ are counterterm coefficients to be determined by a choice of renormalization conditions.

Now our classical tension is
\bea
\rho_0&=&\frac{m_0^3}{3\lambda_0}=\frac{m^3}{3\lambda}\left(1-\frac{3}{2}\frac{\delta m^2}{m^2}\right)\left(1+2 \frac{\delta \sl}{\sl}\right)+O(\lambda)\nonumber\\
&=&\frac{m^3}{3\lambda}+m^3\left(-\frac{1}{2}\frac{\delta m^2}{\lambda m^2}+\frac{2}{3} \frac{\delta \sl}{\lambda \sl}\right)+O(\lambda).
\eea
The order $O(1/\lambda)$ term is written in terms of the renormalized $m$ and $\lambda$ which, with any reasonable renormalization condition, will be finite.  We will not attempt to compute the $O(\lambda)$ corrections, which arise at two loops.  Instead, we will ask whether the order $O(\lambda^0)$ terms in the tension are finite.

Needless to say, $\rho_1$ also contributes at order $O(\lambda^0)$.  Therefore, the total contribution to the tension at order $O(\lambda^0)$ is 
\beq
\rho_1+m^3\left(-\frac{1}{2}\frac{\delta m^2}{\lambda m^2}+\frac{2}{3} \frac{\delta \sl}{\lambda \sl}\right). \label{o1}
\eeq
The question is, for any renormalization condition leading to a finite $m$ and $\lambda$, is this quantity finite?

This question now has nothing to do with solitons.  The mass and coupling must anyway be renormalized in the vacuum sector.  The renormalization there is achieved by fixing a renormalization condition and then enforcing it on $\delta\sl$ and $\delta m^2$ by computing one loop diagrams in ordinary, vacuum sector, perturbation theory.  Once the model is renormalized in the vacuum sector, $\lambda_0$ and $m_0$ are fixed as a function of our regulator.  We then have no more freedom to remove divergences in the soliton sector.  One might hope that the choice of $\df$ gives us more freedom to remove divergences, but in fact any shift in $\df$ will be compensated with a shift in $\vac$ once we impose that our state $\df\vac$ is a Hamiltonian eigenstate, and so such a shift cannot help to remove divergences.

Now choose your favorite vacuum sector renormalization condition. For example, impose that the rest mass of the meson is $m$ including all loop corrections, and that, for some choice of external momenta, the four-point coupling with all loop corrections is $\lambda/4$.  For any such renormalization condition which leads to observables that are finite in terms of $m$ and $\lambda$, if $\Lambda$ is a momentum cutoff, then in the large $\Lambda$ limit the counterterm coefficients in the $\phi^4$ model will be
\beq
\frac{\delta m^2}{\lambda m^2}=-\frac{9}{8\pi^2}{\rm{ln}}\left(\Lambda\right)+{\rm{finite}}\hsp \frac{\delta \sl}{\lambda \sl}=-\frac{9}{16\pi^2}{\rm{ln}}\left(\Lambda\right)+{\rm{finite}}. \label{ct}
\eeq

These lead to a positive and divergent contribution to the $O(\lambda^0)$ tension in Eq.~(\ref{o1}).  This divergence was observed by Coleman in his Erice lectures \cite{erice}.  He used it along with other divergences that arise at 2 loops in 2+1 dimensions, to argue that, above 1+1 dimensions solitons are not described by coherent states.  While the two-loop divergences have not yet been studied in the present framework, now we see that at one loop the divergence in fact vanishes.  Inserting Eqs.~(\ref{ct}) and (\ref{r1}) into the $O(\lambda^0)$ tension (\ref{o1}) we find that the divergent terms cancel, leaving a finite result.  For various choices of renormalization condition, this finite result has been calculated several times in the literature \cite{fin1,fin2,fin3,fin4,fin5}.  In Ref.~\cite{noifin} it was recalculated using LSPT.  Refs.~\cite{gw25} and \cite{noiweigel} showed that all of these computations led to mutually consistent results except for Refs.~\cite{fin4,fin5}.

Let us summarize our results.  The usual renormalization (\ref{ct}) led to a positive, infinite contribution to the tension.  However, the one-loop correction (\ref{r1}) due to the squeezing led to a negative, infinite contribution that removed the divergence.  In less dimensions, the squeeze led to $Q_1$ or $\rho_1$ which was always negative, however it was finite.  If one created a kink in a coherent state with no squeeze, it would emit a meson or two and relax to its ground state, perhaps with some stable shape mode excitations.  On the other hand, in 3+1 dimensions the excitation energy density of the unsqueezed state is infinite, and so one is forced to squeeze.  Indeed, it is common in the study of solitons that corrections which are finite in low dimensions become infinite as the dimension increases.

In summary, up to corrections suppressed by powers of the coupling, in 1+1 dimensions, coherent states describe slightly excited solitons which, in time $O(1/m)$, relax to the lower energy squeezed, coherent states, which better approximate the true ground states.  The same is likely true in 2+1 dimensions once one includes the perturbative renormalization which is anyway needed in the vacuum sector.  In 3+1 dimensions, on the other hand, the squeeze is mandatory to arrive at a finite energy density.

\section{Summary}

We have presented a novel formalism, LSPT, and shown how it may be used to obtain novel results, whose derivation would have been prohibitively complicated with older techniques.  

The key step was the decomposition (\ref{decprin}) of the fields in terms of normal modes.  A decomposition  of the fields in terms of the collective coordinates such as the kink position $x_0$, and in terms of the amplitudes of oscillations, is very complicated, consisting of an infinite series of terms.  Here we have instead provided a linear decomposition which is exact and simple.  The price of this linear decomposition is that $\phi_0$ is only proportional to the kink position $x_0$ at leading order, and $\pi_0$ to its momentum at leading order.  Similarly, $\phi_S$ is the shape mode amplitude only at leading order, and $\phi_k$ are the unbound meson perturbation amplitudes only at leading order.  

The connection between each of these and the corresponding collective coordinate or excitation amplitude will have quadratic and higher corrections.  However these corrections only rarely caused complications.  For example, they caused a correction to the reduced inner product, but this correction consisted of a single term.  On the other hand, the efficiency gained by the simple and exact decomposition (\ref{decprin}) was considerable, as the decomposed operators on the right hand side obey canonical commutation relations.

Another critical component of our formalism was the displacement operator $\df$, constructed from a classical soliton solution $f$, which we used to introduce a soliton Hamiltonian
\beq
H\p=\df^\dag H\df
\eeq
where $H$ is the regularized Hamiltonian that defines the theory.  We pulled the $\df$ off of each state, which was the most nonperturbative part, and the price was that the Hamiltonian $H$ is replaced by $H\p$.  With $\df$ pealed off of the state, $H\p$ generates time evolution and its eigenvalue is the energy.  Since $H$ and $H\p$ are unitarily equivalent and so have the same spectrum as a function of the regulator, we were able to reliably compare quantities computed in the soliton and vacuum sectors, to compute quantum corrections to the soliton mass, which are the difference between eigenvalues of $H\p$ and $H$.

\section{What Next?}

LSPT is a novel tool which allows many robust computations that would have been impractical using older formalisms.  So far it has only been formulated in a rather restrictive case of the semiclassical expansion about time-independent solutions in scalar quantum field theories.  In fact, beyond 1+1 dimensions, these have only been understood at one loop.

The Sine-Gordon breather, which is a time-dependent soliton, was quantized nearly half a century ago in Ref.~\cite{dhnsg}.  Formally, we have described how LSPT can be applied to time-dependent solutions in Ref.~\cite{kehinde}.  In the near future we would like to use this formalism to quantize the oscillon, with the hope of demonstrating that its rapid radiation emission, claimed in Refs.~\cite{hertzberg} and \cite{tanmay}, is simply a small but short burst of energy as it relaxes to its ground state rather than a decay into pure radiation.  In other words, we hope to show that the quantum oscillon's life time is much longer than is currently believed.


\subsection{What Can Be Done Now?}

With the formalism in its current state, many interesting generalizations and applications are already within reach.

First of all, one needs to address the concerns of Refs.~\cite{erice} and more recently \cite{cocorr23} that coherent states will fail at two loops.  The naive divergences at one loop were canceled by the squeeze.  At two loops, the divergences are suppressed by powers of the coupling and so, following the argument of Ref.~\cite{fk78} should be removed by the same counterterms that remove them in the vacuum sector.  Thus we believe that our two-loop corrections will be finite, but it remains to test this.  An ideal testing ground is the domain wall string in 2+1 dimensions, which has a logarithmic divergence requiring mass renormalization that one may hope will cancel the divergent $|V|^2$ term in Eq.~(\ref{q2pad}).

What about two loops in 3+1 dimensions?  Here there are quadratic divergences and so we are not sure how to proceed.  Thus far the only regulator that we have used is a momentum cutoff, which is sufficient for logarithmic divergences but can lead to counterterms that are not Lorentz-invariant in the case of polynomial divergences.  In the literature one usually relies on dimensional regularization in such cases.  Dimensional regularization has been shown to produce the correct one-loop quantum corrections to domain wall masses \cite{dimreg}.  However, following the general arguments of Ref.~\cite{lit}, such an approach may lead to errors as one does not know how the classical solution should be modified in a noninteger dimension.  Therefore, we do not know how the displacement operator should be treated in dimensional regularization.  

Our approach, for now, is to stick with logarithmic divergences.  After all, Nature appears to be kind.  In the Standard Model, most quadratic divergences are eliminated by gauge-invariance.  The only one that remains is that resulting from the quartic Higgs coupling, but this coupling in fact has never been measured.  There are linear divergences arising from anomalies, but these are rather special. 

Another obvious generalization is the inclusion of fermions.  Kinks in theories with fermions have a rich phenomenology \cite{ferm1,ferm2,ferm3,ferm4}.  These include supersymmetric kinks, which have two advantages.  First, one and two-loop corrections are available \cite{shifsusy} in the literature, providing a sanity test for our results.  Next, it would be interesting to understand how nonrenormalization theorems are encoded in the structure of the states.  Just as we saw that translation-invariance fixes the descendant parts of the states, each supercharge will fix a number of coefficients.  The question is, as the amount of supersymmetry increases \cite{nastase} how many coefficients need to be fixed before we can say something meaningful about strong coupling?

Beyond adding spin $1/2$ particles, clearly it is of interest to include gauge fields.   The holy grail in this field is the quantization of the 't Hooft-Polyakov monopole.  And one may proceed further, with the inclusion of spin two fields one may quantize the gravitating kink \cite{gk1,gk2,gk3}.

Besides pushing to more general quantum field theories, one may also work towards removing the perturbative crutch.  The first step in this direction is to understand instanton corrections at weak coupling.  The simplest instanton corrections are the mixing between degenerate kinks in the same topological sector.  We have calculated the leading correction to the corresponding splitting in Ref.~\cite{mstbinst}, which required finding the instanton that interpolates between the kinks.  This result was generalized to the supersymmetric case in Ref.~\cite{shiffinst}.  If the normal modes of the instanton can be found, then one can compute the subleading correction.  Incidentally, this instanton in 2+0 dimensions is also the kink-in-a-kink solution of the (2+1)-dimensional MSTB model.  Thus, a calculation of its normal modes could also be used to compute the quantum corrections to the mass of this rather novel solution.

\subsection{Dreams}

The research projects mentioned in the previous subsection can be completed now, by the interested reader of these lecture notes.

Yet our motivation lies in two very long term applications.  One of the most important problems in quantum field theory is understanding the vacua of Yang-Mills theory and QCD.  It is believed that the confinement in these two theories results from some unbeknownst order in these vacua, which saves energy but then is destroyed by the presence of net color charge, requiring the energy to be paid back.  One of the most popular candidates for this order \cite{thooftconf,mandconf} is the condensation of 't Hooft-Polyakov monopoles.

Now, we have spent these entire lecture notes explaining that quantum solitons consist of classical solitons plus quantum corrections.  However neither Yang-Mills nor QCD possesses any classical solution which resembles a monopole.  So what could this mean?

The meaning is clear in a cousin of QCD, with adjoint quarks and some additional fields and couplings tuned so as to give an enhanced symmetry.  This toy model is known as $\mathcal{N}=2$ superQCD.  Here, Seiberg and Witten have shown \cite{sw2} that indeed the vacuum contains a condensate of monopoles which causes confinement.  And so the idea is not so crazy.

Now the interesting observation is that the theory which confines here also has no classical monopole solutions.  So what does it mean that a monopole condenses?  It means that the theory has a parameter, called the bare hypermultiplet mass.  When the parameter is large, there is a semiclassical monopole.  However, as all semiclassical monopoles are, it is very heavy and so does not condense.  As the parameter is tuned to zero, the monopole becomes massless, and then a small, soft perturbation\footnote{Several distinct perturbations can be used for this purpose, such as a Fayet-Iliopoulos term or a polynomial superpotential.} causes it to condense, leading to confinement.

In other words, we have one theory with monopoles and one with confinement.  However, one can interpolate between these theories by varying the bare mass and critically, for each value of the bare mass, the monopole is BPS.  This means that the monopole is created by an operator which commutes, or anticommutes, with half of the supercharges.  This BPS property is critical.  It means that the operator is uniquely defined for all values of $m$ from large $m$ where it is essentially the displacement operator which creates a monopole, to small $m$, where it condenses and causes confinement.

At large $m$ we know the operator, indeed these lecture notes more or less describe how to construct it.  The flow to small $m$ is unique, although the coupling is not always small.  Thus the challenge is to construct this operator at each $m$.  If one can do it, one will know that one has done it because it will (anti)commute with half of the supercharges, at least up to terms annihilating the vacuum.  In other words, acting it on a state which is annihilated by all of the supercharges, one arrives at a state annihilated by half of the supercharges.

Can this be done?  We have seen that translation invariance greatly simplifies the construction of states.  The question is whether $\mathcal{N}=2$ supersymmetry simplifies it enough to allow us to go beyond perturbation theory.  Even if it does not, then supersymmetry will fix the state up to a primary part.  The nonrenormalization theorem tells us that this primary part does not affect the monopole mass, since we know that this is only affected by instanton corrections.  And so perhaps this situation will be sufficient.

If one constructs a monopole operator in $\mathcal{N}=2$ QCD, then one will have constructed a monopole operator in a theory with no monopoles.  This is just what is needed in Yang-Mills and QCD.  And so one may try to understand its topology in the space of operators to see how it works.  More naively, since Yang-Mills is a subsector of superQCD, one may use this operator, removing the fields that do not feature in Yang-Mills or QCD, to create an Ansatz for a monopole operator in Yang-Mills and QCD.  Then it remains to test on the lattice to see whether this operator actually condenses.

While the above may seem like a hopeless longshot, there are yet more remote dreams.  For example, the claimed oscillon decay in Ref.~\cite{hertzberg} is a decay via pairs of fundamental particles.  The radiation arises from the fact that the oscillon state considered in the literature was coherent but not squeezed.  Once the squeeze is included, the radiation turns off.  Conversely if the squeeze is not included, one can wait until the radiation finishes and then the oscillon will be relaxed to its squeezed ground state.  In other words, the radiation was an artifact of neglecting the one-loop correction to the oscillon ground state.  A decay into pairs of elementary quanta of course is well-known in another context, from various horizons.  One may therefore attempt to compute one-loop corrections to states in such cases to see whether, also in some such cases, the one-loop corrected states are similar to the uncorrected states and yet do not radiate.  Thus one may wonder whether, like unsqueezed solitons,  the usual black hole states radiate only for a moment and then reach their true, stable ground states.

\appendix
\section{The Contraction} \label{capp}

In Eq.~(\ref{idef}) we introduced the contraction factor
\beq
\I(x)= \ :\phi^2(x):-:\phi^2(x):_b. \label{idefb}
\eeq
In this Appendix, we will show that $\I(x)$ is given by Eq.~(\ref{ival}).

This needs to be done carefully as the two terms on the right hand side of Eq.~(\ref{idefb}) are potentially individually ill-defined, or more precisely some of their matrix elements contain divergent momentum integrals.  Indeed, inserting the decompositions (\ref{adec}) and (\ref{dec}) one finds respectively
\bea
:\phi^2(x):&=&\int\frac{d^2p}{(2\pi)^2}e^{-i(p_1+p_2)x}\left(A^\ddag_{p_1}A^\ddag_{p_2}+\frac{A^\ddag_{p_1}A_{-p_2}}{\omega_{p_2}}+\frac{A_{-p_1}A_{-p_2}}{4\omega_{p_1}\omega_{p_2}}
\right)\label{phi2}\\
:\phi^2(x):_b&=&\phi_0^2 \g_B^2(x)+\g_B(x)\ppin{k}\g_k(x)\left(2\phi_0B^\ddag_k+\frac{\phi_0B_{-k}}{\ok{}}\right)\nonumber\\
&&+\ppink{2}\g_{k_1}(x)\g_{k_2}(x)\left( 
B^\ddag_{k_1}B^\ddag_{k_2}+\frac{B^\ddag_{k_1}B_{-k_2}}{\omega_{k_2}}+\frac{B_{-k_1}B_{-k_2}}{4\omega_{k_1}\omega_{k_2}}
\right).\nonumber
\eea

Our strategy to evaluate $\I(x)$ will be to insert the Bogoliubov transform (\ref{ibog}) into this expression for $:\phi^2(x):$, so that $\I(x)$ is an integral over continuous values of $k$ plus a sum over discrete shape modes, also indexed by $k$.  As the two normal ordering prescriptions differ by commutators, they are $c$-numbers.  The final expression will therefore be an integral and sum over $k$, and any potential divergences will arise in this integral or sum.

As $\I(x)$ arises entirely from commutators, this calculation can be performed quickly by noting that, after writing $:\phi^2(x):$ in the $\{\phi_0,\pi_0,B^\ddag,B\}$\ basis, the only contributions to $\I(x)$ will arise from those terms that are not normal ordered.  In other words, we need only compute the $\pi_0\phi_0$ and $B B^\ddag$ terms in $:\phi^2(x):$, and these contributions to $\I(x)$ will be proportional to $[\pi_0,\phi_0]$ and $[B,B^\ddag]$ respectively.  We therefore decompose
\beq
\I(x)=\I_B(x)+\I_C(x)
\eeq
where $\I_B(x)$ consists of contributions from $[\pi_0,\phi_0]$ and $\I_C(x)$ of the contributions from $[B,B^\ddag]$.

Substituting the Bogoliubov transform (\ref{ibog}) into the first line of Eq.~(\ref{phi2}) and keeping the $\pi_0\phi_0$ terms one finds
\bea
\I_B(x)&=&[\pi_0,\phi_0]\int\frac{d^2p}{(2\pi)^2}e^{-i(p_1+p_2)x}\tilde{\g}_B(-p_1)\tilde{\g}_B(-p_2)\left[
\left(-\frac{i}{2\omega_{p_1}}\right)\left(\frac{1}{2}\right)\right.\\
&&\left.+
\left(-\frac{i}{2\omega_{p_1}}\right)
\left(\omega_{p_2}\right)
\left(\frac{1}{\omega_{p_2}}\right)
+
\left(i\right)
\left(\omega_{p_2}\right)
\left(\frac{1}{4\omega_{p_1}\omega_{p_2}}\right)\right]\nonumber\\
&=&-\int\frac{d^2p}{(2\pi)^2}e^{-i(p_1+p_2)x}\frac{\tilde{\g}_B(-p_1)\tilde{\g}_B(-p_2)}{2\omega_{p_1}}
.\nonumber
\eea
Similarly, the $B^\ddag B$ terms yield
\bea
\I_C(x)&=&\int\frac{d^2p}{(2\pi)^2}e^{-i(p_1+p_2)x}\ppink{2}\tilde{\g}_{k_1}(-p_1)\tilde{\g}_{k_2}(-p_2)\left[ B_{-k_1},B^\ddag_{k_2}\right] \\
&&\times\left[
\left( \frac{\omega_{p_1}-\ok1}{4\ok 1\omega_{p_1}} \right) 
\left( \frac{\ok 2+\omega_{p_2}}{2\omega_{p_2}} \right)
+
\left( \frac{\omega_{p_1}-\ok1}{4\ok 1\omega_{p_1}} \right) 
\left( \omega_{p_2}-\ok 2 \right)
\left( \frac{1}{\omega_{p_2}} \right)\right.\nonumber\\
&&\left.
+
\left( \frac{\omega_{p_1}+\ok 1}{2\ok 1} \right) 
\left( \omega_{p_2}-\ok 2 \right)
\left( \frac{1}{4\omega_{p_1}\omega_{p_2}} \right)
\right] \nonumber\\
&=&\int\frac{d^2p}{(2\pi)^2}e^{-i(p_1+p_2)x}\ppin{k}\frac{\tilde{\g}_{k}(-p_1)\tilde{\g}_{-k}(-p_2)}{4\ok{}\omega_{p_1}\omega_{p_2}} \left[
\left( {\omega_{p_1}-\ok{}}{} \right) 
\left( \frac{\ok {}+\omega_{p_2}}{2} \right)\right.\nonumber\\
&&\left.
+
\left( {\omega_{p_1}-\ok{}} \right) 
\left( { \omega_{p_2}-\ok {}} \right)
+
\left( \frac{\omega_{p_1}+\ok {}}{2} \right) 
\left({\omega_{p_2}-\ok {}} \right)
\right] \nonumber\\
&=&\int\frac{d^2p}{(2\pi)^2}e^{-i(p_1+p_2)x}\ppin{k}{\tilde{\g}_{k}(-p_1)\tilde{\g}_{-k}(-p_2)} \left[\frac{1}{2\ok{}}-\frac{1}{2\omega_{p_1}}
\right]. \nonumber
\eea

The Fourier transform of the completeness relation (\ref{cr}) is the completeness relation
\beq
\tilde{\g}_B(p_1)\tilde{\g}_B(p_2)+\ppin{k}\tilde{\g}_{k}(p_1)\tilde{\g}_{-k}(p_2)=2\pi\delta(p_1+p_2).
\eeq
Adding $\I_B(x)$ and $\I_C(x)$ and using the completeness relation one concludes
\bea
\I(x)&=&\int\frac{d^2p}{(2\pi)^2}e^{-i(p_1+p_2)x}\left[ \ppin{k}\frac{{\tilde{\g}_{k}(-p_1)\tilde{\g}_{-k}(-p_2)}}{2\ok{}}
-\frac{2\pi\delta(p_1+p_2)}{2\omega_{p_1}}
\right]\nonumber\\
&=&\ppin{k}\frac{{\g}_{k}(x){\g}_{-k}(x)}{2\ok{}}-\int\frac{dp}{2\pi}\frac{1}{2\omega_p}. \label{idiff}
\eea
This is the difference between two divergent integrals.  So is it meaningless?

If the kink solution is smooth, then at high momenta the kink will have little effect on the normal modes.  As a result, the normal modes $\tilde{\g}_k(p)$ are supported when $p$ is close to $k$, or more precisely when $k-p$ is of order the meson mass $m$.  This means that the integrals over $k$ and $p$ appearing in $\I_B$ and $\I_C$ are supported where $k$ and $p$ are close.  We conclude that both integrals may be handled with the same cutoff, $\Lambda$, in the sense that any other double-scaling limit of the two cutoffs would give the same answer as the region where $|k-p|$ is large has an exponentially-suppressed integrand.  Once $k$ and $p$ are integrated over the same range, the integrals may be combined, by renaming the dummy variable $p$ to the continuum variable $k$, so that the second term contributes to the continuum part of the first term and one finds
\beq
\I(x)=\pin{k}\frac{{g}_{k}(x){g}_{-k}(x)-1}{2\ok{}}+\sum_S\frac{{g}^2_{S}(x)}{2\omega_S}. \label{ifin}
\eeq
A more careful treatment of such ultraviolet divergences is provided in Ref.~\cite{andy}.  There the theory is made manifestly ultraviolet finite from the beginning with a cutoff in the momentum $p$, equivalent to a lattice regularization.  One finds that this leads not to a cutoff in $k$, but rather to a shift in the normal modes which acts as a gentle cutoff smeared over the momentum scale $m$.

In general, the difference between two logarithmically divergent integrals is rarely problematic so long as the difference between the integration variables in the support of the integration remains finite, because any finite shift in the cutoff $\Lambda$ will shift the integral by $1/\Lambda$ which vanishes.  However, in 2+1 dimensions, the two terms in Eq.~(\ref{idiff}) will instead be linearly divergent.  In any dimension the integrand in Eq.~(\ref{ifin}) scales as $1/k^3$ and so the integral is only divergent in 3+1 dimensions, where anyway one encounters other quadratic divergences.  Therefore, it would be interesting to repeat the analysis of Ref.~\cite{andy} to see whether in 2+1 dimensions $\I(x)$ continues to be described by Eq.~(\ref{ifin}).  

The argument for (\ref{ifin}) in Ref.~\cite{me2loop} does not use the order of the divergence, but simply the fact that the second term in Eq.~(\ref{idiff}) is $x$-independent but $\I(x)$ should vanish as $|x|\rightarrow\infty$ because the kink should not affect the meson interactions far away\footnote{Recently an interesting counterexample to this intuition has appeared in Ref.~\cite{confkink}.}.  As a result, the second term can be fixed by simply shifting the first term so that it vanishes at $|x|\rightarrow\infty$, leading to the integrands in Eq.~(\ref{ifin}) in any dimension. 

\section{Examples of the $\I(x)$, $\Delta$ and $V$ Symbols}
\label{sapp}
Recall that we have introduced two decompositions of fields into operators $A^\ddag_p$ and $A_p$ which create and destroy plane waves and also operators $\{B^\ddag_k,B_k,B^\ddag_S,B_S,\pi_0,\phi_0\}$ where the first four create and destroy normal modes, $\pi_0$ is proportional to the center of mass momentum and $\phi_0$ is, at the linear level, proportional to the center of mass position.  We also introduced corresponding normal ordering prescriptions.  Plane wave normal ordering places all $A^\ddag_p$ to the left of $A_p$ while normal mode normal ordering places all $B^\ddag_k$,\ $B^\ddag_S$ and also $\phi_0$ to the left. The contraction factor $\I(x)$, which intuitively can be represented by a loop beginning and ending at the same vertex, is defined to be the difference between the plane wave and normal mode normal-ordered forms of $\phi^2(x)$.  It is given in general in Eq.~(\ref{ival}).

In Eq.~(\ref{deldef}) we introduced the symbol $\Delta_{k_1k_2}$, where $k_1$ and $k_2$ run over all discrete and continuum modes.  It represents the matrix elements of the momentum operator $P$ written in the $\{\phi_k,\pi_k\}$ basis.  Later, in Eq.~(\ref{vdef}), we introduced the symbol $V_{\I^m k_1\cdots k_n}$.  This is the interaction vertex factor for an interaction at a point relating the modes $k_i$, with $m$ internal loops beginning and ending at the vertex.  Again, here $k_i$ may be a shape mode, a meson or even a zero mode.  However, in the case of zero modes, the Ward Identities of Subsec.~\ref{wardsez} determine the interaction vertices $V$ in terms of lower order interaction vertices with no zero modes or the translation matrix elements $\Delta$.

These three symbols can be calculated given the normal modes and the potential of any model.  In the present Appendix we will provide them in the case of the popular Sine-Gordon and $\phi^4$ double-well models.  We will not give formulas for the interaction vertices involving zero modes as these are easily computed from the Ward Identities.

The results below can be easily derived using the integrals
\bea
\int dx e^{-ikx}\sech^{2n}(\b x)&=&\left\{
\begin{array}{cl}
2\pi\delta(k) &  {\rm{\ \ \ if}}\  n=0 \\ \frac{\pi}{(2n-1)!k}\left[\prod_{j=0}^{n-1}\left(\frac{k^2}{\b^2}+(2j)^2\right)\right]\ck   & {\rm{\ \ \ if}}\ n>0
\end{array}
\right.\nonumber\\
\int dx e^{-ikx}\sech^{2n+1}(\b x)&=& \frac{\pi}{(2n)!\b}\left[\prod_{j=0}^{n-1}\left(\frac{k^2}{\b^2}+(2j+1)^2\right)\right]\sk \nonumber\\
\int dx e^{-ikx}\sech^{2n}(\b x)\tanh(\b x)&=&-i\frac{\pi}{(2n)!\b}\left[\prod_{j=0}^{n-1}\left(\frac{k^2}{\b^2}+(2j)^2\right)\right]\ck\nonumber  \\
\int dx e^{-ikx}\sech^{2n+1}(\b x)\tanh(\b x)\,\,&=&\,\,\, -i \frac{\pi k}{(2n+1)!\b^2}\left[\prod_{j=0}^{n-1}\left(\frac{k^2}{\b^2}+(2j+1)^2\right)\right]\sk .
\eea

\subsection{The Sine-Gordon Model}

The Sine-Gordon model (\ref{sgpot}) enjoys a zero mode (\ref{sgz}) and continuum modes (\ref{sgm}).  The results below are largely taken\footnote{In these notes, $\g_B$ and $\g_k$ are equal to, in the notation of \cite{me2loop}, $-g_B$ and $ig_k{\rm{sign}}(k)$ respectively.} from Ref.~\cite{me2loop}.  The contraction factor $\I(x)$ is
\beq
\I(x)=-\frac{\sech^2(mx)}{2\pi}.
\eeq
The momentum matrix elements are
\bea
\Delta_{kB}&=&\frac{\pi\ok{}}{\sqrt{8m}}{\rm{sech}}\left(\frac{k\pi}{2m}\right)\label{dkk}\\
\Delta_{k_1k_2}&=&i(k_1-k_2)\pi\delta(k_1+k_2)+\frac{i\pi}{2}\frac{(k_1^2-k_2^2)}{\omega_{k_1}\omega_{k_2}}{\rm{csch}}\left(\frac{\pi\left(k_1+k_2\right)}{2m}\right). \nonumber
\eea

Using the third and fourth derivatives of the potential evaluated at the Sine-Gordon soliton solution $f(x)$ in Eq.~(\ref{sgs}), one finds the cubic and quartic interactions in the kink Hamiltonian
\beq
V^{(3)}[\sqrt{\lambda} f(x)]=2 m  \tanh (m x)\sech (m x)\hsp
V^{(4)}[\sl f(x)]=  -1+2\sech^2(mx).
\eeq
The independent interaction vertices appearing at cubic order are then
\bea
V_{\I k}&=&-\frac{\ok{}^3}{8m^3} \sech\left( \frac{k\pi}{2m}\right)\\
V_{k_1k_2k_3}&=&\frac{\pi (\ok 1+\ok 2+\ok 3)}{4m\ok 1\ok 2\ok 3}(\ok 1+\ok 2-\ok 3)(\ok 1-\ok 2+\ok 3)\nonumber\\
&&\times(-\ok 1+\ok 2+\ok 3)
\sech\left( \frac{(k_1+k_2+k_3)\pi}{2m}\right).\nonumber
\eea
Note that on-shell scattering using the three-point function will lead to one $\omega_k$ being the sum of the other two, and so the corresponding factor in parentheses will vanish, leading to a vanishing three-point coupling.  This is the tree-level manifestation of the familiar fact that meson multiplication, corresponding to the process
\beq
{\rm{soliton\ +\ meson}}\rightarrow {\rm{soliton\ +\ 2\ mesons}} 
\eeq
is forbidden in the Sine-Gordon model as a result of its integrability.  To compute leading order elastic scattering, one also needs the four-point coupling 
\beq
V_{\I kk}=\frac{k(4k^2+m^2)}{15m^4}\csch\left( \frac{k\pi}{m}\right).
\eeq

\subsection{The $\phi^4$ Double-Well Model}

The $\phi^4$ model enjoys a zero mode (\ref{p4z}), a shape mode (\ref{p4s}) and continuum modes (\ref{p4m}).  The following results are taken\footnote{The normal modes $\g_B$ and $\g_S$ here are equal to $-g_B$ and $ig_S$ in Ref.~\cite{mephi4}.} from Ref.~\cite{mephi4}.

The general formula (\ref{ival}) for the contraction factor $\I(x)$ yields
\beq
\I(x)=\frac{1}{4\sqrt{3}}\sech^2\mx\tanh^2\mx-\frac{3}{8\pi}\sech^4\mx. \label{ieq}
\eeq
The momentum matrix elements are
\bea
\Delta_{SB}&=&\pi \frac{3m}{16\sqrt{2}}\hsp
\Delta_{kB}=-i\pi {\sqrt{\frac{3}{2}}}{}\frac{k^2\ok{}}{m^{3/2}\sqrt{m^2+4k^2}}\ckm\\\Delta_{kS}&=&\pi \frac{\sqrt{3}}{16}\frac{(3m^2+4k^2)\sqrt{m^2+4k^2}}{m^{3/2}\ok{}}\skm\nonumber\\
\Delta_{k_1k_2}&=&i \pi (k_1-k_2)\delta(k_1+k_2)+3i \pi \left(\frac{\ok{1}}{\ok{2}}-\frac{\ok{2}}{\ok{1}}\right)\frac{m^2+k_1^2+k_2^2}{\sqrt{m^2+4k_1^2}\sqrt{m^2+4k_2^2}}\csch\left(\frac{\pi(k_1+k_2)}{m}\right).\nonumber
\eea

Using the derivatives of the potential evaluated at the $\phi^4$ kink solution $f(x)$ of Eq.~(\ref{ksol}), one obtains the cubic and quartic interactions of the kink Hamiltonian
\beq
\V3=3\sqrt{2\lambda}m\ \tanh\mx\hsp
\V4=6\lambda.
\eeq
With these in hand, one may evaluate the various $n$-point interactions $V$.  The simplest at each order contain a loop beginning and ending at the same interaction
\bea
V_{\I\I}&=&\frac{\lambda}{35m}\left(1-\frac{4\sqrt{3}}{\pi}+\frac{54}{\pi^2}\right)\\
V_{\I k}&=&i\frac{\sqrt{\lambda}}{2\sqrt{6}}\frac{k^2\omega_k}{m^4\sqrt{m^2+4k^2}}\left[2\pi(-m^2+2k^2)+6\sqrt{3}\omega_k^2\right]\ckm\nonumber\\
V_{\I S}&=&-\frac{3\sqrt{\lambda}}{64\sqrt{2}}\sqrt{m}(3\sqrt{3}-2\pi).\nonumber
\eea

While those containing the zero mode are related to lower interactions or momentum matrix elements by the Ward identities, those containing the shape mode are finite and independent
\bea
V_{SSS}&=&\pi\frac{9\sqrt{3\lambda}}{32\sqrt{2}}m^{3/2}\hsp
V_{kSS}=-i\pi\frac{{3\sqrt{\lambda}}}{\sqrt{2}}\frac{k^2\ok{}(m^2-2k^2)}{m^3\sqrt{m^2+4k^2}}\ckm\\
V_{k_1k_2S}&=&\pi \frac{3\sqrt{3\lambda}}{32\sqrt{2}}\frac{\left(17m^4-16(\ok1^2-\ok2^2)^2\right)(m^2+4k_1^2+4k_2^2)+128m^2k_1^2k_2^2}{m^{3/2}\ok1\ok2\sqrt{m^2+4k_1^2}\sqrt{m^2+4k_2^2}}\sech\left(\frac{\pi(k_1+k_2)}{m}\right).\nonumber
\eea

The three meson coupling is divergent when the momenta sum to zero, in other words when no momentum is exchanged with the kink.  It therefore needs to be handled carefully.  Let us first decompose it into terms which are $x$ integrals with integrands that have $2I$ powers of $\sech$ and $J$ powers of $\tanh$
\beq
V_{k_1k_2k_3}=\sum_{I=0}^3\sum_{J=0}^1 V_{k_1k_2k_3}^{IJ}.
\eeq
Of these, only those terms with $I=0$ are divergent, and so the others can be summed into $V_{k_1k_2k_3}^{F}$
\beq
V_{k_1k_2k_3}=V_{k_1k_2k_3}^{00}+V_{k_1k_2k_3}^{01}+V_{k_1k_2k_3}^{F}.
\eeq
The first term contains a Dirac $\delta$ function divergence, the second a simple pole.  The finite term is
\bea
V_{k_1k_2k_3}^F
&=&\cc_{k_1k_2k_3}\frac{4\pi\kkk}{m^2}\csk\\
&&\times\sum_{I=1}^3\frac{2^{2I-2}}{(2I-1)!}\left(\Phi_{k_1k_2k_3}^{I0}-i\frac{\kkk}{I m}\Phi_{k_1k_2k_3}^{I1}\right)\prod_{j=1}^{I-1}\left(\frac{\left(\kkk\right)^2}{m^2}+j^2\right)\nonumber
\eea
where we have defined the coefficient
\beq
\cc_{k_1k_2k_3}=\frac{3\sqrt{2\lambda}m}{\ok1\ok2\ok3\sqrt{m^2+4k_1^2}\sqrt{m^2+4k_2^2}\sqrt{m^2+4k_3^2}}.
\eeq
Defining the symmetrized products
\bea
S_1^n&=&k_1^n+k_2^n+k_3^n\hsp 
S_2^n=(k_1k_2)^n+(k_1k_3)^n+(k_2k_3)^n\hsp
S_3^n=(k_1k_2k_3)^n\nonumber\\
S_2^{mn}&=&k_1^mk_2^n+k_1^mk_3^n+k_2^mk_3^n+k_1^nk_2^m+k_1^nk_3^m+k_2^nk_3^m
\eea
the $\Phi$ symbols are
\bea
\Phi_{k_1k_2k_3}^{00}&=&{3im}{}\left[-m^4S_1^1+m^2\left(2S_2^{21}+9S_3^1\right)-4S_3^1S_2^1\right]\\
\Phi_{k_1k_2k_3}^{10}&=&{3im}{}\left[4m^4S_1^1+m^2\left(-5S_2^{21}-18S_3^1\right)+4S_3^1S_2^1\right]\nonumber\\
\Phi_{k_1k_2k_3}^{20}&=&\frac{9im^3}{4}\left[-7m^2S_1+4S_2^{21}+12S_3^1\right]\hsp \Phi_{k_1k_2k_3}^{30}=\frac{27im^5}{4}S_1^1\nonumber\\
\Phi_{k_1k_2k_3}^{01}&=&-{m^6}{}+{m^4}{}(9S_2^1+2S_1^2)+2{m^2}{}(-2S_2^2-9S_3^1S_1^1)+8S_3^2
\nonumber\\
\Phi_{k_1k_2k_3}^{11}&=&\frac{3m^2}{2}\left[3m^4+m^2(-15S_2^1-4S_1^2)+(4S_2^2+12S_3^1S_1^1)\right]
\nonumber\\
\Phi_{k_1k_2k_3}^{21}&=&\frac{9m^4}{4}\left[-3m^2+(6S_2^1+2S_1^2)\right]
\hsp
\Phi_{k_1k_2k_3}^{31}=\frac{27m^6}{8}.
\nonumber
\eea
The case $\sum_i k_i=0$ is often useful to consider separately, even for the finite terms.  In this case they reduce to
\beq
\hat{V}_{k_1k_2}^F=V_{k_1,k_2,-k_1-k_2}^F={\cc}_{k_1,k_2,-k_1-k_2}\frac{4}{m}\hp_{k_1k_2}^{10}.
\eeq
Defining the symmetrized products
\beq
\hat{S}_2=\frac{k_1^2+k_2^2+(k_1+k_2)^2}{2}\hsp
\hat{S}_3=k_1k_2(k_1+k_2)
\eeq
the reduced coefficient is
\beq
 \hp_{k_1k_2}^{10}={3im}{} \hat{S}_3 \left(3m^2+4\hat{S}_2\right).
\eeq

The divergent terms are
\beq
V_{k_1k_2k_3}^{00}=\cc_{k_1k_2k_3}\Phi_{k_1k_2k_3}^{00}2\pi\delta(k_1+k_2+k_3)
\eeq
and
\beq
V_{k_1k_2k_3}^{01}=-i\cc_{k_1k_2k_3}\Phi_{k_1k_2k_3}^{01}\frac{2\pi}{m}\csk .
\eeq

\section*{Acknowledgements}

\noindent
This work was supported by the Higher Education and Science Committee of the Republic of Armenia (Research Project No. 24RL-1C047).  We thank Christoph Adam, Bilguun Bayarsaikhan, Sujoy Mahato, Kehinde Ogundipe and in particular Hui Liu for proofreading these lecture notes.

\end{document}